\documentclass[traditabstract]{aa}

\usepackage{paralist}
\usepackage{rotating}

\usepackage{graphicx,natbib}
\usepackage{txfonts}

\begin{document}

\title{\emph{WISE} morphological study of Wolf-Rayet nebulae}

\author{J.A. Toal\'{a}\inst{1}, M.A. Guerrero\inst{1}, G. Ramos-Larios\inst{2},
  \and V. Guzm\'{a}n\inst{2}}

\institute{
Instituto de Astrof\'{\i}sica de Andaluc\'{\i}a, IAA--CSIC, 
Glorieta de la Astronom\'\i a s/n, E-18008, Granada, Spain;
\underline{toala@iaa.es} \and 
Instituto de Astronom\'{\i}a y Meteorolog\'{\i}a, 
Av.\ Vallarta No.\ 2602, Col.\ Arcos Vallarta, C.P. 44130, 
Guadalajara, Jalisco, Mexico.
}

\abstract{We present a morphological study of nebulae around
  Wolf-Rayet (WR) stars using archival narrow-band optical and {\it
    Wide-field Infrared Survey Explorer} (\emph{WISE}) infrared
  images. The comparison among \emph{WISE} images in different bands
  and optical images proves to be a very efficient procedure to
  identify the nebular emission from WR nebulae, and to disentangle it
  from that of the ISM material along the line of sight. In
  particular, WR nebulae are clearly detected in the \emph{WISE} W4
  band at 22 $\mu$m. Analysis of available mid-IR {\it Spitzer}
  spectra shows that the emission in this band is dominated by thermal
  emission from dust spatially coincident with the thin nebular shell
  or most likely with the leading edge of the nebula. The WR nebulae
  in our sample present different morphologies that we classified into
  well defined WR bubbles (bubble ${\cal B}$-type nebulae), clumpy
  and/or disrupted shells (clumpy/disrupted ${\cal C}$-type nebulae),
  and material mixed with the diffuse medium (mixed ${\cal M}$-type
  nebulae).  The variety of morphologies presented by WR nebulae shows
  a loose correlation with the central star spectral type, implying
  that the nebular and stellar evolutions are not simple and may
  proceed according to different sequences and time-lapses.  We report
  the discovery of an obscured shell around WR\,35 only detected in
  the infrared.}


\authorrunning{Toal\'a et al.}

\titlerunning{\textit{WISE} morphological study of WR nebulae}

\maketitle

\section{Introduction}
Massive stars ($M_{\mathrm{i}}\gtrsim$\,30\,$M_{\odot}$) end up their
lives as Wolf-Rayet\,(WR) stars. Before entering the WR phase, these
stars evolve through the red or yellow supergiant (RSG or YSG) or
luminous blue variable (LBV) phases, depending on the initial stellar
mass, and eject their envelopes via copious slow winds expanding at
10--100~km~s$^{-1}$. When a fast stellar wind
(1000--2,000~km~s$^{-1}$) develops during the WR phase, it sweeps up
the previously ejected slow RSG/LBV wind material to form a WR
nebula. The swept-up circumstellar material is photoionized by the
central star and has temperatures of $\sim$10$^{4}$~K.

WR nebulae present different morphological features such as bubbles,
blowouts, clumps, filaments, and diffuse emission. These nebulae are
observable at all frequencies of the electromagnetic spectrum: radio
\citep[e.g.,][]{Arnal1996,Arnal1999,Cappa2002,Cappa2008,Cappa2009},
infrared \citep[IR;
  e.g.,][]{vanBuren1988,Gv2010,Mauerhan2010,Wachter2010,Flagey2011,Stringfellow2012,Wachter2011},
optical
\citep[e.g.,][]{Chu1982,Treffers1982,Chu1983,Gruendl2000,Stock2010,Fernandez-Martin2012},
and X-rays
\citep[e.g.,][]{B1988,Wrigge1994,Wrigge1999,Chu_etal2003,Wrigge2002,Wrigge2005,Zhekov2011,Toala2012,Toala2015}.

\citet{Chu1981} started a series of works proposing a coherent
morphological classification of nebulae associated with WR stars 
as: 
$R$ - radiatively excited H~{\sc ii} regions, 
$E$ - stellar ejecta, and 
$W$ - wind-blown bubbles 
\citep[see][for an updated review of the morphology of WR nebulae]{Chu2003}. 
The $R$-type nebulae present optical emission lines with widths 
comparable to those in ordinary H~{\sc ii} regions, and are 
further divided into amorphous H~{\sc ii} regions ($R_{\mathrm{a}}$) 
and shell-structured H~{\sc ii} regions ($R_{\mathrm{s}}$).  
The $E$-type nebulae, characterized by their highly clumpy appearance 
and irregular velocity field, were proposed to form through sudden 
episodes of mass ejection.  
Finally, the $W$-type nebulae present a thin sheet of gas and filaments 
curving around a WR star which is close to the geometric center of the 
nebula or offset toward the brightest rim.  
$W$-type nebulae are predicted by numerical simulations of the circumstellar 
medium (CSM) around WR stars
\citep[e.g.,][]{GS1996a,GS1996b,Freyer2003,Freyer2006,Toala2011}.
Using this classification scheme, \citet{Stock2010} extended and
revised the morphologies of a sample of southern WR nebulae.

\citet{Chu1981} found a correlation, later confirmed by \citet{Chu1983}, 
between nebular morphology and spectral type of the central WR star: 
WC stars are mostly associated with $R_{\mathrm{s}}$ nebulae, whereas the 
central stars of $W$ nebulae are mostly early WN stars and those of $E$ 
nebulae are exclusively of spectral type WN8. 
This correlation was interpreted in terms of nebular evolution, so that 
$W$ nebulae, at an early nebular phase, will precede $R_{\mathrm{s}}$ 
nebulae \citep{Chu1983}.

The classification scheme described above is very demanding
observationally.  It requires both direct line imaging and
spectroscopic campaigns to investigate the morphology, kinematics, and
abundances of the ionized material in order to disentangle the
emission of the WR nebulae from that of the ISM. In the recent years,
mid-IR observations have proved very useful to study WR nebulae or to
discover new nebulae around evolved stars
\citep[e.g.,][]{Wachter2010}. In particular, \textit{Spitzer}
observations in the 24~$\mu$m band have been found to be sensitive to
the WR nebular emission, whereas the near-IR $K$ and mid-IR 8~$\mu$m
bands trace mainly the emission of material along the line of sight
\citep[e.g.,][]{Wachter2010,Stringfellow2012}. Mid-IR observations of
WR nebulae are thus a promising tool to study these objects.

Here we present \textit{Wide-field Infrared Survey Explorer}
\citep[\emph{WISE};][]{WISE2010} images of a sample of 31 WR stars
and compare them with narrow-band optical images (Section~2).  
This observational approach provides a straightforward method to identify
nebulae around WR stars, clearly separating its emission from that of the 
ISM.  
The nature of the emission of WR nebulae in different \emph{WISE} bands is 
investigated using mid-IR \emph{Spitzer} spectra (Section~3).  
Their morphologies have been broadly grouped into three morphological 
types that can be interpreted in the framework of the evolution of the 
previously ejected dense wind through different evolutive paths 
(Section~4). 

\section{Observations}

We have searched the \textit{WISE} Preliminary Release Database made
public in April 2011 for IR images of stars in the VIIth catalogue of
Galactic Wolf-Rayet stars \citep{vdH2001} and found a sample of 31 WR
stars with available data\footnote{The mid-IR \emph{WISE} images of
  the nebulae around WR\,43abc, WR91, and WR\,93 (NGC\,3603, RCW\,122,
  and NGC\,6357, respectively) were also examined. These stars belong
  to young stellar clusters, and thus their nebulae are influenced not
  only by the WR star, but by all stars in the cluster. Their
  resulting morphologies is complex and difficult to classify.
  Furthermore, their \emph{WISE} W3 and W4 band images were saturated
  at their central regions.  Accordingly, we discarded these nebulae
  from our analysis.}. To compare their IR and optical morphologies,
we have selected H$\alpha$ line images of these nebulae in the Super
COSMOS Sky Survey \citep{Parker2005}, or STScI Digitized Sky Survey
(DSS)\footnote{https://archive.stsci.edu/cgi-bin/dss\_form.} from the
Second Palomar Observatory Sky Survey (POSS-II).  For a few cases, we
have used the [O~{\sc iii}] and H$\alpha$ images presented in
\citet[][]{Gruendl2000} that have been kindly provided by Y.-H.\ Chu
and R.A.\ Gruendl. The names and coordinates of the sources in our
sample are listed in Table~\ref{tab:spectral} together with the
distance, height over the Galactic Plane, and the spectral type and
wind terminal velocity of their central
stars. Table~\ref{tab:spectral} also identifies the telescope or
public database used to acquire the optical images used in this
paper. Note that for the specific cases of WR\,7 and WR\,136 we have
also used the existing DSS blue band to complement the morphology seen
in the DSS red band\footnote{The rest of the WR nebulae do not have
  their corresponding DSS blue band images.}.

The \emph{WISE} images of the WR nebulae in Table~\ref{tab:spectral}
were downloaded from the \emph{WISE} Image search tool at the
NASA/IPAC Infrared Science Archive\,(IRSA).  
To study the IR morphology of these nebulae, we will use the 
W2 ($\lambda_\mathrm{c}$=4.6\,$\mu$m), 
W3 ($\lambda_\mathrm{c}$=12\,$\mu$m), and 
W4 ($\lambda_\mathrm{c}$=22\,$\mu$m) bands. 
The spatial resolutions for the W2, W3, and W4 bands are 
6\farcs1, 6\farcs4, 6\farcs5, and 12\farcs0, respectively, 
with an astrometric accuracy for bright sources better than 
0\farcs15.

%\begin{sidewaystable}
\begin{table*}
%\begin{table}
%\tiny
\scriptsize
\caption{Stellar and Nebular Parameters of the WR Sample$^{\mathrm{a}}$}
\centering
\begin{tabular}{lllcrcrcrlc}
\hline\hline\noalign{\smallskip}
\multicolumn{1}{c}{WR~~~}&
\multicolumn{1}{c}{Nebula$^{\mathrm{b}}$}&
\multicolumn{1}{c}{R.A.}&
\multicolumn{1}{c}{Dec.}&
\multicolumn{1}{c}{Galactic Coord.}&
\multicolumn{1}{c}{$d$}&
\multicolumn{1}{c}{$z$}&
\multicolumn{1}{c}{Spectral}&
\multicolumn{1}{c}{$v_{\infty}$}&
\multicolumn{1}{c}{Optical}&
\multicolumn{1}{c}{Nebular} \\
\multicolumn{1}{c}{} &
\multicolumn{1}{c}{} &
\multicolumn{1}{c}{} &
\multicolumn{1}{c}{} &
\multicolumn{1}{c}{$l,b$} &
\multicolumn{1}{c}{} &
\multicolumn{1}{c}{} &
\multicolumn{1}{c}{Type} &
\multicolumn{1}{c}{} &
\multicolumn{1}{c}{Observation$^{\mathrm{c}}$} &
\multicolumn{1}{c}{Type} \\
\multicolumn{1}{c}{} &
\multicolumn{1}{c}{} &
\multicolumn{2}{c}{(J2000)} &
\multicolumn{1}{c}{($^\circ$)} &
\multicolumn{1}{c}{(kpc)} &
\multicolumn{1}{c}{(pc)} &
\multicolumn{1}{c}{} &
\multicolumn{1}{c}{(km~s$^{-1}$)}&
\multicolumn{1}{c}{} &
\multicolumn{1}{c}{} \\
%\midrule
\hline
\noalign{\smallskip}
 ~~6    & S\,308       & 06 54 13.05 & $-$23 55 42.1 & 234.76$-$10.08~~ & 1.50 & $-$262 & WN4-s     & 1700  ~~& CTIO (H$\alpha$, [O~{\sc iii}]) & ${\cal B}$\\
 ~~7    & NGC\,2359    & 07 18 29.13 & $-$13 13 01.5 & 227.75$-$0.13~~~ & 3.67 &   $-$8 & WN4-s     & 1600  ~~& SuperCOSMOS/DSS &${\cal B}$\\
 ~~8    &              & 07 44 58.22 & $-$31 54 29.6 & 247.07$-$3.79~~~ & 3.47 & $-$229 & WN7/WCE+? & 1590  ~~& SuperCOSMOS &${\cal C}$\\
 ~16    & Anon         & 09 54 52.91 & $-$57 43 38.3 & 281.08$-$2.55~~~ & 2.37 & $-$105 & WN8h      &  650  ~~& SuperCOSMOS &${\cal B}$\\
 ~18    & NGC\,3199    & 10 17 02.28 & $-$57 54 46.9 & 283.57$-$0.97~~~ & 2.20 &  $-$37 & WN4-s     & 1800  ~~& DSS        &${\cal C}$\\
 ~22    & Anon         & 10 41 17.52 & $-$59 40 36.9 & 287.17$-$0.85~~~ & 3.24 &  $-$48 &WN7h+O9III-V& 1785 ~~& DSS        &${\cal M}$ \\
 ~23    & Anon         & 10 41 38.33 & $-$58 46 18.8 & 286.78$-$0.03~~~ & 3.24 &   $-$2 & WC6       & 2340  ~~& DSS        &${\cal B}$ \\
 ~30    & Anon         & 10 51 06.01 & $-$62 17 01.8 & 284.44$-$2.61~~~ & 5.83 & $-$263 & WC6+O6-8  & 2100  ~~& SuperCOSMOS &${\cal M}$\\
 ~31a   &              & 10 53 59.66 & $-$60 26 44.3 & 288.94$-$0.81~~~ & 8.0  & $-$113 & WN11h     & 365   ~~& SuperCOSMOS &${\cal B}$\\
% ~31b   &              & 10 56 11.58 & $-$60 27 12.8 & 289.18$-$0.70~~~ & 6.1  &  $-$75 & WN11h     &90-400 ~~& SuperCOSMOS &${\cal M}$\\
 ~35    & Anon         & 11 00 22.10 & $-$61 13 51.0 & 289.97$-$1.19~~~ & 17.87& $-$371 & WN6h-w    & 1100  ~~& SuperCOSMOS &${\cal C}$\\
 ~35b   & Anon         & 11 00 02.30 & $-$60 14 01.0 & 289.63$-$0.24~~~ & 2.19 &   $-$9 & WN4       &\dots  ~~& SuperCOSMOS &${\cal M}$\\
 ~38    & Anon         & 11 05 46.52 & $-$61 13 49.1 & 290.57$-$0.92~~~ & 5.83 &  $-$94 & WC4       & 3200  ~~& SuperCOSMOS &${\cal M}$\\
 ~40    & RCW\,58      & 11 06 17.21 & $-$65 30 35.2 & 292.31$-$4.83~~~ & 2.26 & $-$190 & WN8h      &  650  ~~& CS (H$\alpha$, [O~{\sc iii}]) &${\cal C}$\\
% ~43abc & NGC\,3603    & 11 15 07.60 & $-$61 15 38.0 & 291.62$-$0.52~~~ & 10.0 &  $-$92 & WN6ha     & 2700  ~~& DSS        & \dots \\
 ~52    & Anon         & 13 18 28.00 & $-$58 08 13.6 & 306.50$+$4.54~~~ & 1.51 &    120 & WC4       & 3225  ~~& SuperCOSMOS &${\cal M}$ \\
 ~54    & Anon         & 13 32 43.79 & $-$65 01 27.9 & 307.27$-$2.50~~~ & 7.53 & $-$321 & WN5-w     & 1500  ~~& SuperCOSMOS &${\cal M}$ \\
 ~55    & RCW\,78      & 13 33 30.13 & $-$62 19 01.2 & 307.80$+$0.16~~~ & 6.03 &     17 & WN7 (WNE-w)& 1200 ~~& SuperCOSMOS &${\cal C}$ \\
 ~68    & G320.5$-$1.4 & 15 18 21.0~ & $-$59 38 10.0 & 320.54$-$1.88~~~ & 3.27 & $-$107 & WC7       & 2100  ~~& SuperCOSMOS &${\cal M}$ \\
 ~75    & RCW\,104     & 16 24 26.23 & $-$51 32 06.1 & 332.84$-$1.48~~~ & 2.18 &  $-$56 & WN6-s     & 2300  ~~& CS (H$\alpha$, [O~{\sc iii}]) &${\cal B}$\\
 ~85    & RCW\,118     & 17 14 27.13 & $-$39 45 47.0 & 347.43$-$0.61~~~ & 4.66 &  $-$50 & WN6h-w (WNL)&1400 ~~& SuperCOSMOS &${\cal C}$ \\
 ~86    & RCW\,130     & 17 18 23.06 & $-$34 24 30.6 & 352.25$+$1.85~~~ & 2.86 &     92 & WC7(+B0III-I)&1855 ~~& DSS        &${\cal M}$ \\
% ~91    & RCW\,122     & 17 20 22.00 & $-$38 56 47.0 & 348.76$-$1.07~~~ & 7.18 & $-$134 & WN7 (WNE-s)& 1700 ~~& SuperCOSMOS & \dots \\
% ~93    & NGC\,6357    & 17 25 08.88 & $-$34 11 12.8 & 353.23$+$0.83~~~ & 1.74 &     25 & WC7+O7-9  & 2600  ~~& DSS        & \dots \\
 ~94    & Anon         & 17 33 07.14 & $-$33 38 23.7 & 354.60$-$0.25~~~ & 3.12 &  $-$14 & WN5-w     & 1300  ~~& SuperCOSMOS &${\cal M}$ \\
 ~95    &              & 17 36 19.76 & $-$33 26 10.9 & 355.13$-$0.70~~~ & 2.09 &  $-$26 & WC9d      & 1900  ~~& SuperCOSMOS &${\cal C}$ \\
 101    & Anon         & 17 45 09.10 & $-$31 50 16.0 & 357.47$-$1.43~~~ & 3.18 &  $-$79 & WC8       &\dots  ~~& SuperCOSMOS &${\cal C}$ \\
 102    & G2.4+1.4     & 17 45 47.00 & $-$26 10 29.0 &   2.38$+$1.41~~~ & 5.56 &    137 & WO2       & 5000  ~~& SuperCOSMOS &${\cal B}$ \\
 113    & RCW\,167     & 18 19 07.36 & $-$11 37 59.2 &  18.91$+$1.75~~~ & 1.79 &     55 & WC8d+O8-9IV& 1700 ~~& DSS        &${\cal M}$ \\
 116    & Anon         & 18 27 04.28 & $-$12 22 52.3 &  19.16$-$0.32~~~ & 2.48 &  $-$14 & WN8h      & 800   ~~& SuperCOSMOS &${\cal C}$ (${\cal M}$?) \\
 124    & M1-67        & 19 11 30.88 & $+$16 51 38.2 &  50.20$+$3.31~~~ & 3.36 &    194 & WN8h      & 710   ~~& DSS        &${\cal C}$\\
 128    & Anon         & 19 48 32.20 & $+$18 12 03.7 &  55.62$-$3.79~~~ & 9.37 & $-$619 & WN4(h)-w  & 2050  ~~& MLO (H$\alpha$, [O~{\sc iii}]) &${\cal B}$\\
 131    &              & 20 00 19.12 & $+$33 15 51.1 &  69.90$+$1.71~~~ & 11.78&    352 & WN7h      & 1400  ~~& DSS        &${\cal C}$\\
 134    & Anon         & 20 10 14.20 & $+$36 10 35.1 &  73.45$+$1.55~~~ & 1.74 &     47 & WN6-s     & 1700  ~~& MLO (H$\alpha$, [O~{\sc iii}]) &${\cal C}$\\
 136    & NGC\,6888    & 20 12 06.55 & $+$38 21 17.8 &  75.48$+$2.43~~~ & 1.26 &     53 & WN6(h)-s  & 1600  ~~& DSS (Red, Blue) &${\cal B}$\\
\hline
\hline
\end{tabular}
\begin{list}{}{}
\item{$^{\mathrm{a}}$ All parameters have been adopted from
    \citet{vdH2001}, but the spectral classification and terminal
    velocities have been revised according to \citet{HGL06} and
    \citet{SHT12}.}
\item{$^{\mathrm{b}}$The name Anon is given to those nebulae that have
  not been catalogued as independent objects nebula, as those that are
  a small portion of a larger nebular region
  \citep[see][]{Chu1981,Chu1983}.}
\item{$^{\mathrm{c}}$ CTIO: Cerro Tololo Inter-American Observatory,
    SuperCOSMOS: Super COSMOS Sky Survey, DSS: STScI Digitized Sky
    Survey, CS: Curtis-Schmidt, and MLO: Mount Laguna Observatory.}
\end{list}
\label{tab:spectral}
%\end{table}
\end{table*}
%\end{sidewaystable}

\subsection{IRS spectra}
\label{sec:spectra}

To help us interpret the nature of the IR emission of WR nebulae in
the \emph{WISE} W3 and W4 bands, we have searched for \emph{Spitzer}
InfraRed Spectrograph (IRS) spectroscopic observations in the
10--37~$\mu$m range of the WR nebulae in our sample listed in
Table~\ref{tab:spectral}.  We found available spectroscopic
observations for a sample of 15 WR nebulae. Whereas the detailed
analysis of these spectra will be presented in a subsequent paper
(Toal\'{a} et al., in prep.), here we will focus on the nebulae around
WR\,7 (NGC\,2359), WR\,8, and WR\,31a, because there are available
high-dispersion spectroscopic observations for these nebulae, as well
as for the background emission from apertures located at their
periphery.

The basic calibrated data (BCD) of the IRS observations of these WR
nebulae were downloaded from the \emph{Spitzer} Heritage Archive.
These data are processed with the \emph{Spitzer} Science Center
Pipeline Version S\,18.18.0, and include high-resolution spectroscopic
observations (R$\sim$600) obtained using the Short-High (SH, 9.9--19.6
$\mu$m) and Long-High (LH, 18.7--37.2 $\mu$m) modules.  The targets
were observed using apertures with sizes of 13\farcs6$\times$4\farcs5
in the SH module and 22\farcs3$\times$8\farcs9 in the LH module.  All
the IRS data were reduced using the CUbe Builder for IRS Spectra Maps
(\emph{CUBISM}).  This tool does not only reduce the IRS data, but it
can also be used to analyze the IRS data and to extract
one-dimensional spectra. The various processing steps followed within
\emph{CUBISM}, including the characterization of noise in the data and
the removal of bad pixels, are described by \citet{Smith2007}.  We
note that the \emph{Spitzer} IRS spectra of WR\,8 and WR\,31a include
the stellar continuum from their WR stars, as they were included in
the aperture given the small angular size of these nebulae.

\section{Interpreting the IR emission of WR nebulae}
\label{sec:interpreting}

We present in Figures~\ref{fig:WR6} to \ref{fig:WR35b} and the figures
in Appendix~\ref{sec:appendix} the optical and \emph{WISE} IR images
of the WR nebulae listed in Table~\ref{tab:spectral}. The inspection
of the colour-composite IR pictures of WR nebulae presented in the
figures (see Figures~\ref{fig:WR6} to \ref{fig:WR35b}) reveals the
prevalence of the emission in the \emph{WISE} W4 band at 22 $\mu$m
(red in these colour-composite pictures).  The emission in this band
is mostly coincident with the optical H$\alpha$ emission from ionized
nebular material (e.g., WR\,16 in Fig.~\ref{fig:WR6}-bottom
panels). This is in agreement with \emph{Spitzer} MIPS 24 $\mu$m
imaging of nebulae around evolved massive stars
\citep[e.g.,][]{Gv2009,Gv2010,Wachter2010,Flagey2011}.  On the other
hand, the emission in the W3 band (green in the colour-composite
pictures), which is sensitive to cold gas or low ionization material,
seems to trace mostly ISM gas along the line of sight (e.g.,
Fig.~\ref{fig:WR35b}), with little contribution of emission from the
WR nebula in some cases (e.g., S\,308 around WR\,6,
Fig.~\ref{fig:WR6}).  Some WR nebulae, however, may show bright
emission in this band from discrete knots and clumps interior to the
nebular shell (e.g., NGC\,6888, Fig.~\ref{fig:WR136}).  Finally, the
W2 band (blue in the colour-composite pictures), which is expected to
trace the continuum emission from small grains and the background
stellar component \citep[see][]{Flagey2011}, shows emission from stars
in the background.  For the most dusty cases, the nebulae around
WR\,7, WR\,18, WR\,22, and WR\,23 the W2 band also includes some
nebular emission.

\begin{figure}
\begin{center}
\includegraphics[angle=0,height=0.5\linewidth]{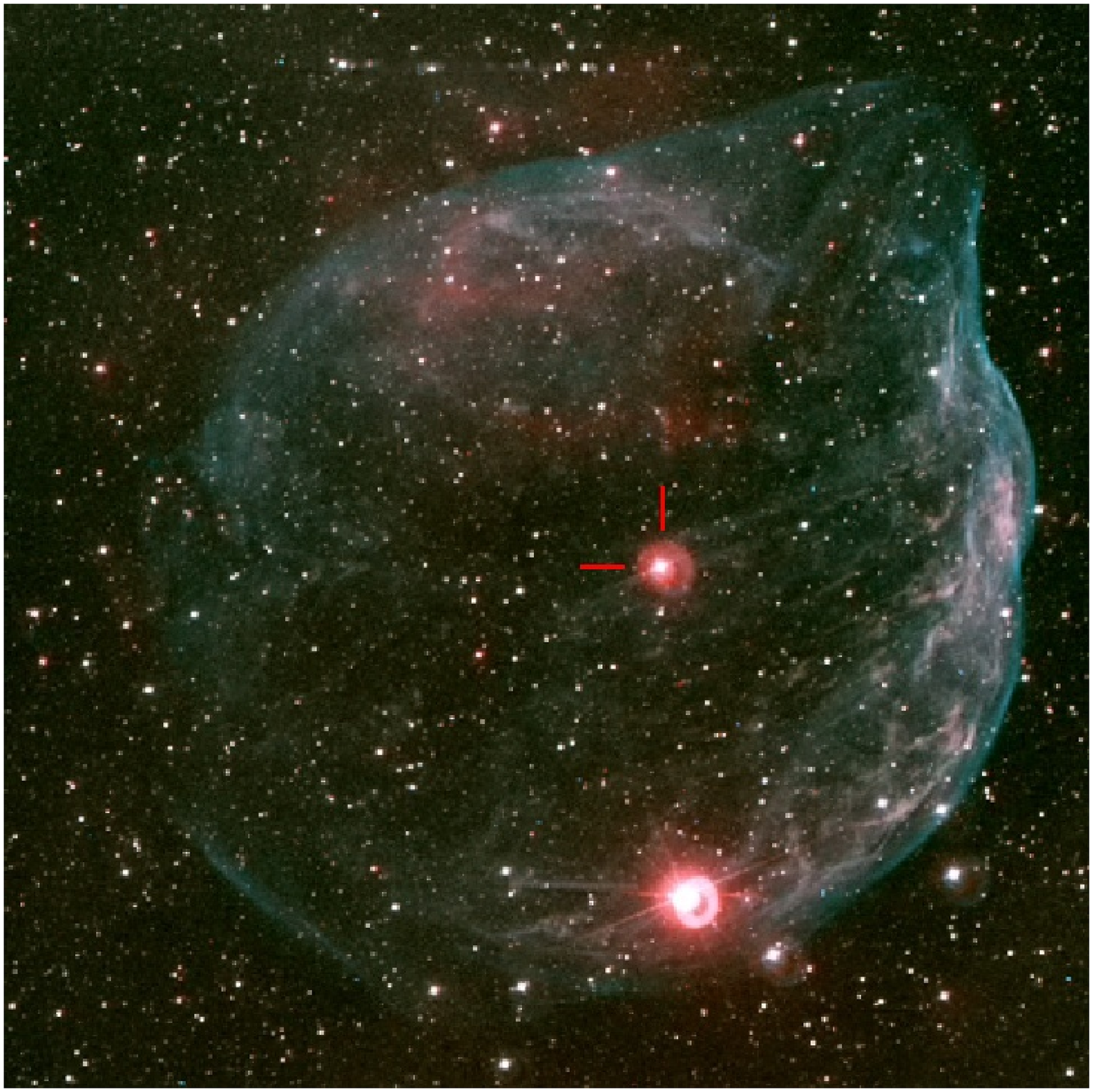}~
\includegraphics[angle=0,height=0.5\linewidth]{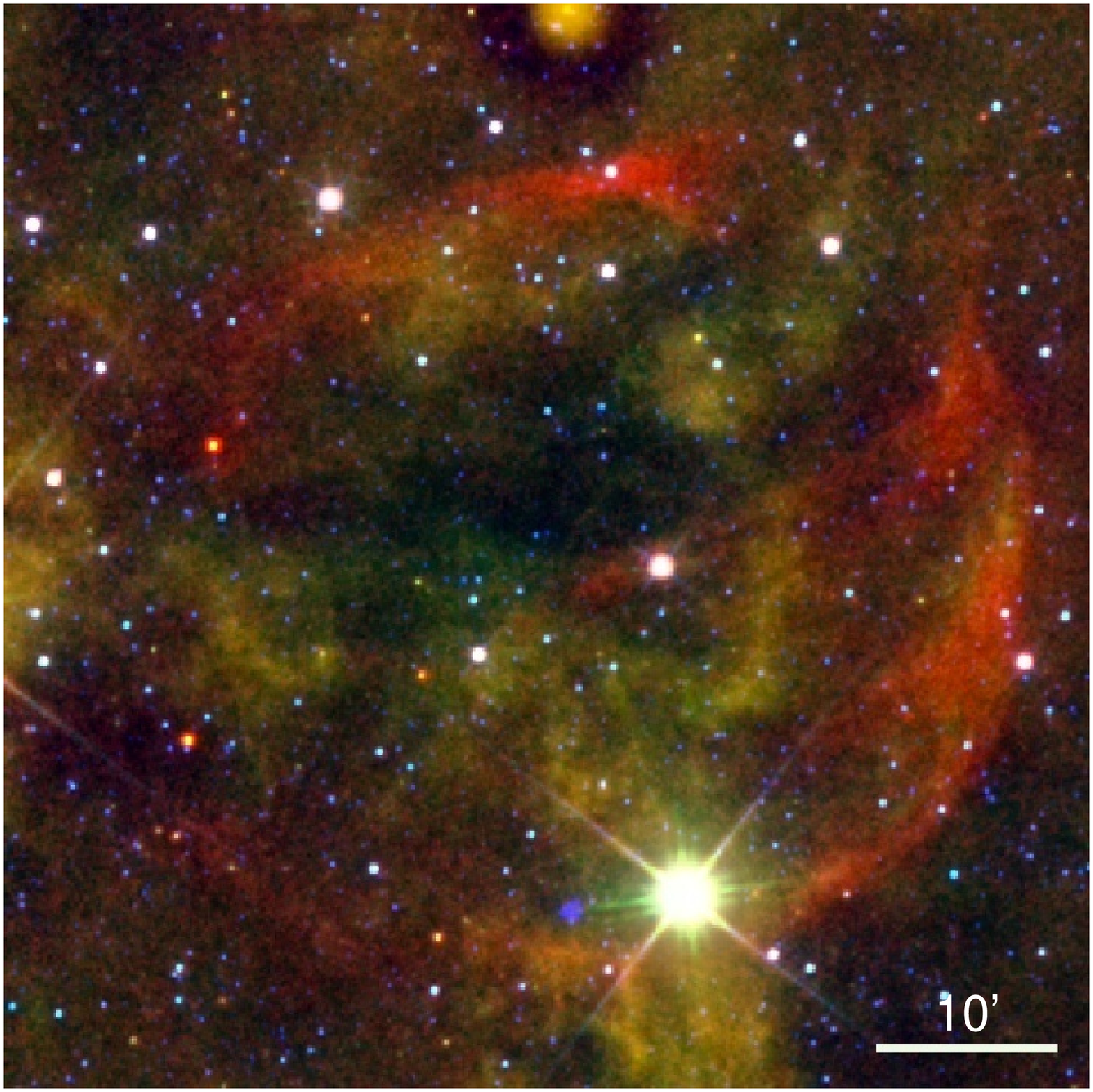}
\\
\vspace*{0.1cm}
\includegraphics[width=0.5\linewidth]{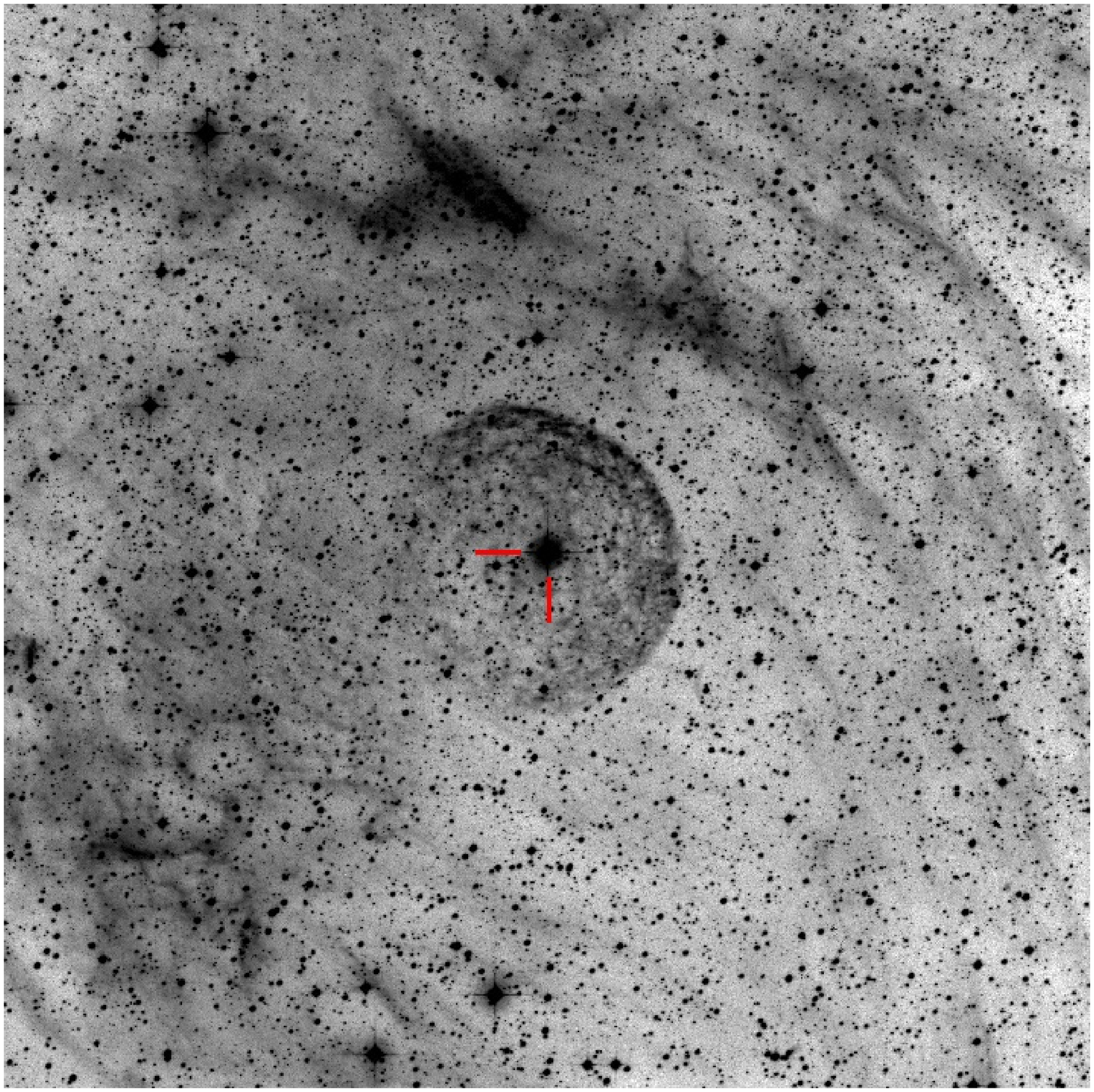}~
\includegraphics[width=0.5\linewidth]{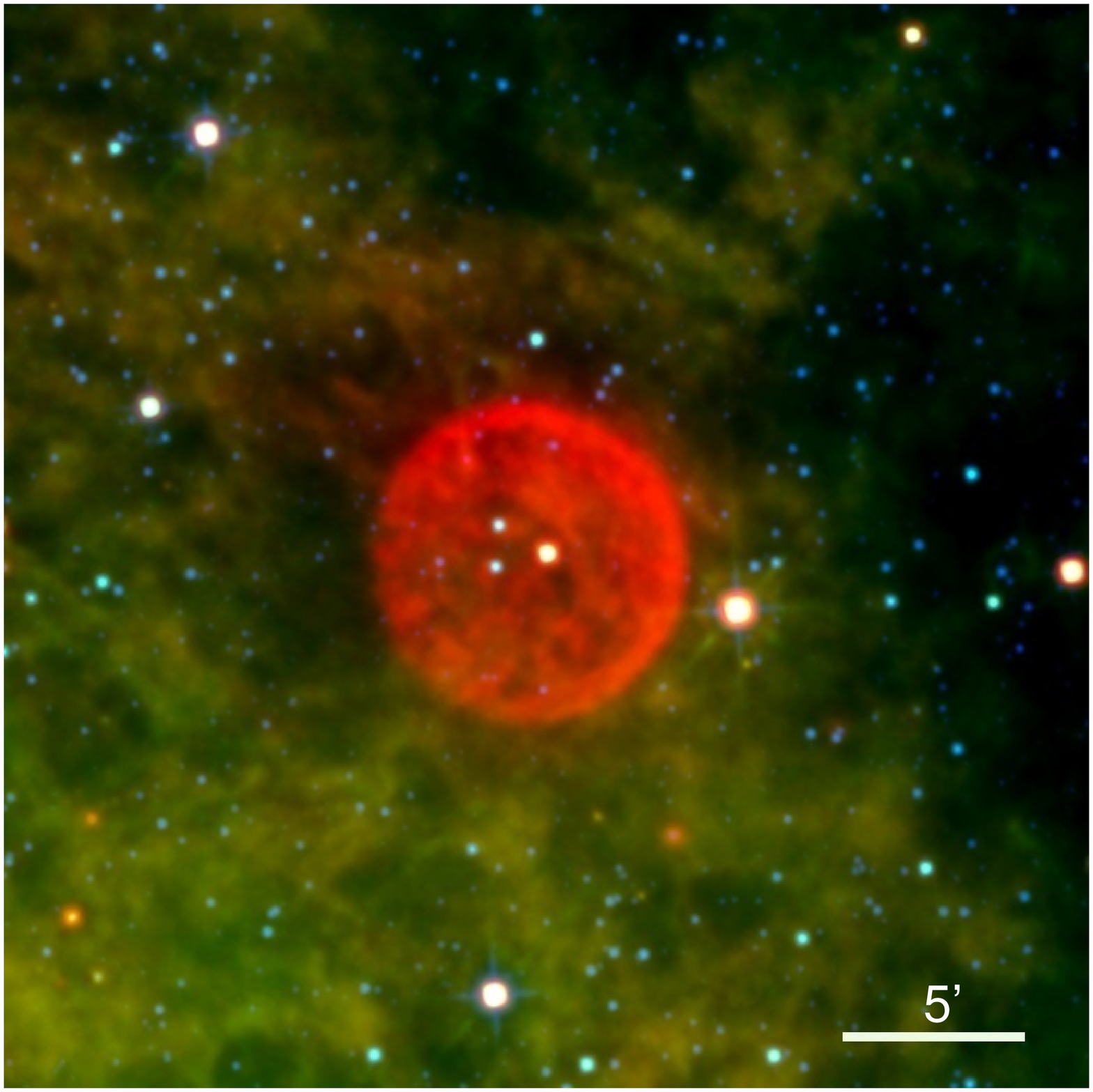}
\end{center}
\caption{Examples of WR nebulae with a ${\cal B}$ type morphology:
  WR\,6 (top panels) and WR\,16 (bottom panels). Left panels show the
  optical morphology while right panels show the {\it WISE} W2 (blue),
  W3 (green), and W4 (red) color-composite picture. The central WR
  stars are marked with red lines in the optical images. 
The colour-composite optical picture of WR\,6 has been done using 
H$\alpha$ (red) and [O\,{\sc iii}] (blue) narrow-band images. 
North is up, east to the left.
}
\label{fig:WR6}
\end{figure}

\begin{figure}
\begin{center}
\includegraphics[width=0.5\linewidth]{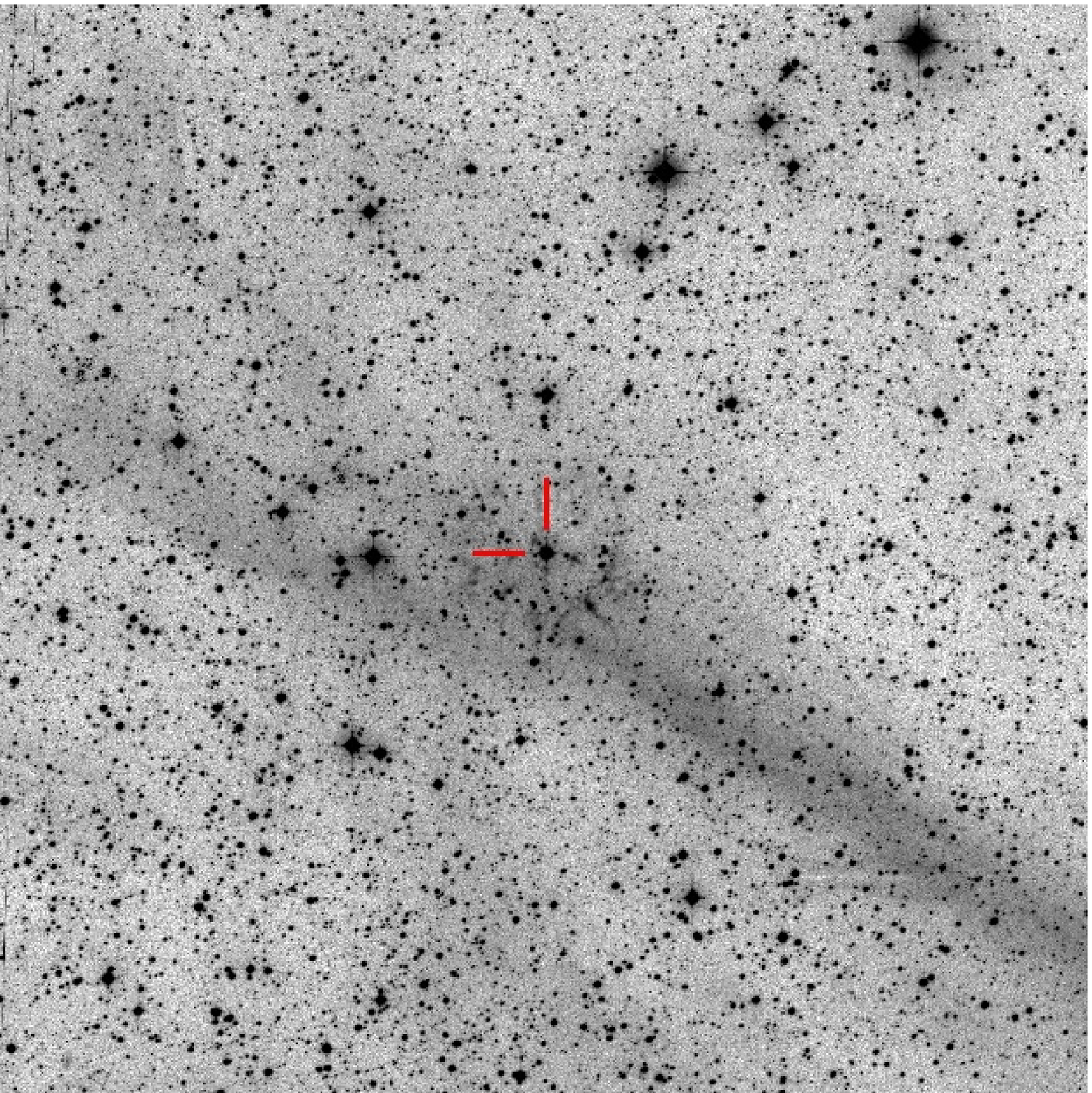}~
\includegraphics[width=0.5\linewidth]{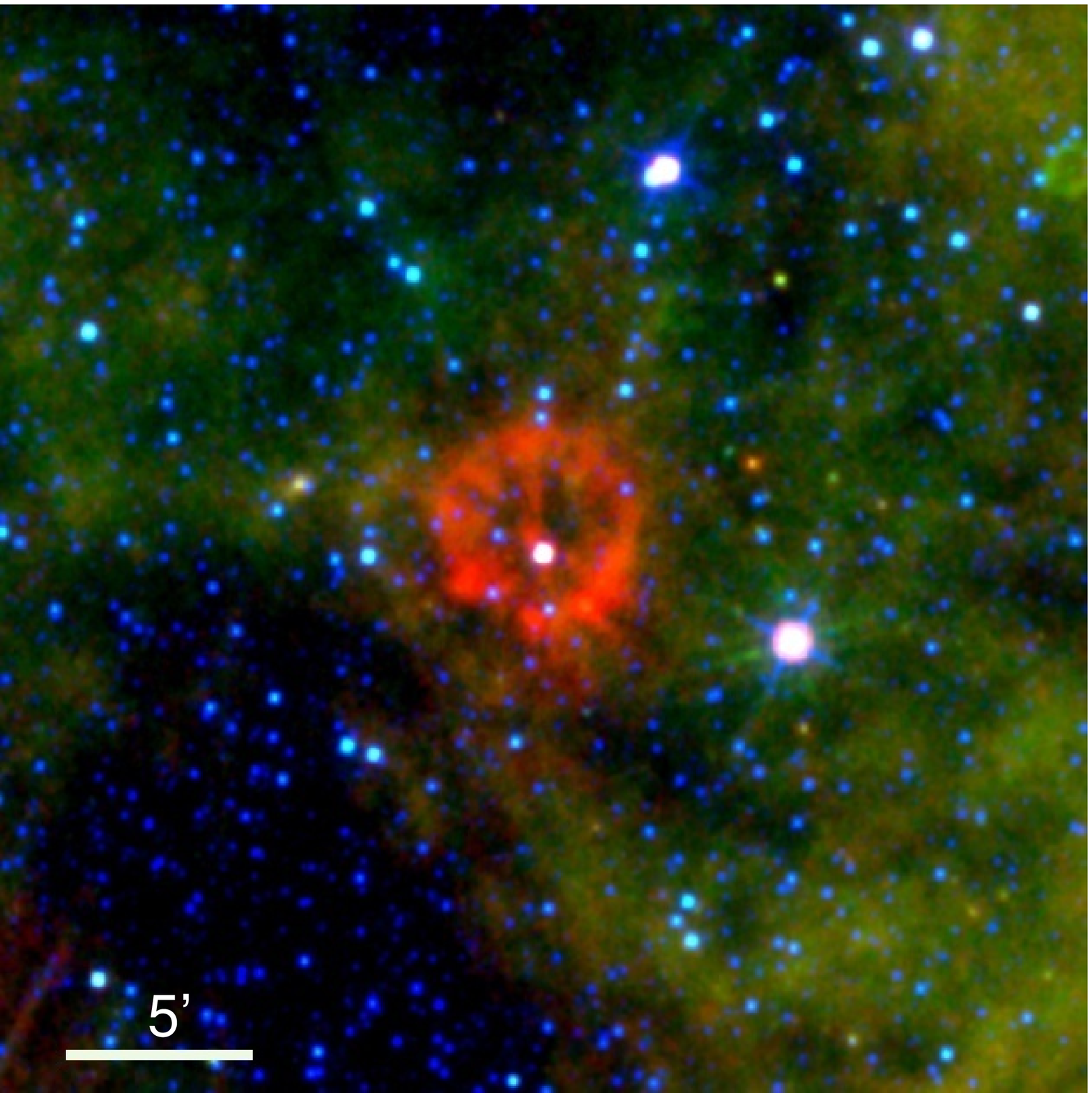}
\\
\vspace*{0.1cm}
\includegraphics[width=0.5\linewidth]{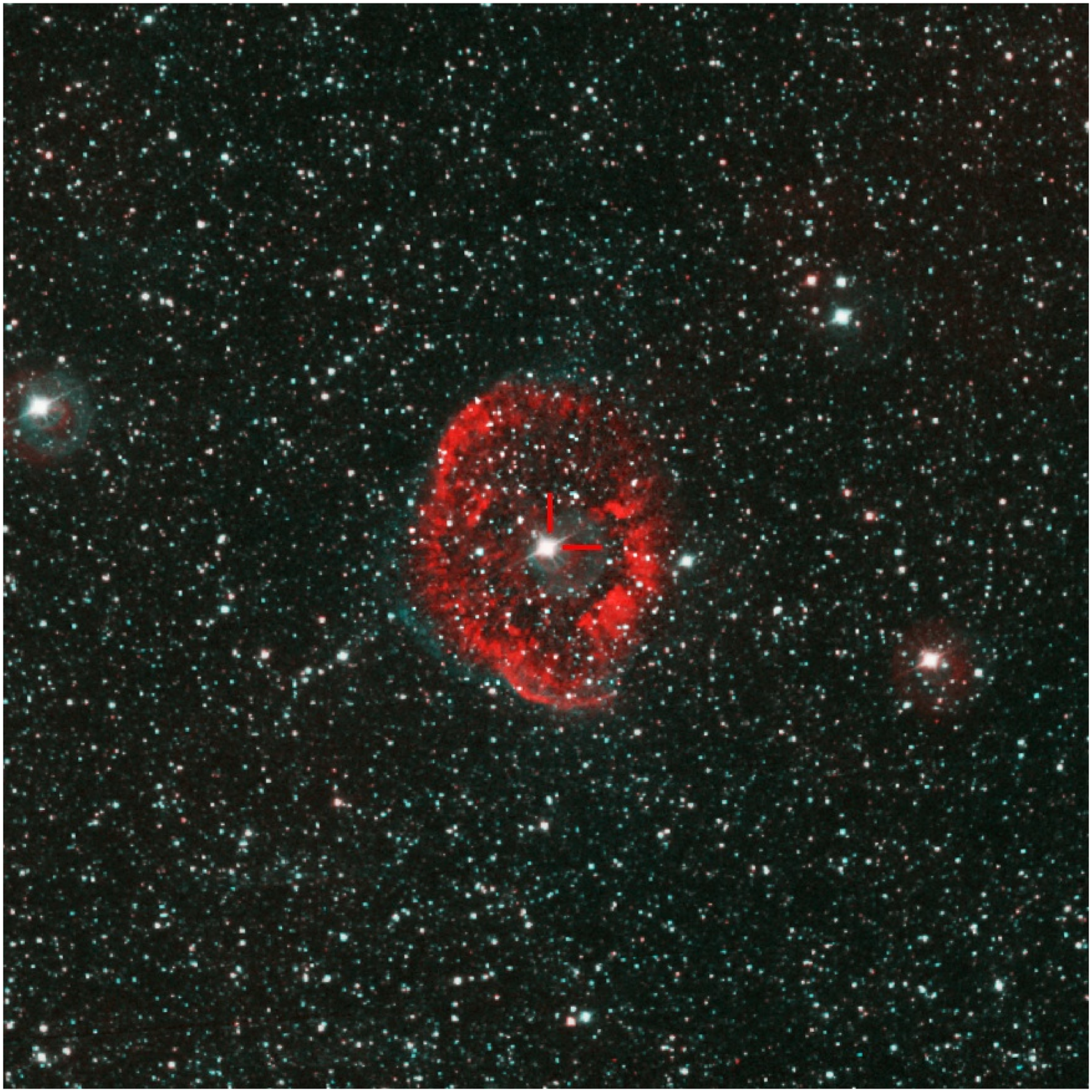}~
\includegraphics[width=0.5\linewidth]{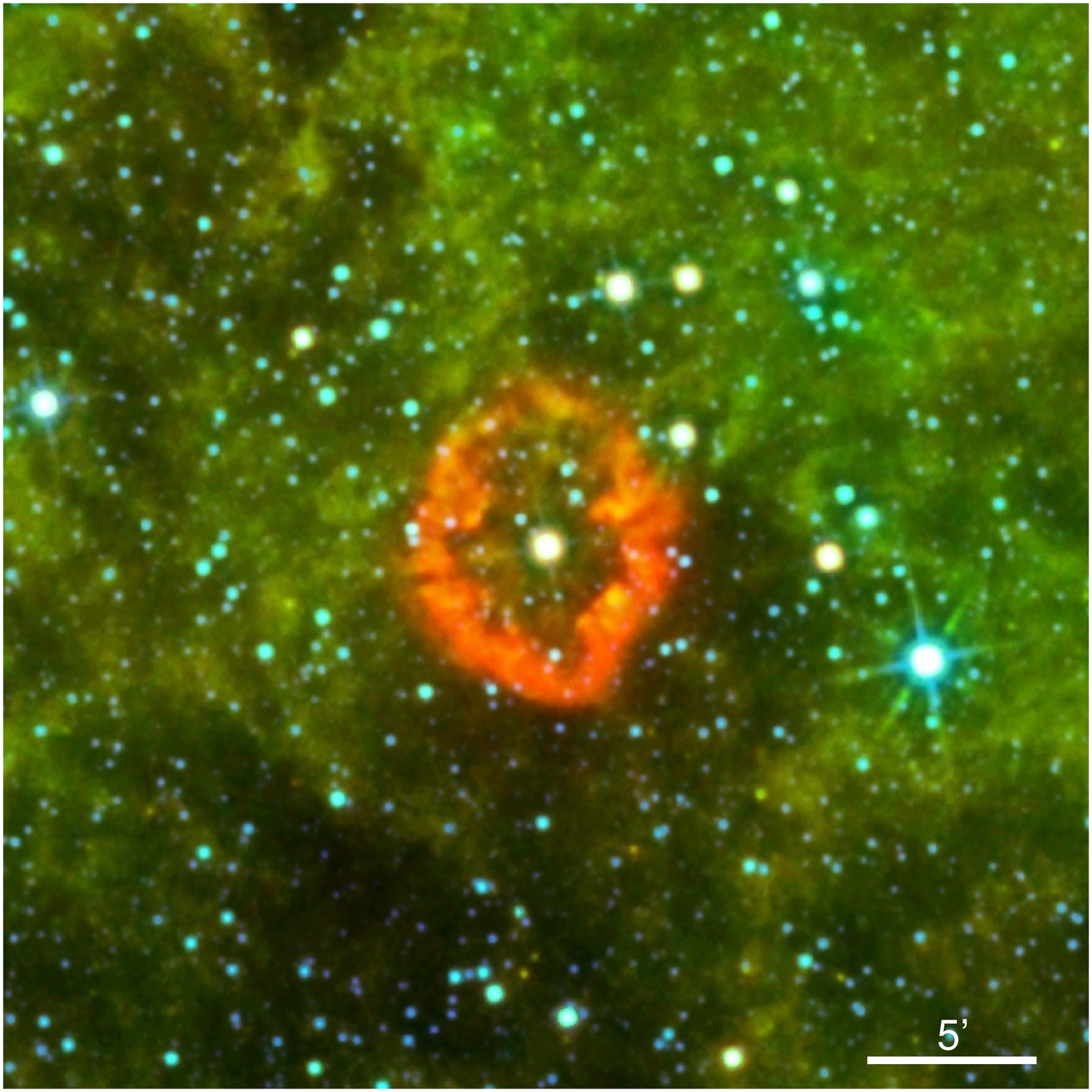}
\end{center}
\caption{Same as Fig.~\ref{fig:WR6} but for the cases of WR nebulae
  with a ${\cal C}$-Clumpy morphology: WR\,8 (top panels) and WR\,40
  (bottom panels). 
The color-composite optical image of WR\,40 has been done using H$\alpha$ 
(red) and [O\,{\sc iii}] (blue) narrow band images. 
}
\label{fig:WR8}
\end{figure}

\begin{figure}
\begin{center}
\includegraphics[width=0.5\linewidth]{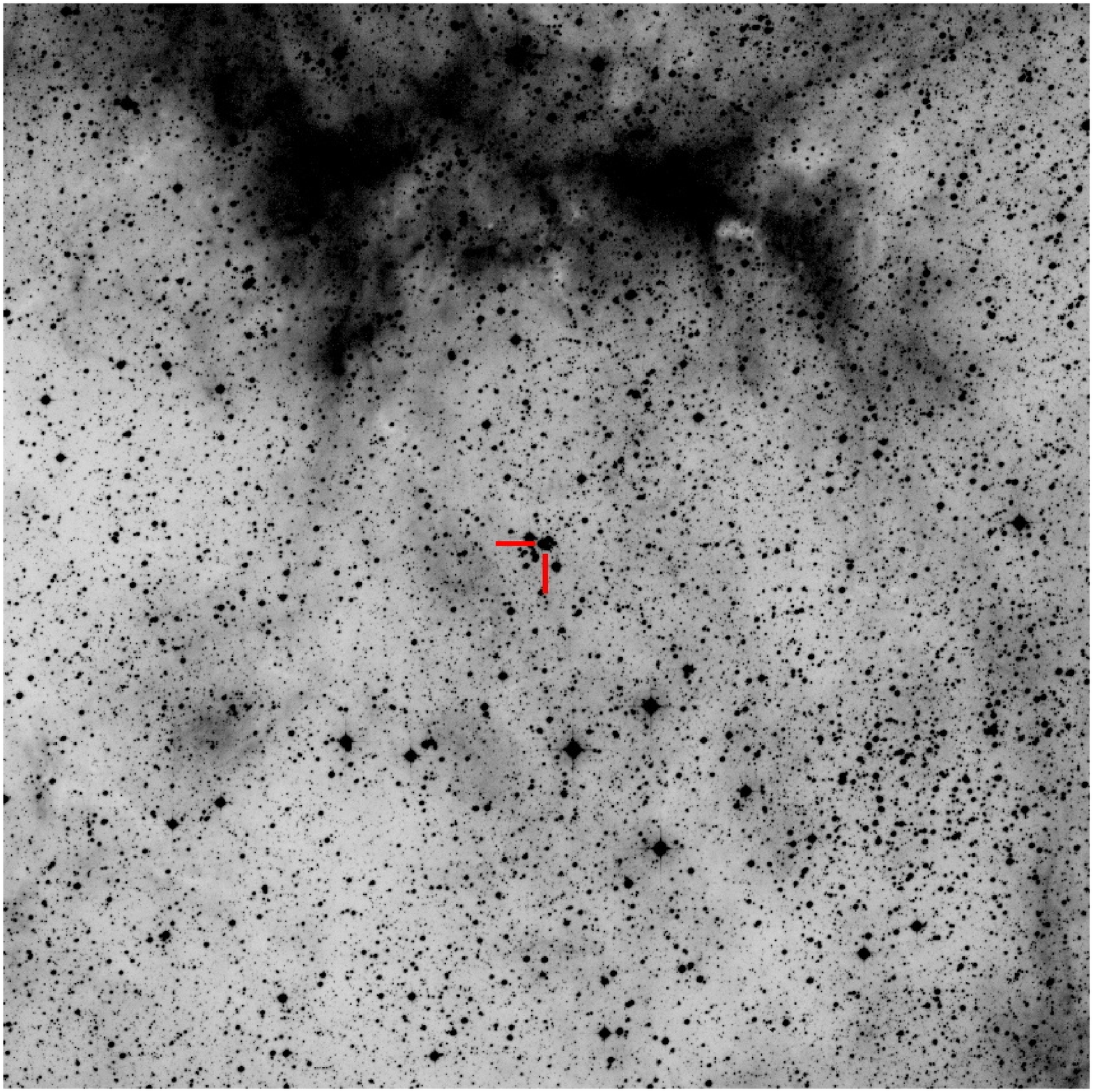}~
\includegraphics[width=0.5\linewidth]{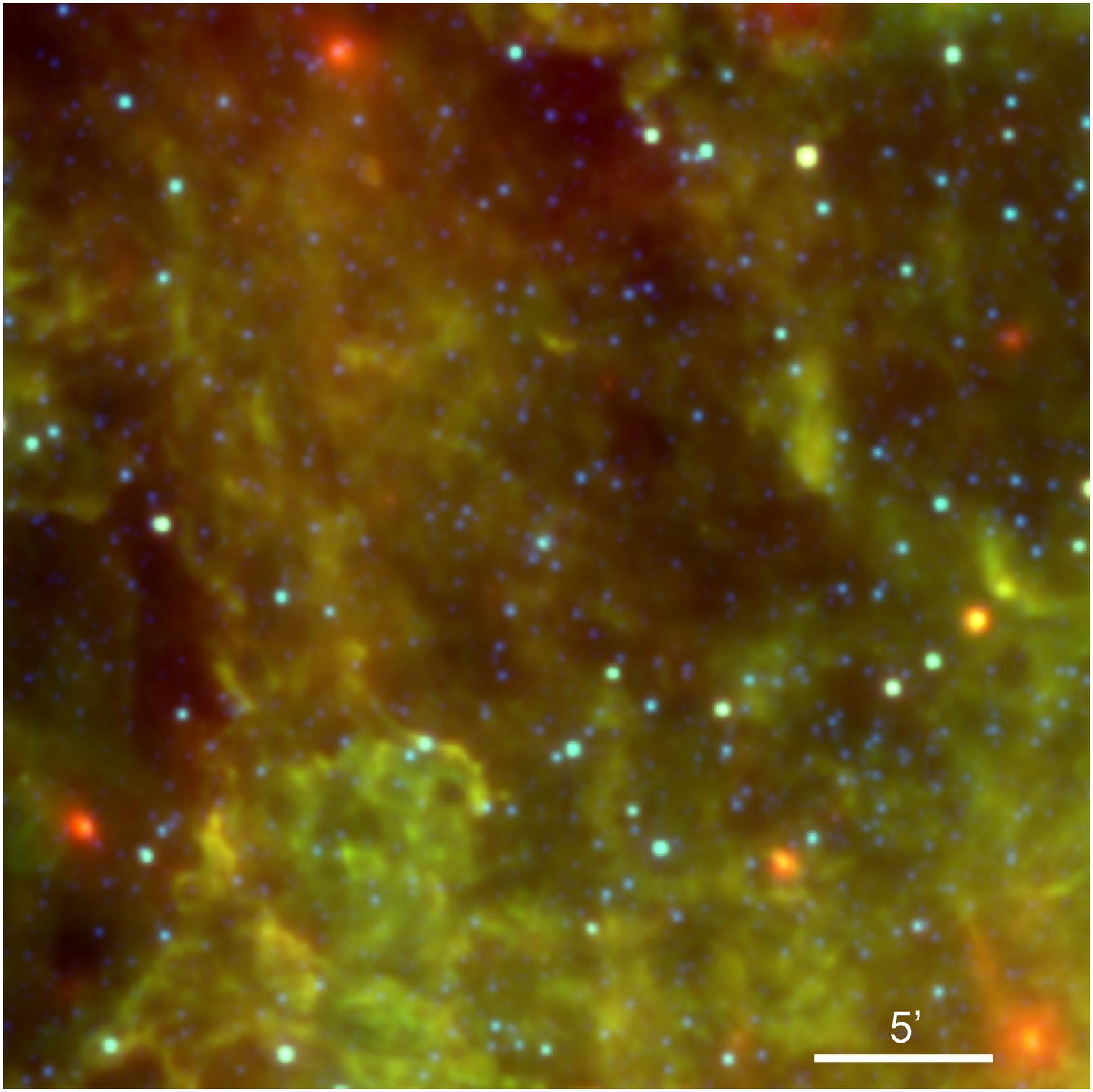}
\\
\vspace*{0.1cm}
\includegraphics[width=0.5\linewidth]{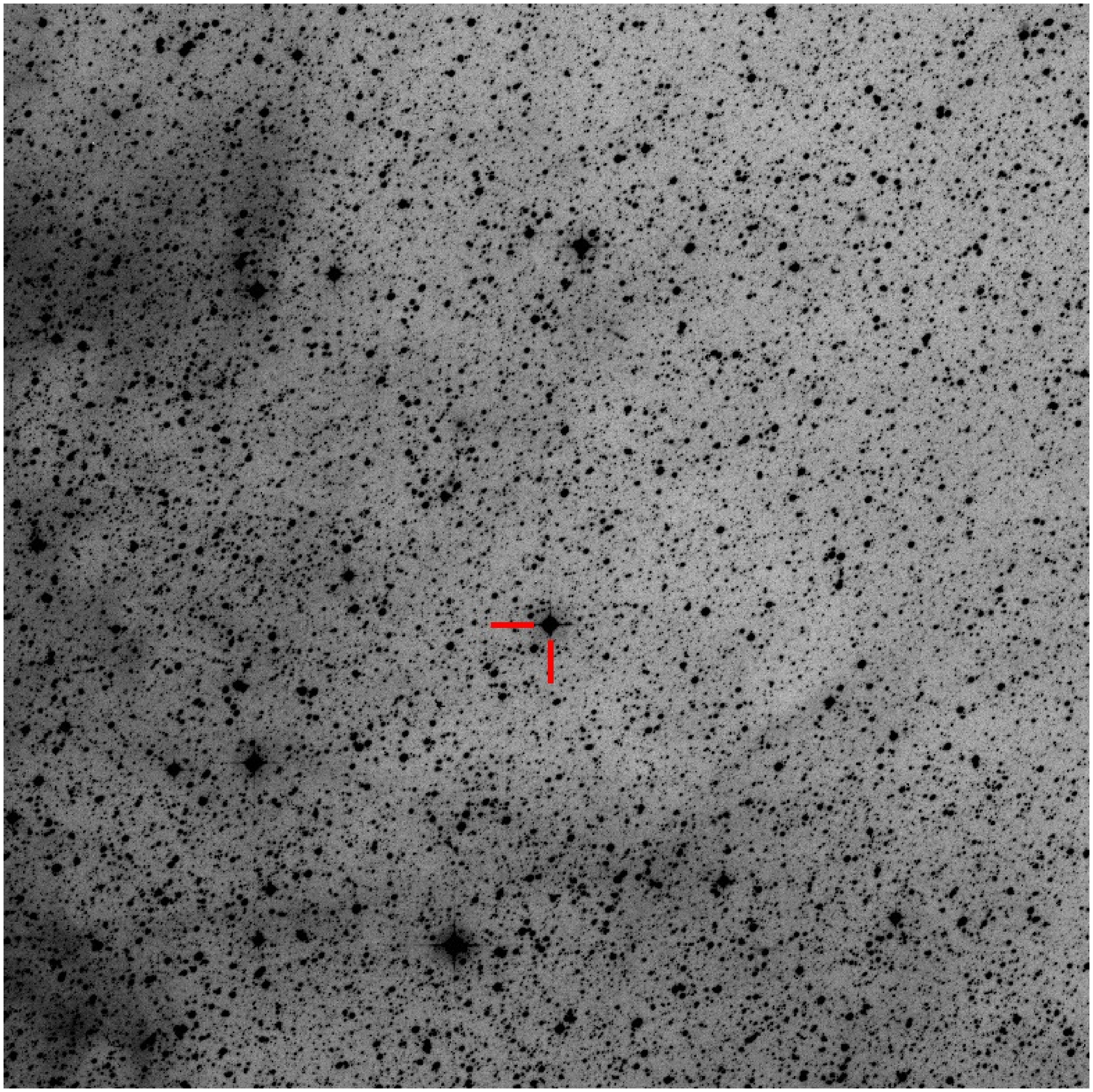}~
\includegraphics[width=0.5\linewidth]{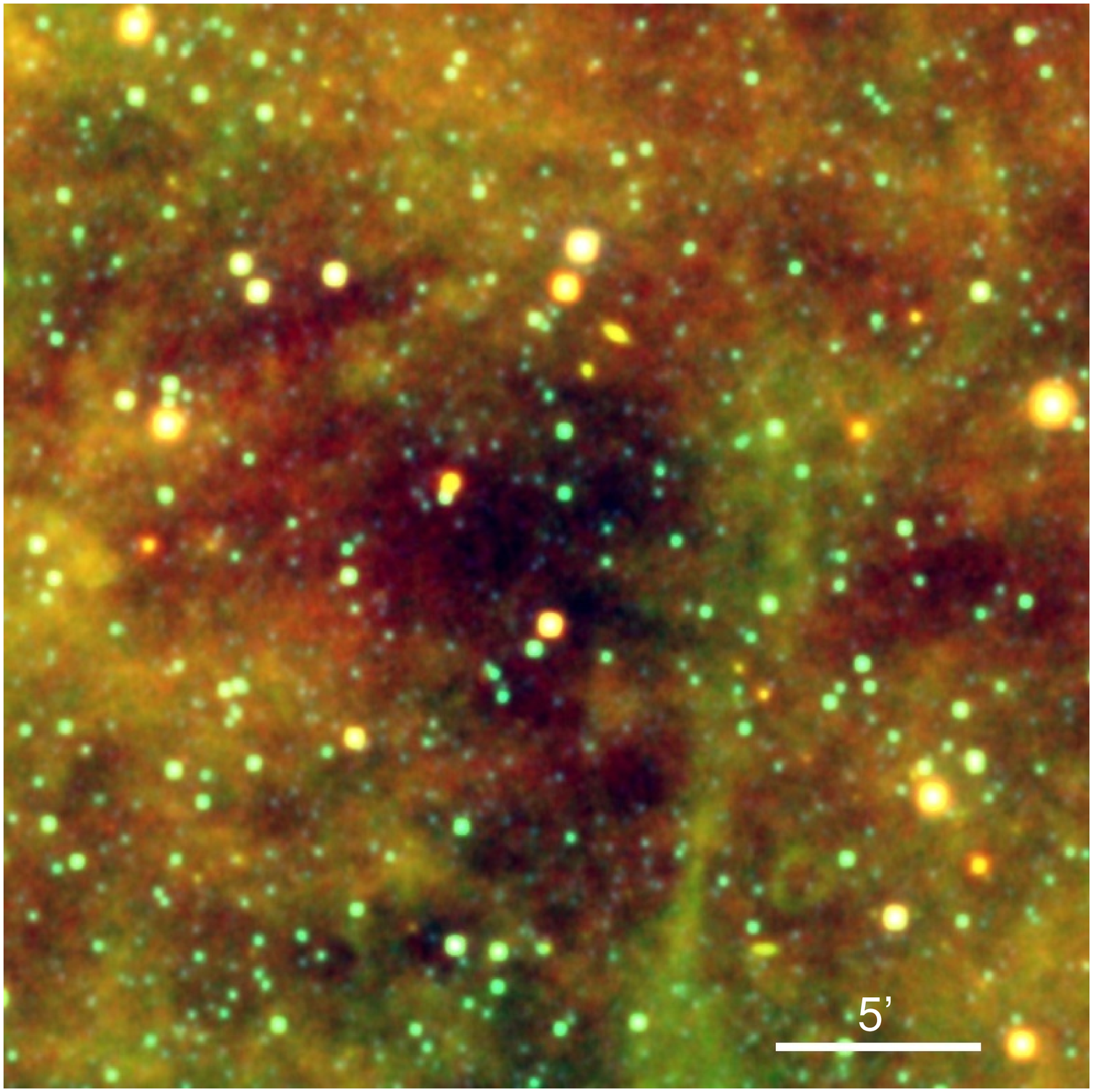}
\end{center}
\caption{Same as Fig.~\ref{fig:WR6} but for the cases of WR nebulae
  with a ${\cal M}$ morphology: WR\,35b (top panels) and WR\,52
  (bottom panels).}
\label{fig:WR35b}
\end{figure}

\begin{figure}
\begin{center}
\includegraphics[width=0.5\linewidth]{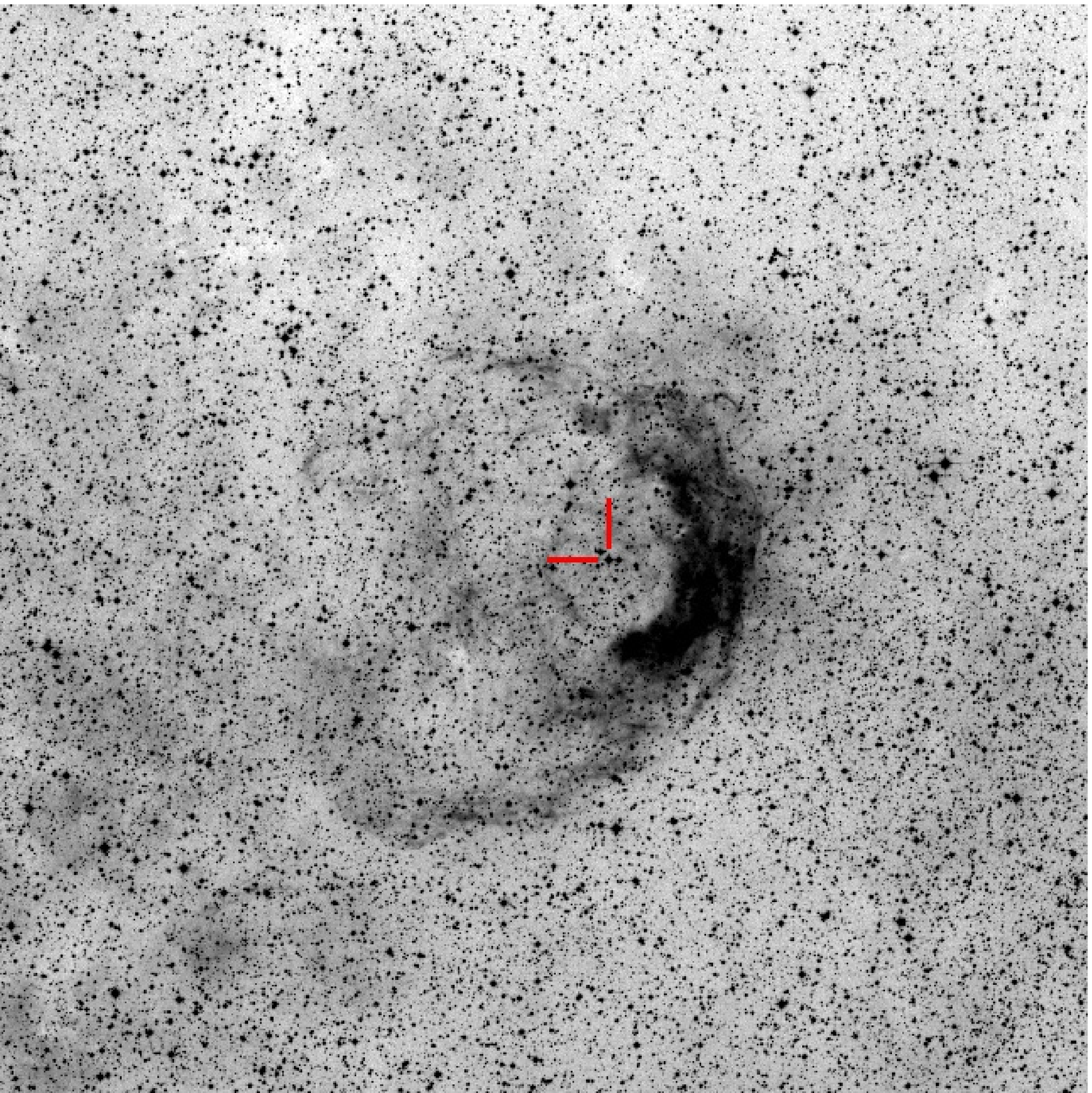}~
\includegraphics[width=0.5\linewidth]{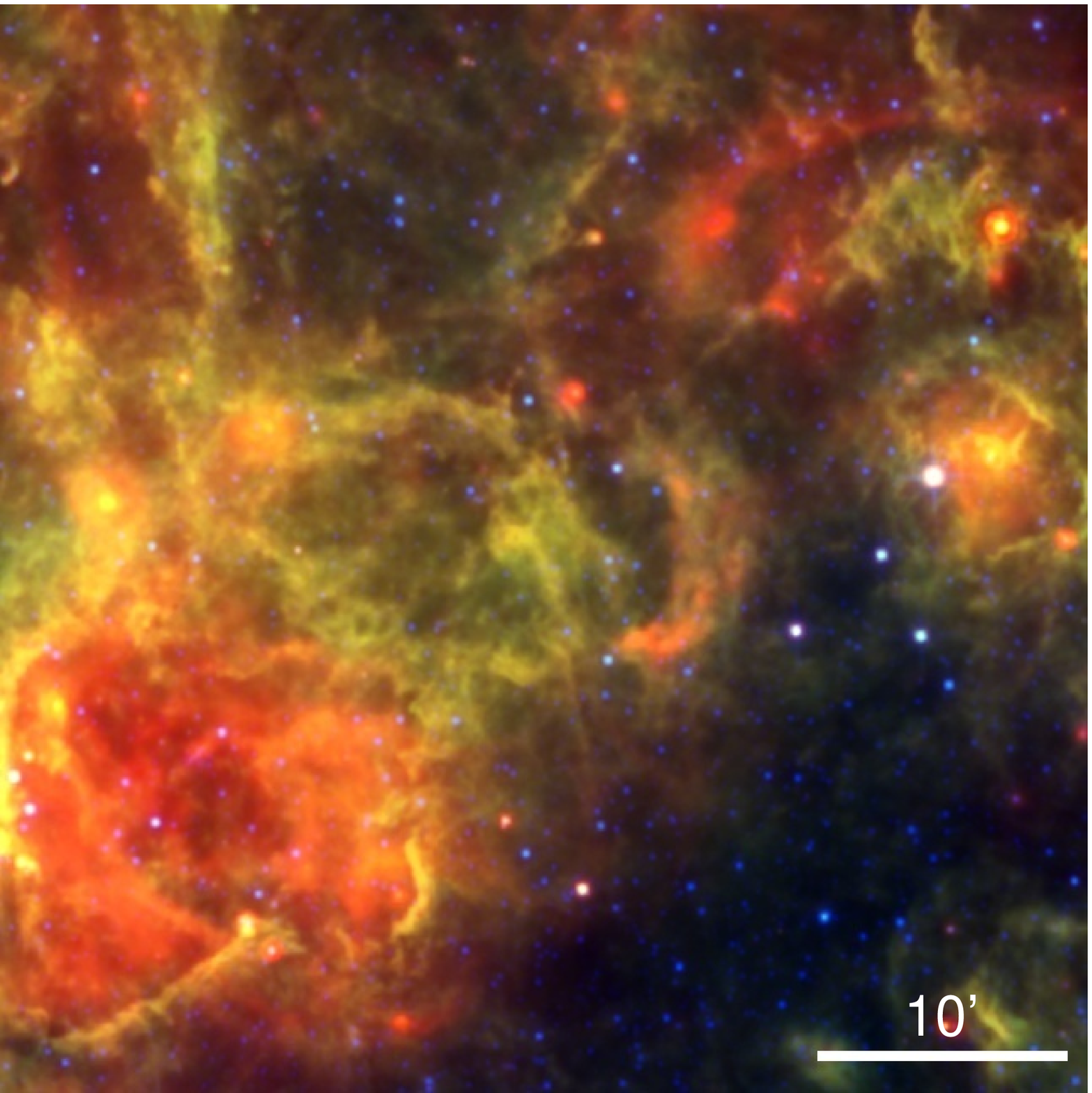}
\\
\vspace*{0.1cm}
\includegraphics[width=0.5\linewidth]{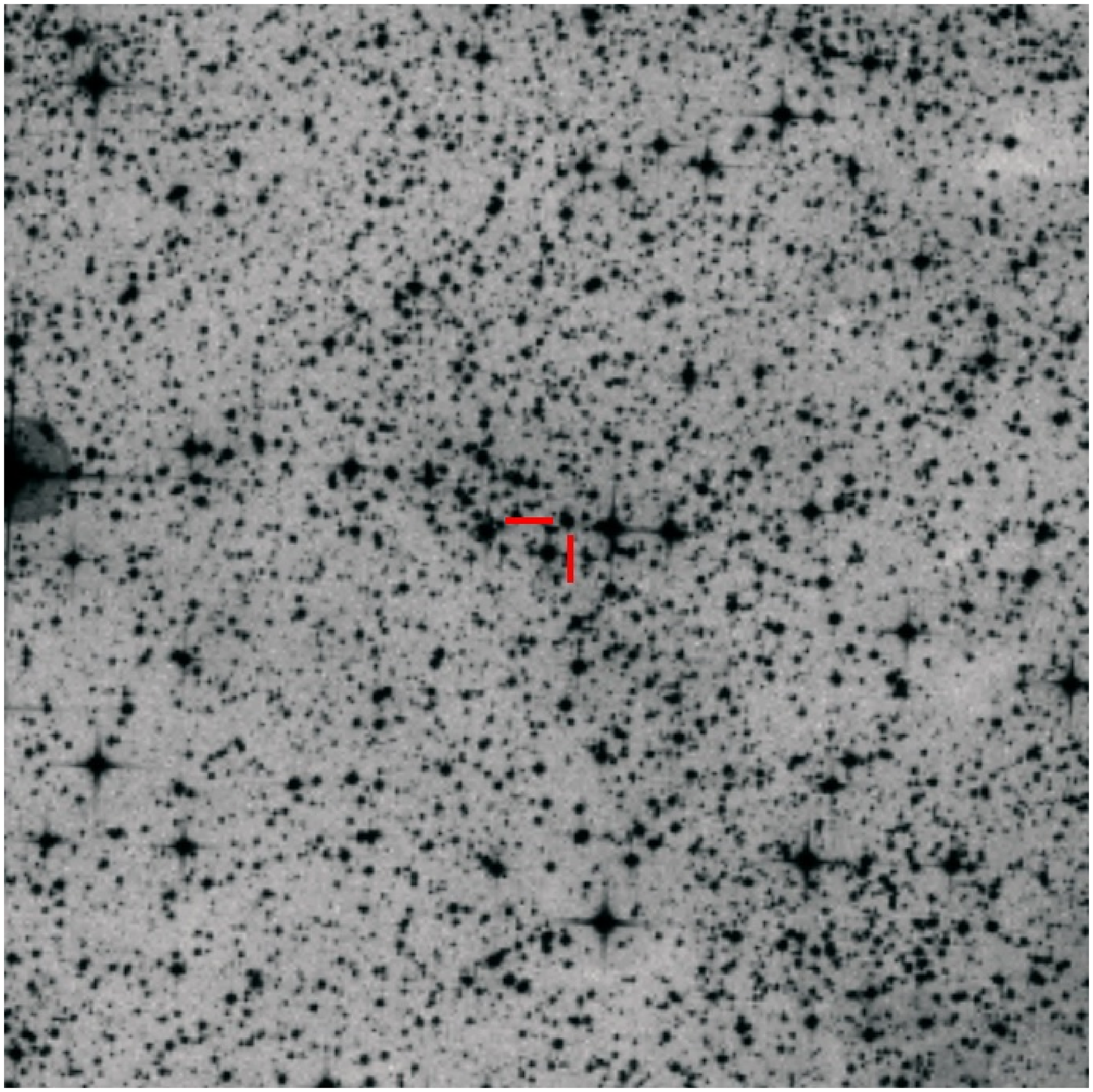}~
\includegraphics[width=0.5\linewidth]{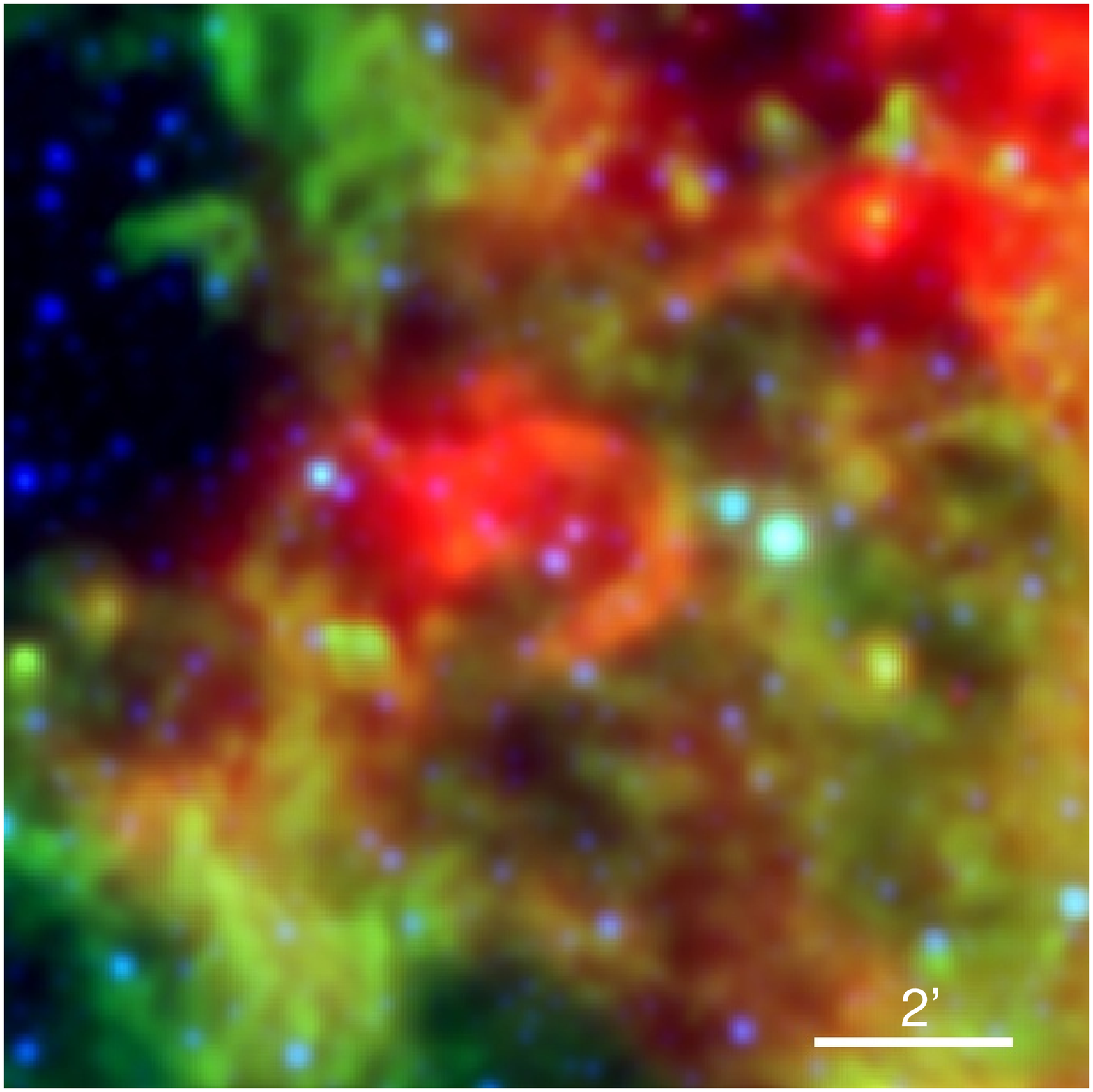}
\end{center}
\caption{Same as Fig.~\ref{fig:WR6} but for the cases of WR nebulae
  with a ${\cal C}$-Disrupted morphology: WR\,18 (top panels) and
  WR\,35 (bottom panels).}
\label{fig:WR18}
\end{figure}

%%%%%%%%%%%%%

\begin{figure*}
\centering
\includegraphics[bb=90 144 515 650,width=0.7\linewidth]{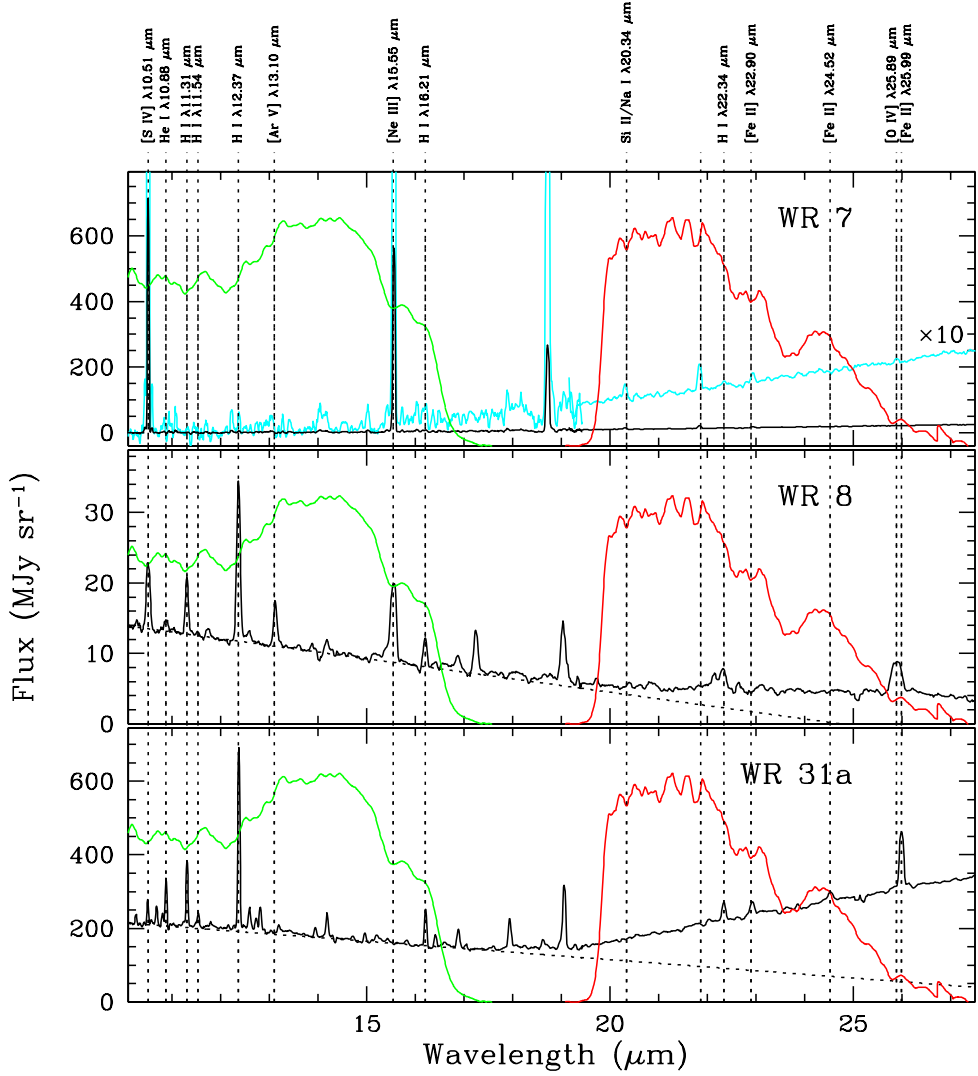}
\caption{ \emph{Spitzer} IRS SH and LH spectra of the nebulae around
WR\,7, WR\,8, and WR\,31a. The spectral responses of the W3
  (green) and W4 (red) \emph{WISE} bands normalized to arbitrary
  values are overplotted on the spectra.  The nebular spectrum of
  WR\,7 corresponds to the patch of bright red emission (the
  \emph{WISE} W4 band) towards the northwest shown in
  Figure~\ref{fig:WR7}.  This spectrum has been plotted at two
  different intensity levels to highlight the low level emission.  The
  contribution of the stellar emission of WR\,8 and WR\,31a to the
  nebular spectra can be roughly reproduced by a black-body model
  (dashed line).}
\label{fig:spec}
\end{figure*}

The nature of the emission in the \textit{WISE} W3 and W4 bands has 
been further investigated using the \emph{Spitzer} IRS high-dispersion 
background-subtracted spectra of the nebulae around % WR\,6 (S\,308), 
WR\,7 (NGC\,2359), WR\,8, and WR\,31a presented in Figure~\ref{fig:spec}.  
These spectra have been overplotted by the spectral responses of the 
\emph{WISE} W3 (green) and W4 (red) bands.  
The spectra of these nebulae imply significant contribution to the 
emission detected in the \emph{WISE} W3 band by a number of emission 
lines, including 
[S~{\sc iv}] $\lambda$10.51 $\mu$m, 
He~{\sc i} $\lambda$10.66 $\mu$m, 
He~{\sc i} $\lambda\lambda$11.31,12.37,16.21 $\mu$m, 
[Ar~{\sc v}] $\lambda$13.10 $\mu$m, 
and [Ne~{\sc iii}] $\lambda$15.55 $\mu$m.  
In sharp contrast, the \emph{WISE} W4 band is mostly dominated by
continuum emission probably from thermal dust continuum emission, with
weak H~{\sc i}, [Fe~{\sc ii}], and [O~{\sc iv}] emission lines.
It is interesting to note that the spectrum of WR\,7, extracted from a
region known to be overabundant in oxygen \citep{Dufour89}, does not
show bright emission from oxygen lines, but a weak [O~{\sc iv}]
$\lambda$25.59 $\mu$m line.

It is thus concluded that the emission from WR nebulae in the
\emph{WISE} W3 band is dominated by H~{\sc i} and forbidden emission
lines, mostly of [S~{\sc iv}] and [Ne~{\sc iii}], with little
contribution to the observed emission from dust thermal continuum.
Interestingly, the spatial correspondence between broad arc-like
patches of emission in the W3 image of S\,308 around WR\,6 and
diminished emission in H$\alpha$ (Fig.~\ref{fig:WR6}) seems to imply
absorption of nebular emission by dust on the foreground.  As
expected, the W3 band traces 
material in the ISM in most cases.  Meanwhile, the emission in the
\emph{WISE} W4 band, tracing mostly the optical nebular shells,
corresponds to thermal continuum emission, with very little
contribution from H~{\sc i}, [Fe~{\sc ii}], and [O~{\sc iv}] emission
lines.  These results are not in disagreement with the IRS spectrum of
the WR bubble MB\,3957, which is dominated by emission lines of iron,
sulfur and neon, but also includes an additional contribution of
thermal dust emission \citep{Flagey2011}.  Otherwise, the
\textit{Spitzer} MIPS 24 $\mu$m images of planetary nebulae include
important contributions from the [O~{\sc iv}] $\lambda$25.89 $\mu$m
and [Ne~{\sc v}] $\lambda$24.3 $\mu$m emission lines
\citep[e.g.][]{Chu2009}, unlike the \emph{WISE} W4 images of WR
nebulae.  The spatial resolution of the \emph{WISE} W4 band images,
$\approx$12\farcs0, does not allow us to determine whether the dust in
WR nebulae is spatially coincident with the ionized material or is
found at the leading edge of the optical WR nebulae.

\section{Morphological classification of WR nebulae}

The images of the WR nebulae presented in Figures~\ref{fig:WR6} to
\ref{fig:WR35b} are examples of the variety of morphologies from
complete shells, arcs (incomplete shells), clumps, and filaments to
diffuse, featureless emission. A close inspection to all WR nebulae in
our sample and the comparison between optical and IR morphologies has
led us to classify them into three broad morphological types:

\begin{itemize}

\item ${\cal B}$ - WR Bubble. \\
  These nebulae present a thin shell or bubble both in optical and IR,
  mostly in the W4 band. Examples of these nebulae are those around
  WR\,6 (S\,308) and WR\,16 (see Fig\ref{fig:WR6}).

\item ${\cal C}$ - Clumpy/Disrupted WR bubbles. \\
  These nebulae present clumpy H$\alpha$ and IR images and/or
  incomplete arcs or shells. The W4 images reveal for a significant
  number of cases that the optical and IR clumps are spatially
  coincident with bow shock-like features.  Archetypes of these WR
  nebulae are those around WR\,8 and WR\,18, WR\,35 and WR\,40 (see
  Fig.~\ref{fig:WR8} and \ref{fig:WR18}).

\item ${\cal M}$ - Mixed WR nebulae. \\
  These nebulae do not present a clear correspondence between the
  optical and IR images. Indeed, they do not show a clearly defined IR
  morphology, with a noticeable lack of emission in the \emph{WISE} W4
  band. Examples of these nebulae are those around WR\,35b
  and WR\,52 (see Fig.~\ref{fig:WR35b}).

\end{itemize}

Our classification of the WR nebulae is listed in the last column of
Table~\ref{tab:spectral}. We note that sometimes the assignation of a
morphological type is difficult given the disparity between the
optical and IR images or due to the ambiguous association between the
star and the nebula. For instance, the nebula around WR\,116
(Fig.~\ref{fig:WR113}) has been assigned a morphological type ${\cal
  C}$ in basis of its optical image, but we reckon that a
morphological type ${\cal M}$ cannot be completely excluded by its IR
counterpart.

\subsection{Notes on individual objects}

\subsubsection{WR\,35 and its obscured shell}
We report the previously unknown partial shell around WR\,35.  This
shell, clearly detected in the \emph{WISE} W4 band
(Fig.~\ref{fig:WR18}-bottom panels), has a size
$\sim$2\farcm0$\times$1\farcm5 and opens towards the southeast of
WR\,35.  In contrast, the H$\alpha$ image from the Super COSMOS Sky
Survey only detects emission from an ionized cloud southwest of WR\,35
which does not coincide spatially with the IR shell.  A similar case
of detection of an obscured shell has been claimed by
\citet{Wachter2011} for WR\,8, although an optical nebulosity was
previously reported by \citet{Stock2010}.  The high frequency of
discovery of obscured shells around massive stars in \emph{Spitzer}
MIPS 24 $\mu$m images has led to the suggestion that such shells may
be ubiquitous among these stars \citep[e.g.,][]{Gv2010,Wachter2010}.

\subsubsection{A bipolar nebula around WR\,85}
The nebula RCW\,118 around WR\,85 has been described as a bipolar nebula 
based on H$\alpha$ images obtained by \citet{MCG1994,Marston1994}.  
This bipolar structure is hinted in the \emph{WISE} W4 image 
by the red patches to the northeast and southwest of WR\,85 
(Fig.~\ref{fig:WR85}-{\it right}).  
This IR image is certainly reminiscent of the optical image 
of the nebula NGC\,6164-5 around the O6 star HD\,148937 
\citep{Bruhweiler_etal1981,DPH1988}.

\subsection{Comparison with previous morphological classifications}

Previous imaging studies of WR nebulae have disclosed a rich
morphological variety among them
\citep[e.g.,][]{Chu1983,Gruendl2000,Stock2010}. The most complete
classification scheme of WR nebulae is that originally introduced by
\citet{Chu1981} and more recently described by \citet{Chu2003}.  
This classification scheme relies on narrow-band optical images, kinematics, 
and information on the chemical abundances to define the presence of bubbles 
and their expansion rate and kinematical age. 
Meanwhile, our classification
scheme of WR nebulae is supported by the comparison between
narrow-band optical (mostly H$\alpha$) and mid-IR \emph{WISE} images,
the latter being very useful to unveil obscured shells and to
disentangle the nebular emission from background diffuse and ISM
emission as discussed in Section~\ref{sec:interpreting}. 

The comparison between these two morphological classifications of WR
nebulae in Table~\ref{tab:comp} reveals a good agreement. 
This is excellent for the wind-blown bubbles (our ${\cal B}$ nebulae and 
the $W$ class), as these are the most easy to spot either in direct images 
or kinematical data.  
Indeed, all our ${\cal B}$ nebulae are classified as $W$, and only 4 $W$ 
objects are classified as ${\cal C}$ nebulae because their mid-IR morphology.

\begin{table}
\caption{Comparison of WR Nebula Morphology Classification}
%\scriptsize
\centering
\begin{tabular}{lccl}
\hline\hline\noalign{\smallskip}
\multicolumn{1}{c}{WR~~~} &
\multicolumn{1}{c}{This paper} & 
\multicolumn{1}{c}{Previous works} & 
\multicolumn{1}{l}{References} \\
\hline
\noalign{\smallskip}
 ~~6    & ${\cal B}$ & $W/E$        & 1,2,3,4 \\
 ~~7    & ${\cal B}$ & $W$        & 1,3,4,5 \\
 ~~8    & ${\cal C}$ & $E$        & 4 \\
 ~16    & ${\cal B}$ & $W$        & 4,7 \\
 ~18    & ${\cal C}$ & $W$        & 1,3,4,8 \\
 ~22    & ${\cal M}$ & $W$        & 3,6 \\
 ~23    & ${\cal B}$ & $W$        & 1,3,7 \\
 ~38    & ${\cal M}$ & Ring?      & 7 \\
 ~40    & ${\cal C}$ & $E$        & 1,3,4,8 \\
 ~52    & ${\cal M}$ & $R_\mathrm{s}$      & 1,3,9 \\
 ~54    & ${\cal M}$ & Ring?      & 6 \\
 ~55    & ${\cal C}$ & $R_\mathrm{a}$      & 1,3,9 \\
 ~68    & ${\cal M}$ & Ring       & 6 \\
 ~75    & ${\cal B}$ & $W/E$        & 1,3,4,8 \\
 ~85    & ${\cal C}$ & $R_\mathrm{s}$      & 1,3,9 \\
 ~86    & ${\cal M}$ & $\dots$    & 6 \\
% ~91    & ${\cal C}$ & $W$        & 6 \\
% ~93    & ${\cal C}$ & $W/R_\mathrm{s}$    & 3,6 \\
 ~94    & ${\cal M}$ & $\dots$    & 7 \\
 ~95    & ${\cal C}$ & $\dots$    & 6 \\
 101    & ${\cal C}$ & Ring?      & 7 \\
 102    & ${\cal B}$ & $W/E$        & 4,7 \\
 124    & ${\cal C}$ & $E$        & 1,3,4,10 \\
 128    & ${\cal B}$ & $W/R_\mathrm{s}$    & 3,11 \\
 131    & ${\cal C}$ & H~{\sc ii} & 1,3,10 \\
 134    & ${\cal C}$ & $W$        & 1,3,12 \\
 136    & ${\cal B}$ & $W/E$        & 1,3,4,12 \\
\hline
\end{tabular}
\begin{list}{}{}
\item{
Refs.-- 
(1) \citet{Chu1981}, (2) \citet{Chu_etal82}, (3) \citet{Chu1983}, 
(4) \citet{Stock2010}, (5) \citet{Treffers1982}, (6) \citet{MCG1994}, (7) \citet{Marston1994}, 
(8) \citet{Chu1982}, (9) \citet{CT81b}, (10) \citet{CT81a}, 
(11) \citet{Gruendl2000}, (12) \citet{Treffers1982a}. 
}
\end{list}
\label{tab:comp}
\end{table}

This consistency is confirmed by the
comparison between optical images in the H$\alpha$ and [O~{\sc iii}]
emission lines that has been shown by \citet{Gruendl2000} to be very
helpful to reveal the expansion of shock fronts outside the main
nebular shell. For example, in the case of RCW\,104 around WR\,75
(Figure~\ref{fig:WR75}), the H$\alpha$ line image does not show any
morphological characteristic of a round shell, thus implying a
classification as a ${\cal C}$-type nebula. Instead, the [O~{\sc
  iii}] morphology implies a ${\cal B}$-type. On the contrary, the
nebula M\,1-67 around WR\,124 shows no evidence of an [O~{\sc iii}]
front \citep[][and references therein]{Fernandez-Martin2013}, but its
kinematics reveals an expanding shell consisting of numerous
condensations \citep{SC1982}. While these findings may raise doubts
for the classification of WR nebulae for which [O~{\sc iii}] images
and kinematics are not available, we note that in all these nebulae
the outer expanding shell is noticeably much fainter than the shells
detected in our ${\cal B}$-type nebulae.  
This points to an intrinsic difference between the two morphological types.

As for the ${\cal M}$ nebulae, only two have previous definite
morphology \citep{Chu1991}.  The nebula around WR\,22 is described by
\citet{Chu1991} as a possible wind-blown bubble, although no evidence
for such a bubble is revealed in \emph{WISE} images
(Figure~\ref{fig:WR22}).  Similarly, the nebula around WR\,52 is
categorized as an $R_{\rm s}$ nebula in basis of kinematical data, but
the nebula is rather inconspicuous in the IR
(Figure~\ref{fig:WR35b}-bottom panels).

\subsection{Interpreting the morphology of WR nebulae}
\label{sec:interpreting2}
 
The main morphological characteristics of WR nebulae result from the
interactions of the WR wind and the intense UV radiation with the slow
wind ejected previously by the progenitor.  The most basic
evolutionary path of a WR nebula, as reproduced by multiple
simulations
\citep[e.g.,][]{GSML1995b,GS1996a,GS1996b,Freyer2003,vMarle2005,Freyer2006,Arthur2007a,Arthur2007b,vMarle2007,Dwarkadas2007,Toala2011},
implies a massive star ejecting a significant fraction of its mass
during a RSG or LBV phase through a spherical, slow (10--100
km~s$^{-1}$) and dense wind. In the final stage of stellar evolution,
during the WR phase, this CSM will be swept up by the WR fast stellar
wind ($v_\infty \gtrsim 10^3$ km~s$^{-1}$), creating at early times a
WR bubble (${\cal B}$-type nebulae). Later on, the WR bubble will be
disrupted due to Rayleigh-Taylor and thin shell instabilities (${\cal
  C}$-type nebulae) and the resulting filaments and clumps will later
mix with the ISM (${\cal M}$-type nebulae).

Additional processes may alter this general scheme.  
For instance, the proper motion of the star or a density gradient in 
the ISM may compress and brighten the nebula as it interacts with its 
surroundings, resulting in incomplete shells and arcs.  
Stellar rotation can also introduce changes in the stellar evolution
\citep[e.g.,][]{MM2003,Eks2012} and produce, for instance, anisotropic
stellar winds.  In the case of non-spherical (e.g., bipolar or clumpy)
ejections of material seen in some LBV nebulae \citep[e.g., MN13 and
MN79;][]{H2010,Gv2010}, the wind-wind interaction will give place to
the immediate formation of clumps or filaments, forming directly a
${\cal C}$-type nebulae with no {\cal B}-type nebula in between.
The large-scale structure of the ISM may also alter the morphology of 
WR nebulae.  

Since WR nebulae can form through different morphological sequences,
we should not expect a tight correlation between nebular morphology
and the star evolutionary stage implied by its WR spectral type
\citep[e.g.,][]{Moffat95,Maeder97}. Certainly, ${\cal M}$-type nebulae
can be expected preferentially around evolved WR stars, whereas the
central stars of ${\cal B}$ and ${\cal C}$-type nebulae can be
expected to have late spectral types \citep{Chu1981,Chu1983}.  This
correspondence is further complicated by the time-lapse spent by a WR
star on the different WR sub-phases which depends strongly on
different factors \citep[initial mass, rotation, initial metallicity;
  e.g.,][]{Maeder1991,MM2005,Georgy2012}.  A star may even miss a WR
sub-phase; for example, the stellar evolution models at solar
metallicity of \citet{MM2003} predict that a rotating 25~$M_{\odot}$
star will not enter the WNE and WC stages. Therefore, for those
particular models, any type of WR nebulae can be found around WNL
stars because the long duration of this phase or even the lack of the
WNE stage. On the other hand, for the case of non-rotating models,
which predict comparable durations of the WNL and WNE phases, we may
expect ${\cal B}$-type nebulae around these two type of stars.

\begin{figure}
\centering
\includegraphics[bb=18 245 592 718,width=1.\linewidth]{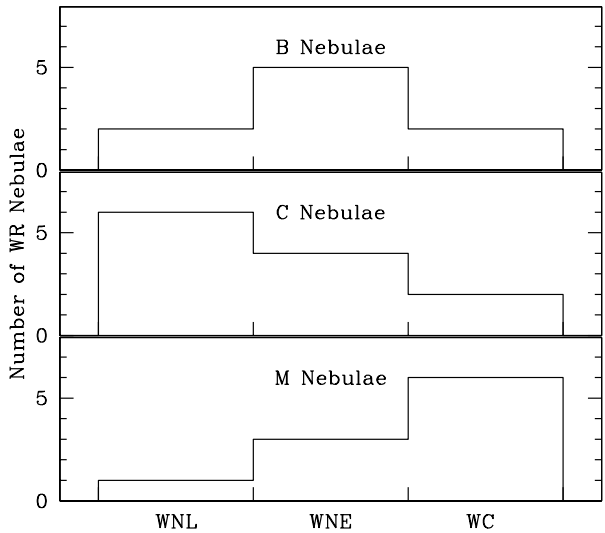}
\caption{
Distribution of the different morphological types of WR nebulae 
depending on the spectral types of their WR central stars.  
}
\label{fig:histo}
\end{figure}

To better quantify these relationships, we show in
Figure~\ref{fig:histo} the distributions of nebular morphology and
stellar spectral type among the WR nebulae in our sample. Five out of
the nine ${\cal B}$ nebulae have WNE stars and two others have WNL
spectral types, whereas only two are early WC stars.  The WNL stars
WR\,16 (WN8) and WR\,31a (WN11) are surrounded by small-sized bubbles
(2.8 and 1.2~pc in radius, respectively), and the WC WR\,23 (WC6) and
WR\,102 (WO2) stars by large bubbles (16 and 4 pc in radius,
respectively). Nebulae with ${\cal C}$-type morphology harbor mostly
late WN7-WN8 stars (6 out of 12) or WNE stars (4 out of 12), with only
two WC stars (WR\,95 and WR\,101).  On the contrary, ${\cal M}$
nebulae have a significant fraction of early WC stars (6 out of 10)
and a very small number of WNL stars (1 out of 10). This confirms
previous suggestions that WR bubbles are associated to late WN stars
and ${\cal C}$-type to WC stars. Most ${\cal C}$ nebulae had been
previously classified either as $W$ or $E$ nebulae, but we notice some
contamination of $R_\mathrm{a}$ and $R_\mathrm{s}$ nebulae, and even
objects of uncertain morphology.  This indicates that ${\cal C}$
nebulae have a variety of kinematical and chemical properties.  This
is also the case for their WR central stars which can be linked to
different evolutionary paths leading to a ${\cal C}$-nebula
morphology.

\section{Comparison with previous mid-IR studies} 

  The extinction by dust in the Galactic Plane makes difficult the
  search for WR nebulae at optical wavelengths.  The detection of the
  mid-IR emission (thermal dust, molecules, PAHs, ...)  of shells
  around evolved stars may provide an alternative method for the
  direct identification of these nebulae.  Mid-IR observations have
  indeed allowed the detection of young PNe
  \citep[e.g.,][]{Morris2006,Fesen2010} and LBV nebulae
  \citep[e.g.,][]{Gv2010b,Wachter2010}. Not unexpectedly,
  \emph{Spitzer} MIPS 24~$\mu$m data have been used to discover new WR
  stars by the direct detection of mid-IR emission from its
  circumstellar nebula \citep[e.g., WR\,121b,][]{Gv2010c}. Two
  outstanding examples are the identification of the interacting
  nebulae around the WN9h type stars, WR\,120bb and WR\,120bc, by
  \citet{Burgemeister2013} and the identification of a WR nebula in
  the Large Magellanic Cloud reported by \citet{Gv2014}. Of course,
  the direct identification of circumstellar nebulae around WR stars
  in the mid-IR 24 $\mu$m band requires a mandatory spectroscopic
  characterization of the nebular emission, but it is important to
  emphasize the simplicity and potential of the mid-IR method.
 
  In this paper, we have taken advantage of the similarities between
  the spectral responses of the \emph{Spitzer} MIPS 24~$\mu$m and
  \emph{WISE} W4 22~$\mu$m bands to investigate the circumstellar
  medium around a significative sample of known Galactic WR stars.  By
  identifying the mid-IR counterparts of these WR nebulae, we have
  assessed their true extent, disentangling the optical emission of
  the circumstellar shells from that of the ISM along the line of
  sight. New mid-IR shells (e.g., WR\,35, Fig.~\ref{fig:WR18}-bottom
  panels) have also been discovered. It is worth mentioning here the
  notable case of NGC\,3199, the nebula around WR\,18 (see
  Fig.~\ref{fig:WR18}-top panels). \citet{Stock2011} derived the
  chemical abundances for a bright H$\alpha$ nebular clump and
  concluded they were consistent with those of the Galactic H~{\sc ii}
  region M\,17 on which NGC\,3199 is projected.  Accordingly, they
  proposed that NGC\,3199 consists mainly of swept-up ISM material.  A
  careful registration of the location of the
  $\sim$15\arcsec\ VLT/UVES slit used by \citet{Stock2011} on the
  \emph{WISE} images shows that this small slit does not include the
  emission detected in the W4 band (i.e., nebular material), but it is
  located on a spot of bright emission in the W3 band (i.e., ISM
  material).  These findings cast doubt on the ISM abundances implied
  by those authors for NGC\,3199 and rather support those reported by
  \citet{Marston2001}.

\section{Summary and conclusions}

We have examined \textit{WISE} IR images of a sample of 31 WR stars to
study their nebular morphologies in mid-IR bands.  We have then
compared the emission in the \emph{WISE} images with optical H$\alpha$
and (in some cases) [O~{\sc iii}] emission line images.  The variety
of morphologies of WR nebulae can be classified as bubble ${\cal
  B}$-type nebulae, clumpy/disrupted ${\cal C}$-type nebulae, and
mixed ${\cal M}$-type nebulae.

We have used \emph{Spitzer} IRS spectra of some WR nebulae to
investigate the nature of the IR emission of WR nebulae in the
different \emph{WISE} bands (Toal\'{a} et al.\ in preparation).
According to previous IR studies of evolved massive stars, we find
that the emission in the \emph{WISE} W4 band of ${\cal B}$ and ${\cal
  C}$-type nebulae most likely traces dust associated to the WR
nebula. On the other hand, the \emph{WISE} W3 band is mainly tracing
the emission of material along the line of sight of the nebula, which
results in diminished emission at optical wavelengths. We emphasize
that the acquisition of mid-IR images of WR nebulae, especially for
wavelengths above 20 $\mu$m, is very useful to disentangle the
emission from the WR nebula from that of the ISM. This provides a
less demanding observational approach for the identification and study
of WR nebulae than those based on optical images and spectra. The
discovery of an obscured shell around WR\,35 which is detected only at
22 $\mu$m in the W4 band \emph{WISE} images or the capability to
distinguish the nebular material of NGC\,3199 around WR\,18 from that
of the Galactic H~{\sc ii} region M\,17 clearly illustrate the
advantages of mid-IR observations of WR nebulae.

We find a loose correlation between nebular morphology and stellar
spectral type, also claimed in the past, but we note that this
correlation is complicated by the different evolutionary
sequences of WR nebulae and the dependence of the stellar evolution in
these phases with initial mass and composition, and stellar rotation.

\section*{Acknowledgments}

\footnotesize 
We would like to thank the referee, Margaret Meixner, for helpful 
comments. 
JAT acknowledges CSIC JAE-PREDOC (Spain) student grant 2011-00189. 
JAT and MAG are partially funded by grants
AYA\,2008-01934 and AYA\,2011-29754-C03-02 of the Spanish MICINN
(Ministerio de Ciencia e Innovaci\'{o}n) and MEC (Ministerio de
Econom\'{i}a y Competitividad) including FEDER funds. GR-L
acknowledges support from CONACyT (grant 177864) and PROMEP (Mexico).
We are grateful to Y.-H.\ Chu and R.A.\ Gruendl for providing us with
the [O~{\sc iii}] and H$\alpha$ optical images of S\,308, RCW\,58,
RCW\,104, WR\,128, and WR\,134.

This paper is based on observations from the {\it Wide-field Infrared
  Survey Explorer}, which is a joint project of the University of
California, Los Angeles, and the Jet Propulsion Laboratory/California
Institute of Technology, founded by the National Aeronautics and Space
Administration. The Digitized Sky Surveys were produced at the Space
Telescope Science Institute under the U.S. Goverment grant NAG
W-2166. The Second Palomar Observatory Sky Survey (POSS-II) was made
by the California Institute of Techonology with funds from the
National Science Foundation, the National Geographic Society, the
Sloan Foundation, the Samuel Oschin Foundation, and the Eastman Kodak
Corporation.

\appendix
\section{Additional Figures}
\label{sec:appendix}
In this appendix we collect the figures of the rest of the WR nebulae
studied in this paper not presented in the main text. The nebulae are
those around the WR stars WR\,7, WR\,22, WR\,23, WR\,30, WR\,31a,
WR\,38, WR\,54, WR\,55, WR\,68, WR\,75, WR\,85, WR\,86, WR\,94,
WR\,95, WR\,101, WR\,102, WR\,113, WR\,116, WR\,124, WR\,128, WR\,131,
WR\,134, and WR\,136.

\clearpage

\begin{figure}
\begin{center}
\includegraphics[width=0.5\linewidth]{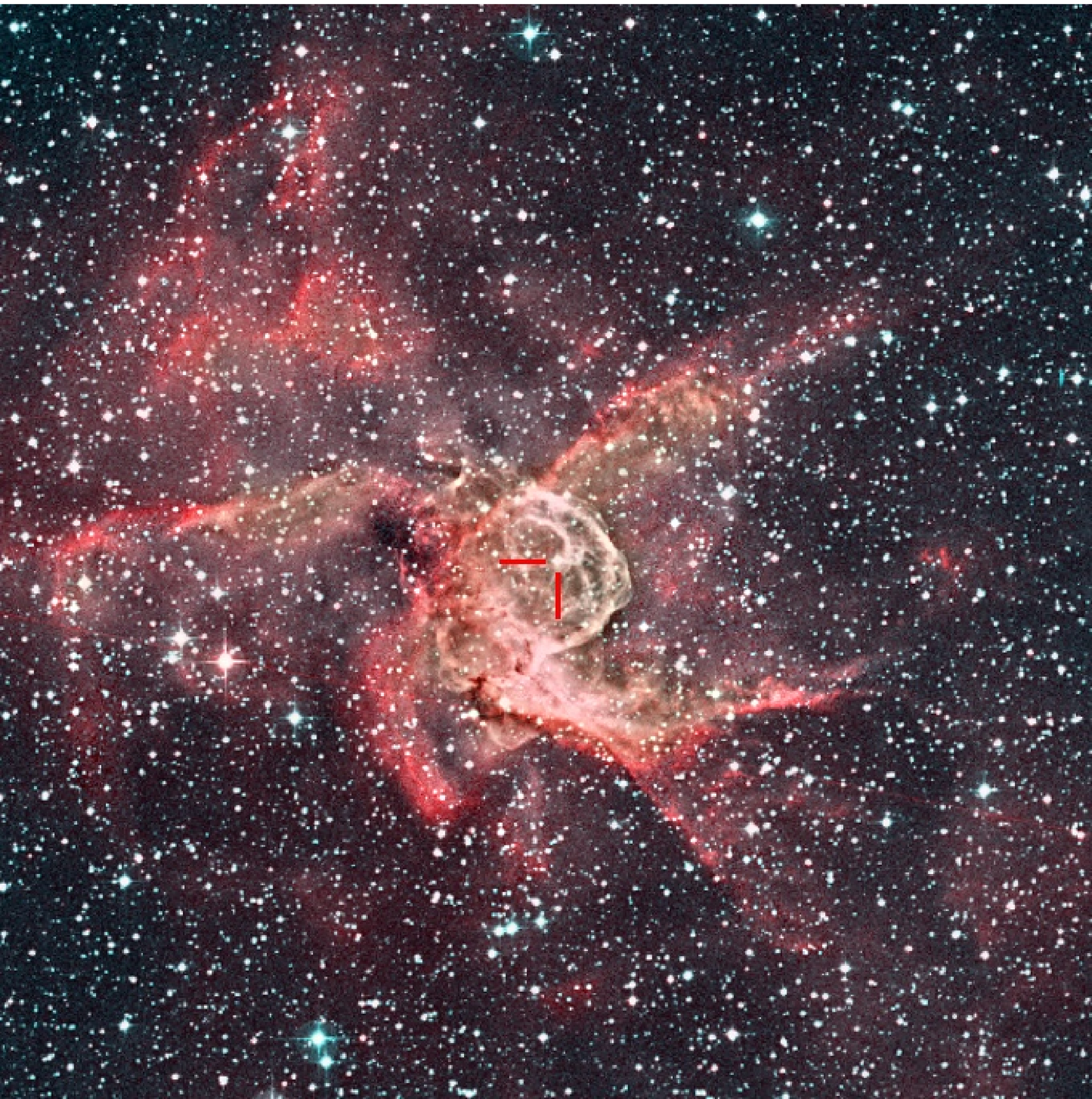}~
\includegraphics[width=0.5\linewidth]{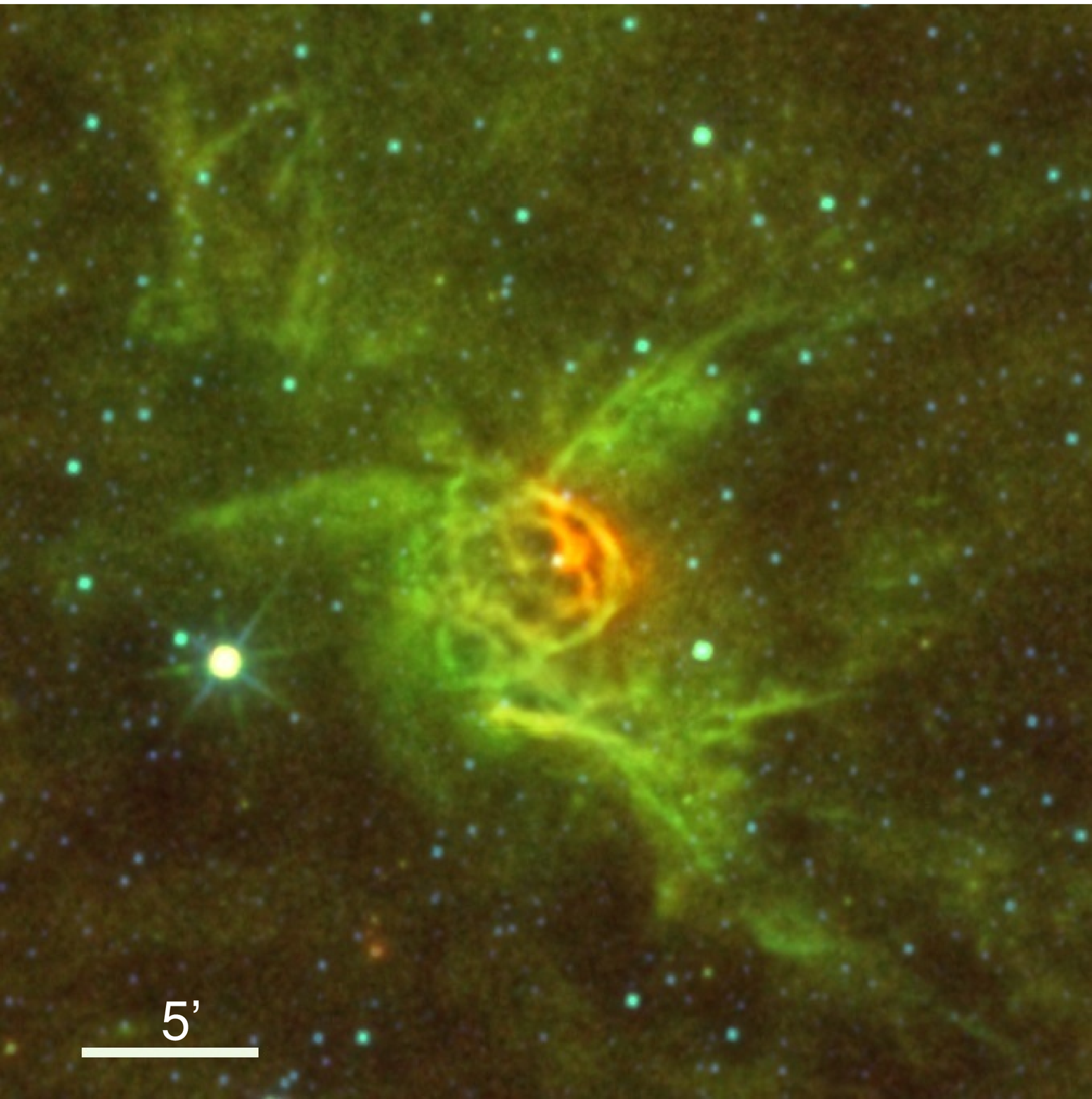}
\end{center}
\caption{Optical (left) and mid-IR {\it WISE} (right) images of WR\,7
  (NGC\,2359). See Table~1 for details of the optical image. The
  central WR star is marked with red lines in the left panel. North is
  up, East to the left.}
\label{fig:WR7}
\end{figure}

\begin{figure}
\begin{center}
\includegraphics[width=0.5\linewidth]{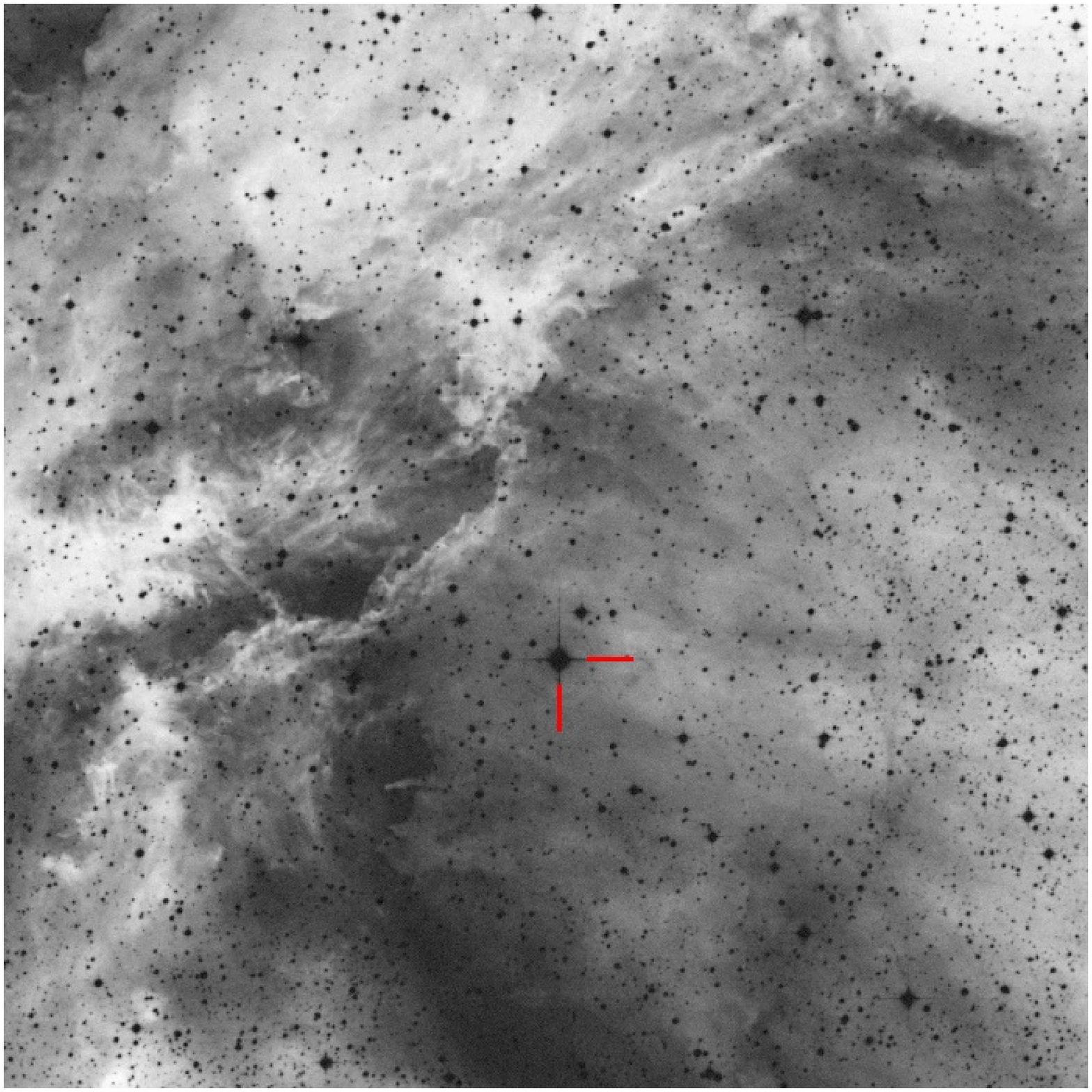}~
\includegraphics[width=0.5\linewidth]{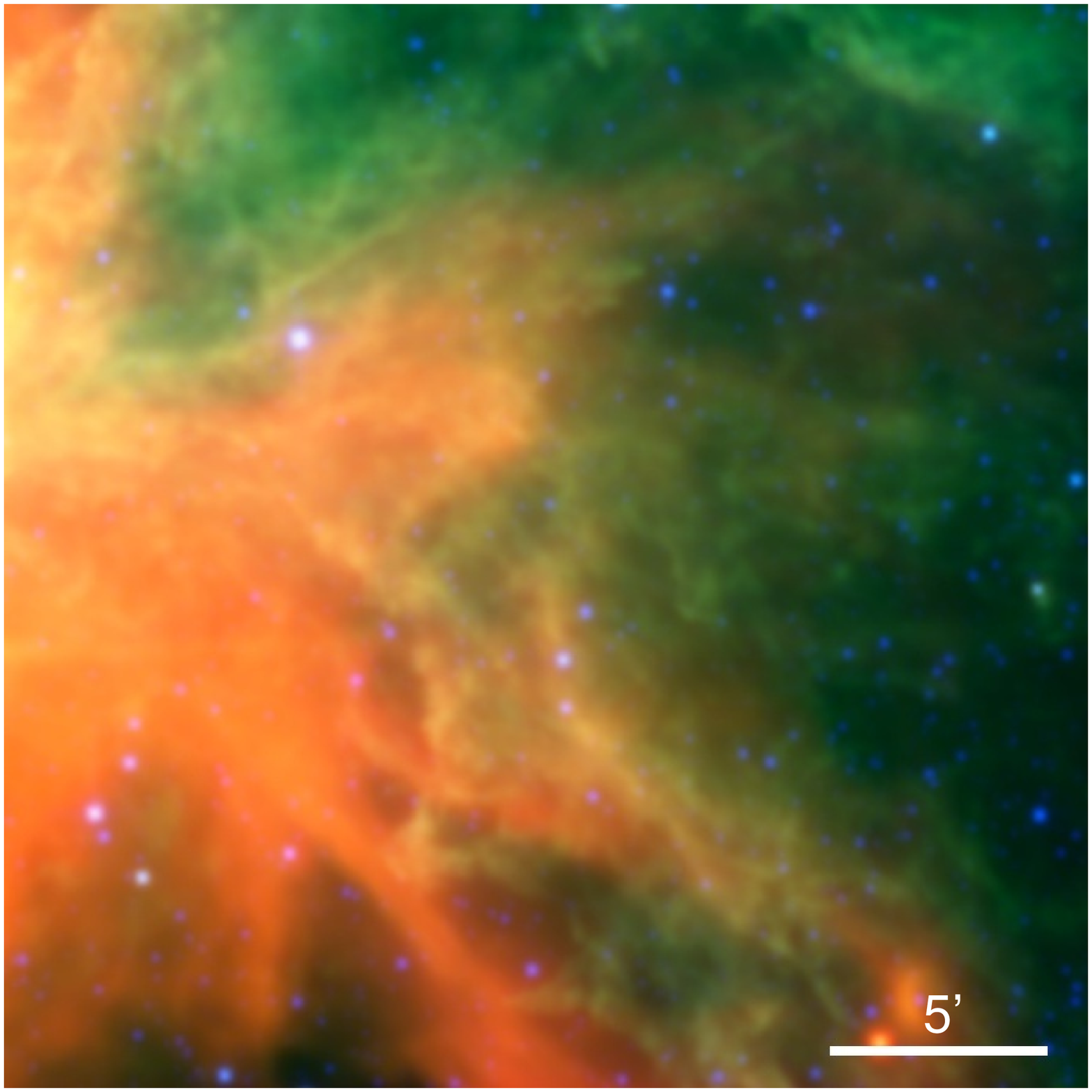}
\end{center}
\caption{Same as Fig.~\ref{fig:WR7} for WR\,22.}
\label{fig:WR22}
\end{figure}

\begin{figure}
\begin{center}
\includegraphics[width=0.5\linewidth]{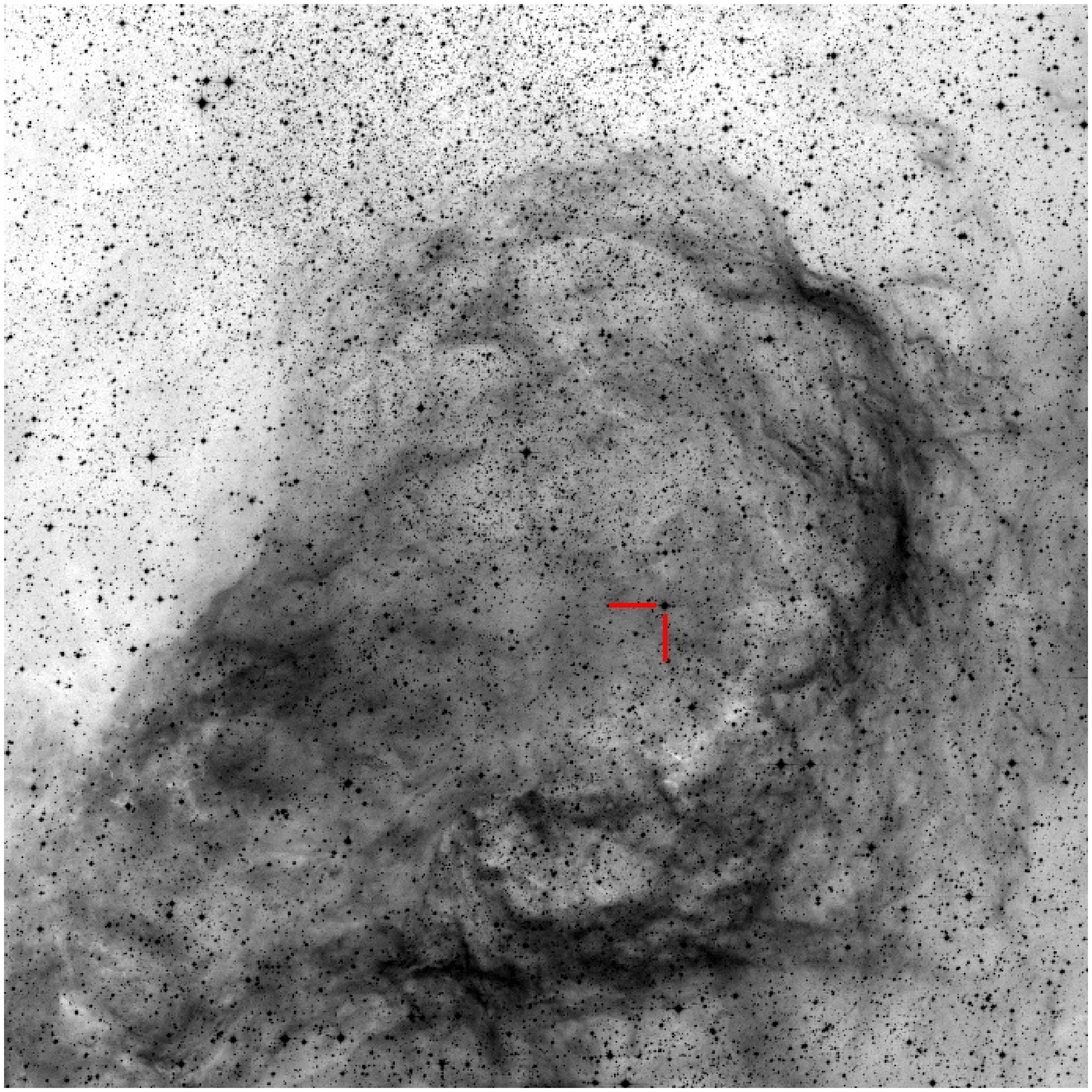}~
\includegraphics[width=0.5\linewidth]{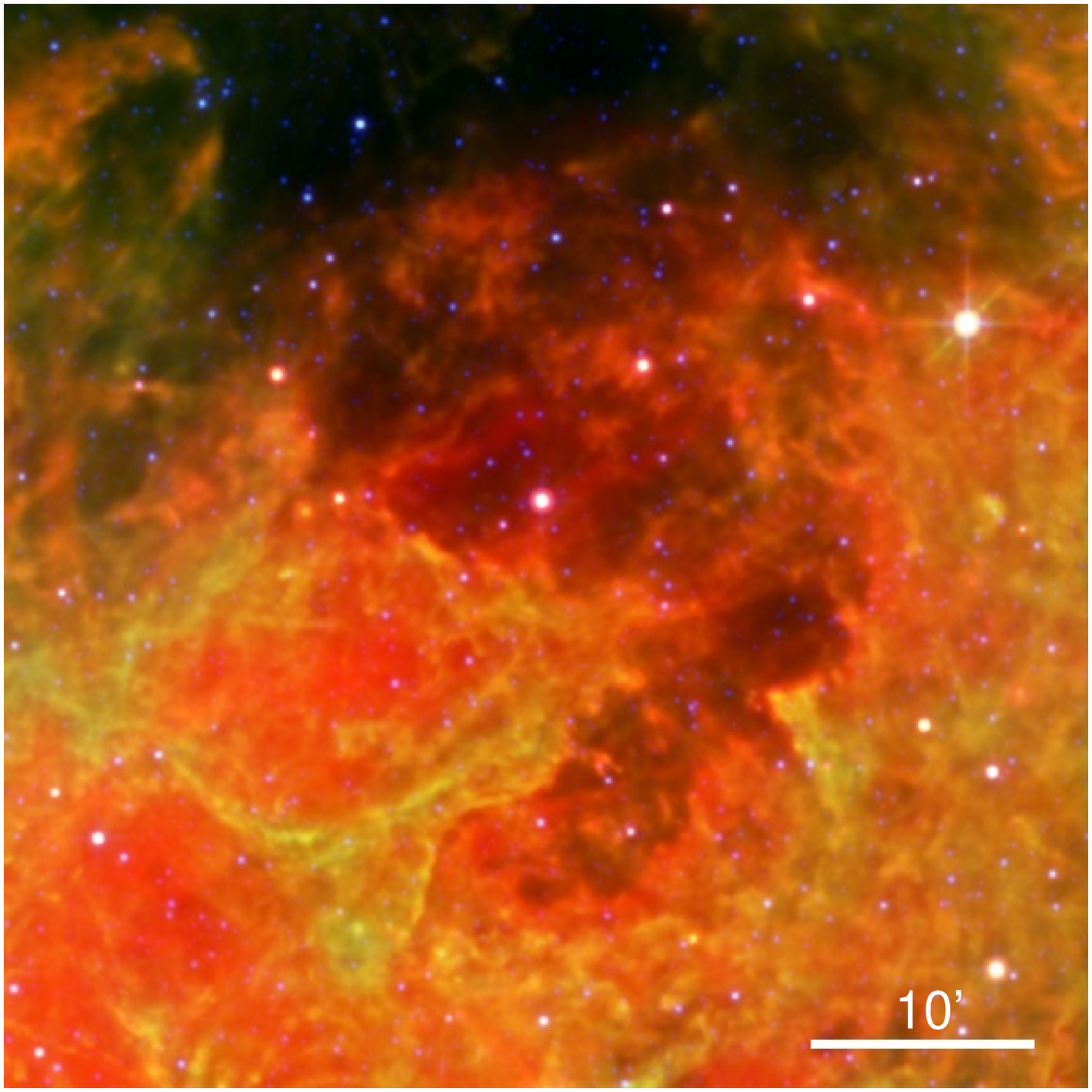}
\end{center}
\caption{Same as Fig.~\ref{fig:WR7} for WR\,23.}
\label{fig:WR23}
\end{figure}

\begin{figure}
\begin{center}
\includegraphics[width=0.5\linewidth]{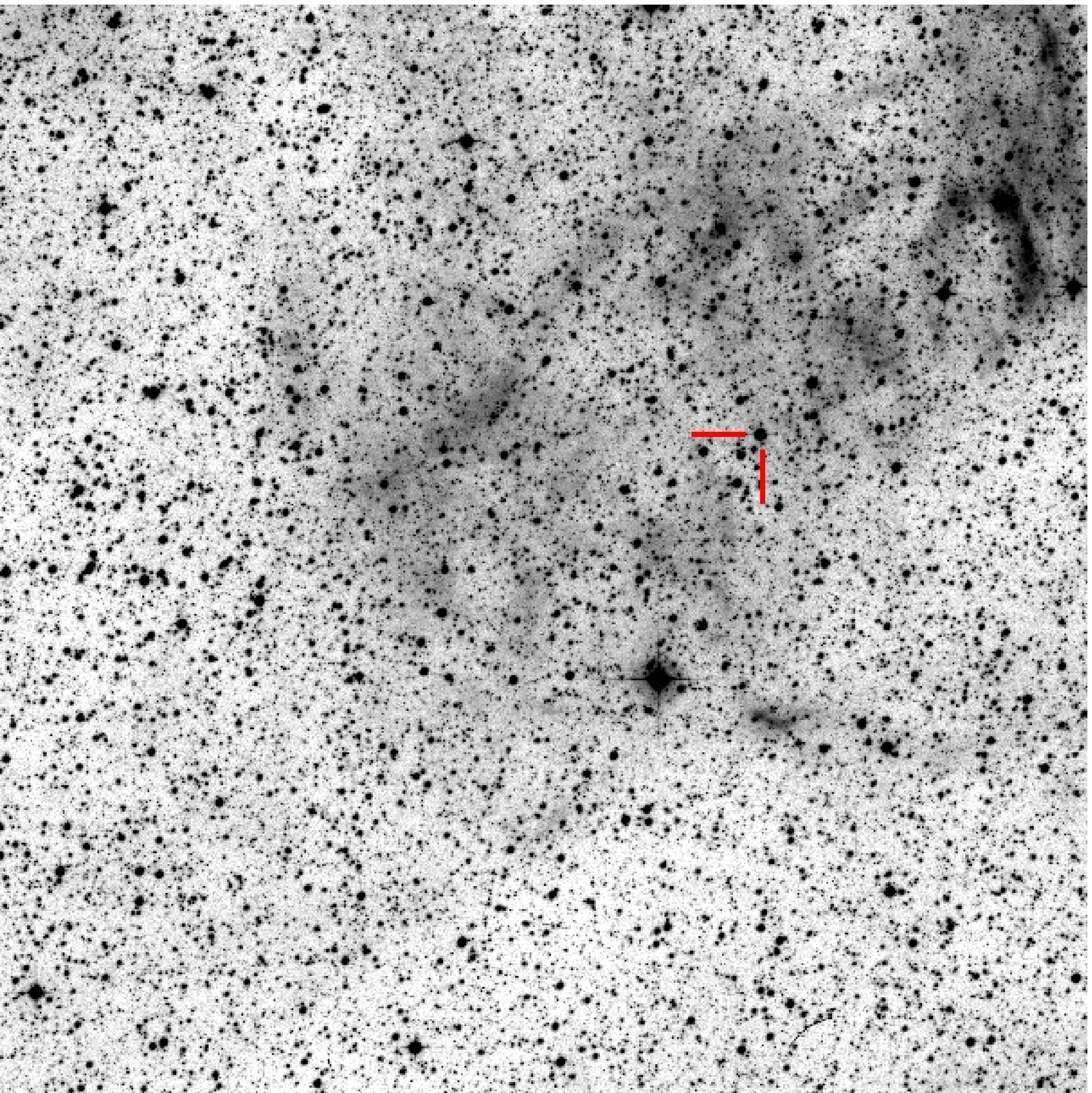}~
\includegraphics[width=0.5\linewidth]{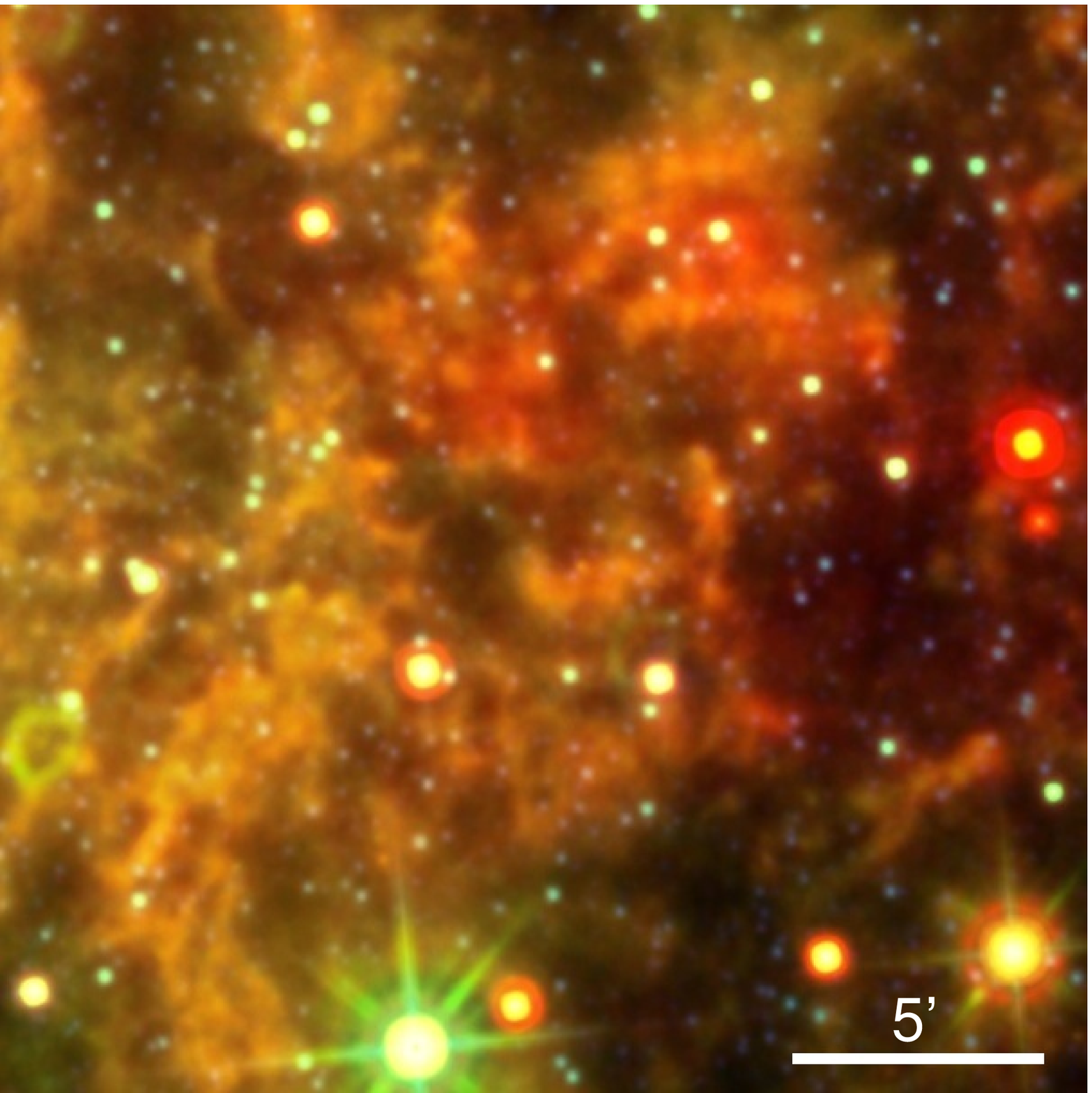}
\end{center}
\caption{Same as Fig.~\ref{fig:WR7} for WR\,30 (Anon).}
\label{fig:WR30}
\end{figure}

\begin{figure}
\begin{center}
\includegraphics[width=0.5\linewidth]{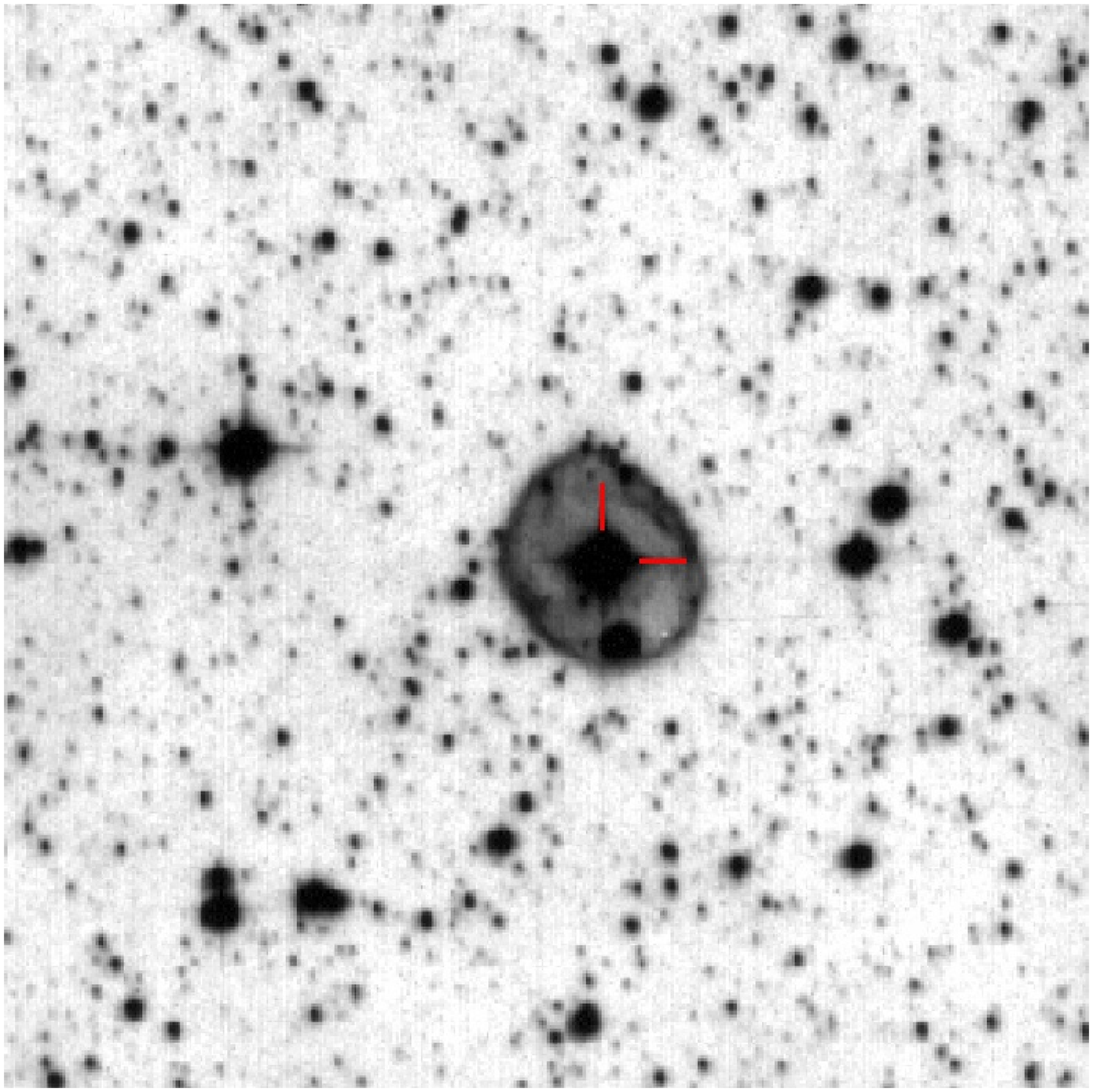}~
\includegraphics[width=0.5\linewidth]{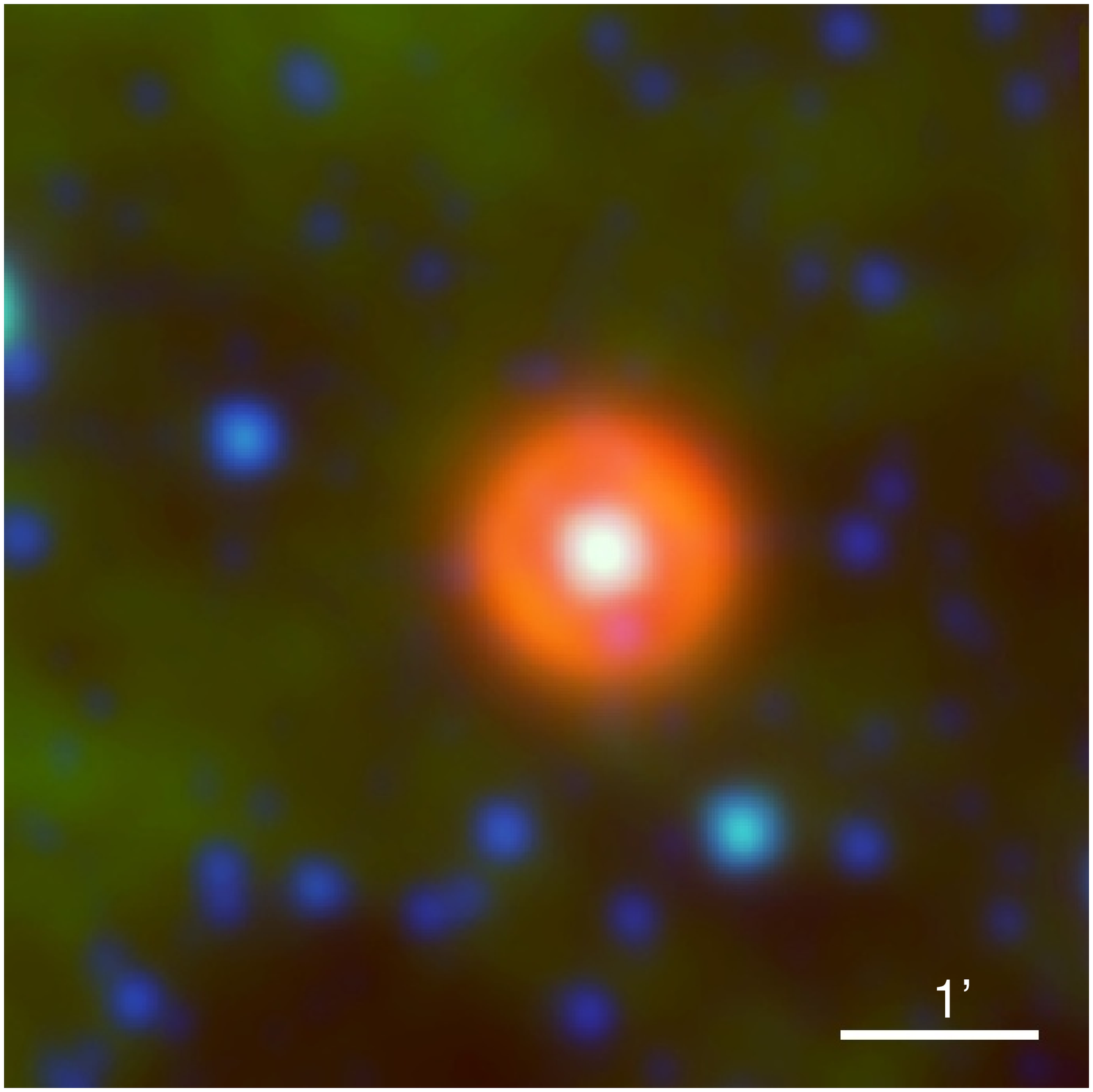}
\end{center}
\caption{Same as Fig.~\ref{fig:WR7} for WR\,31a.}
\label{fig:WR31a}
\end{figure}

\begin{figure}
\begin{center}
\includegraphics[width=0.5\linewidth]{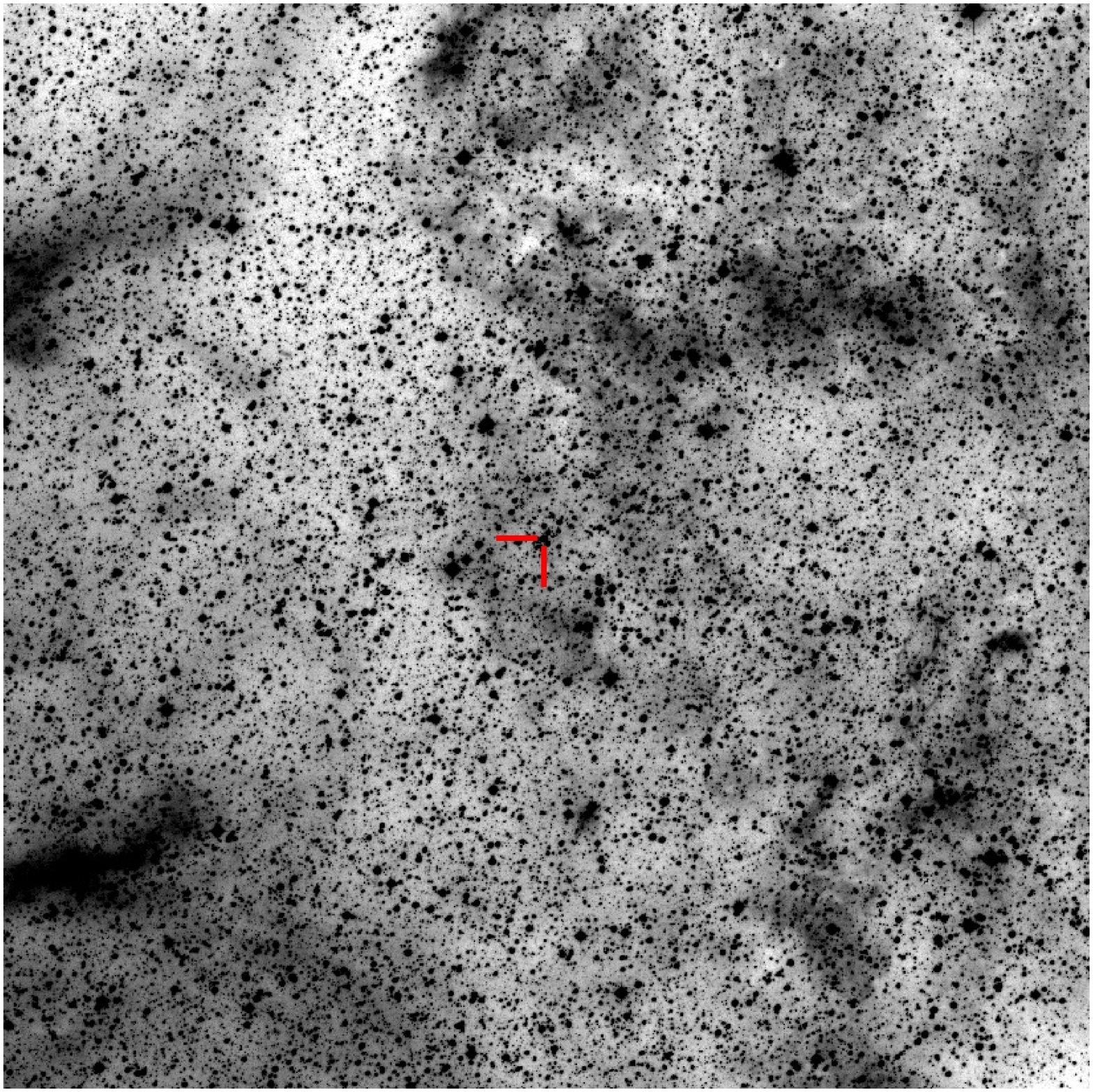}~
\includegraphics[width=0.5\linewidth]{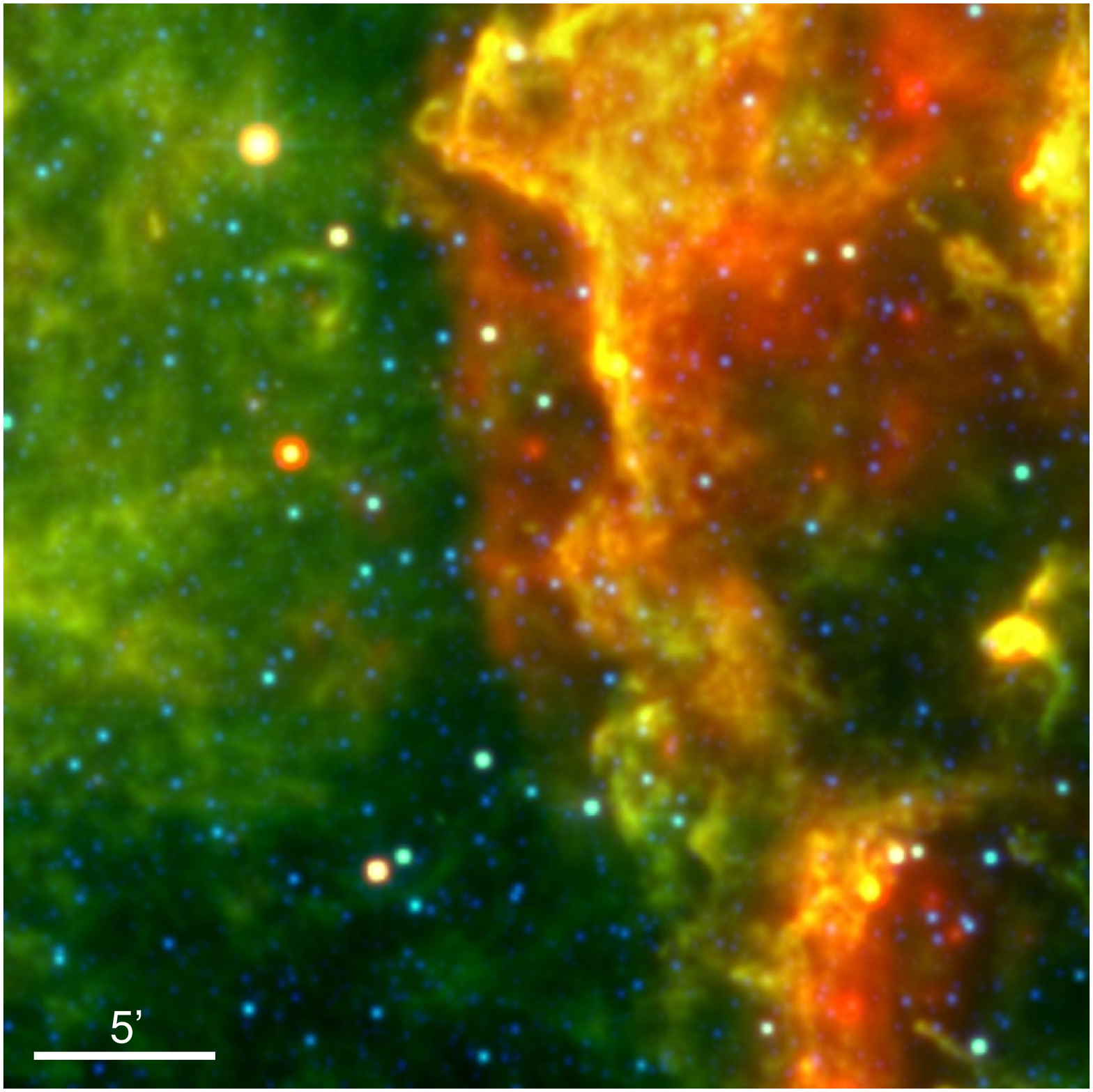}
\end{center}
\caption{Same as Fig.~\ref{fig:WR7} for WR\,38.}
\label{fig:WR38}
\end{figure}

\begin{figure}
\begin{center}
\includegraphics[width=0.5\linewidth]{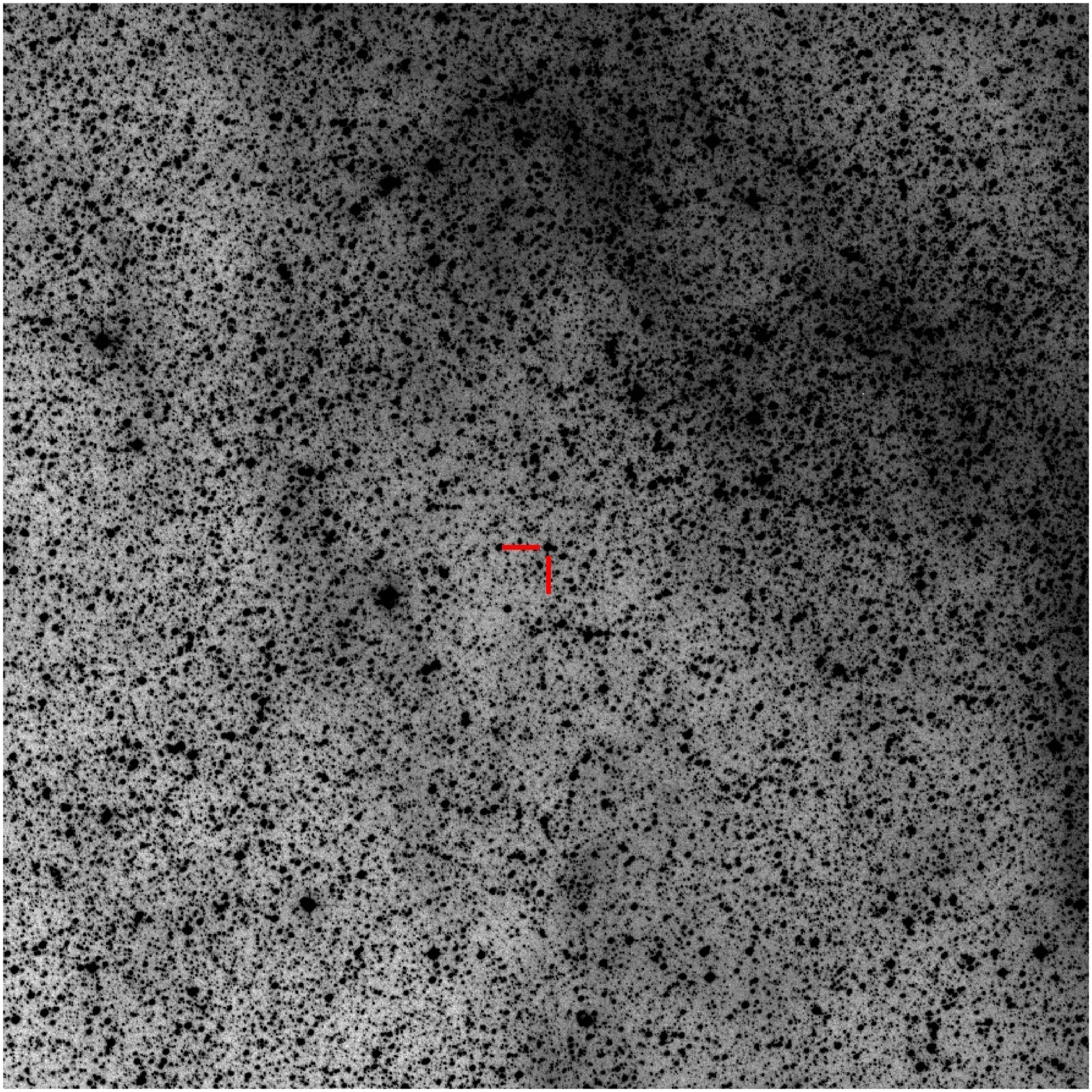}~
\includegraphics[width=0.5\linewidth]{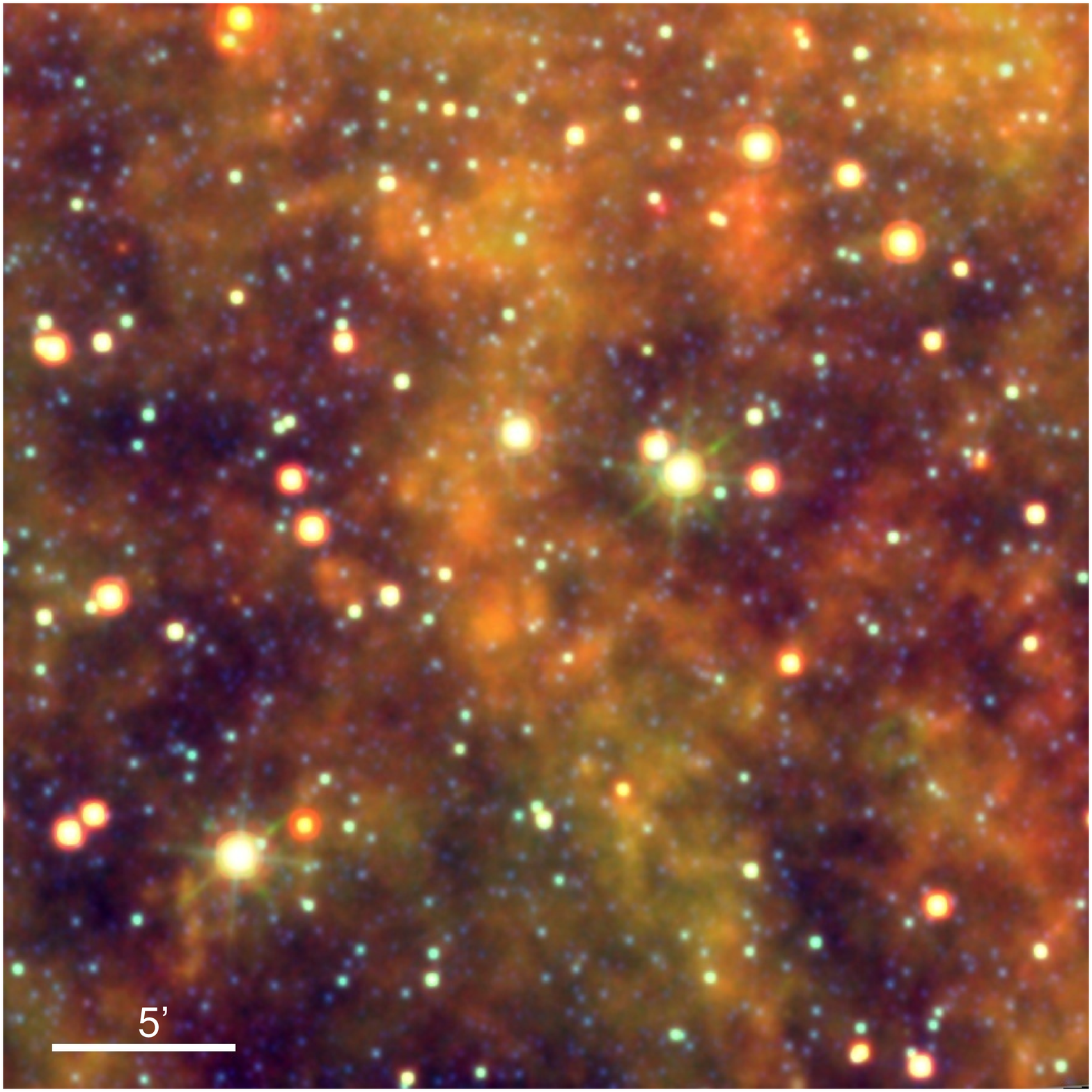}
\end{center}
\caption{Same as Fig.~\ref{fig:WR7} for WR\,54.}
\label{fig:WR54}
\end{figure}

\begin{figure}
\begin{center}
\includegraphics[width=0.5\linewidth]{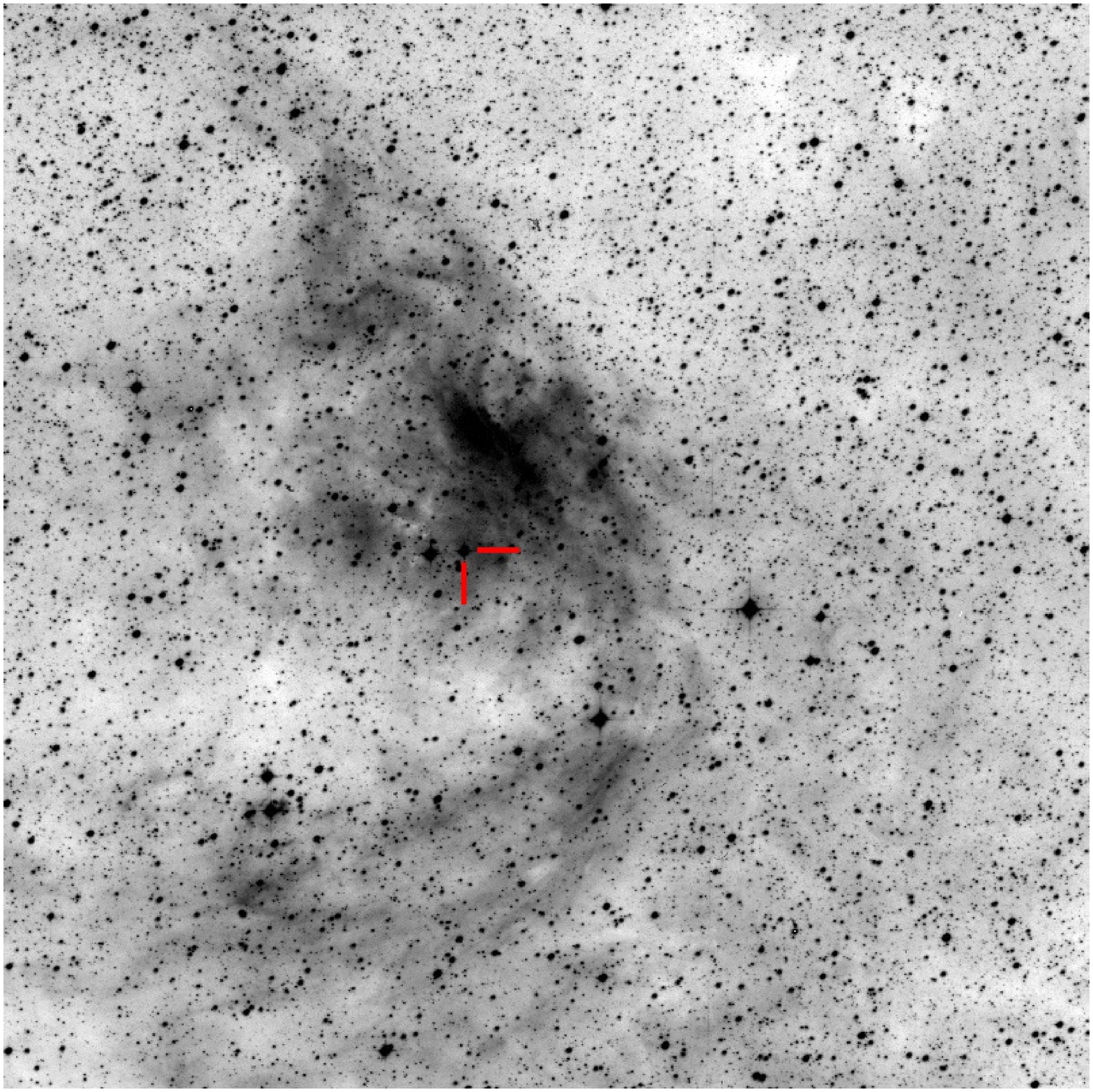}~
\includegraphics[width=0.5\linewidth]{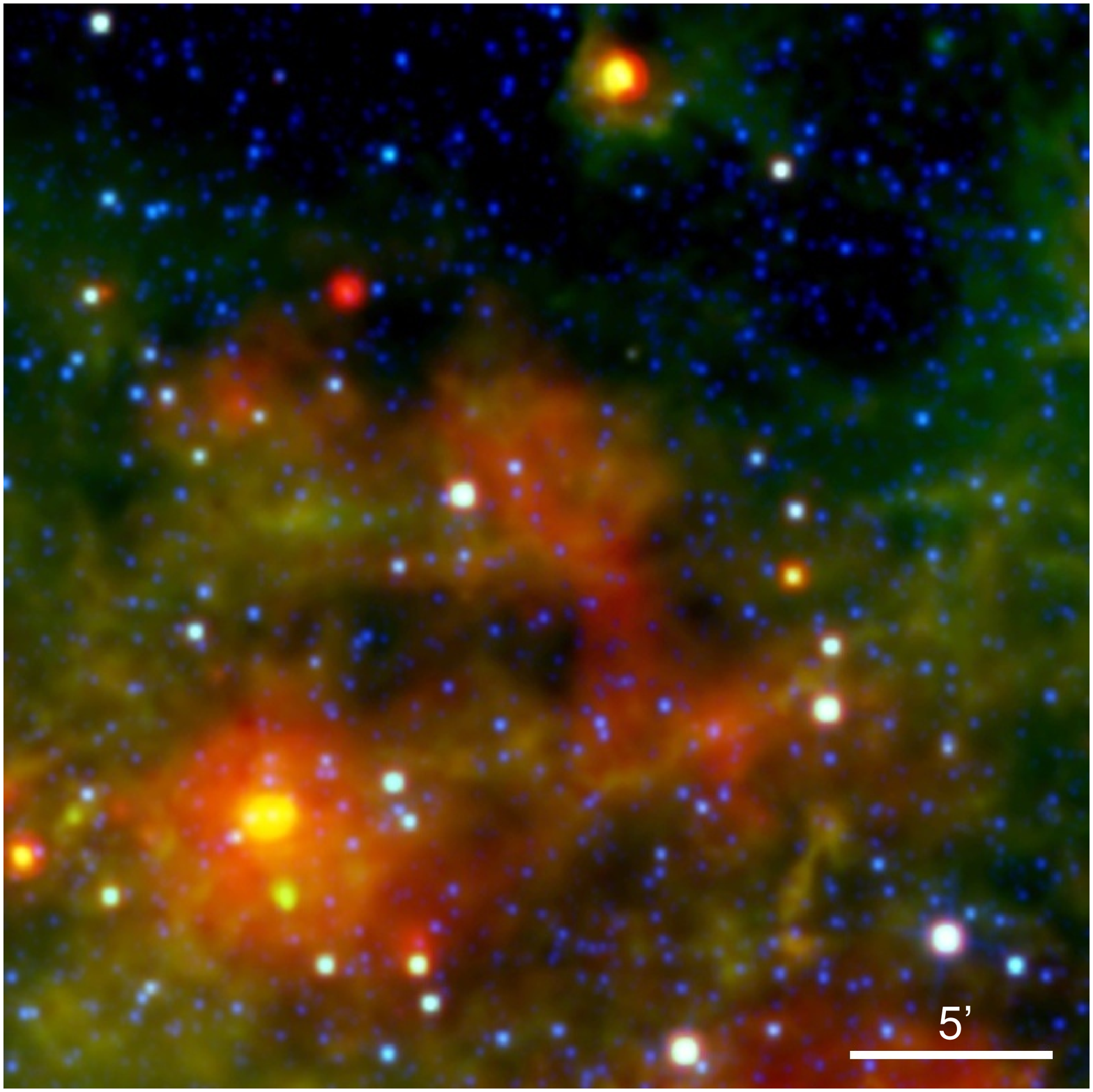}
\end{center}
\caption{Same as Fig.~\ref{fig:WR7} for WR\,55 (RCW\,78).}
\label{fig:WR55}
\end{figure}

\clearpage

\begin{figure}
\begin{center}
\includegraphics[width=0.5\linewidth]{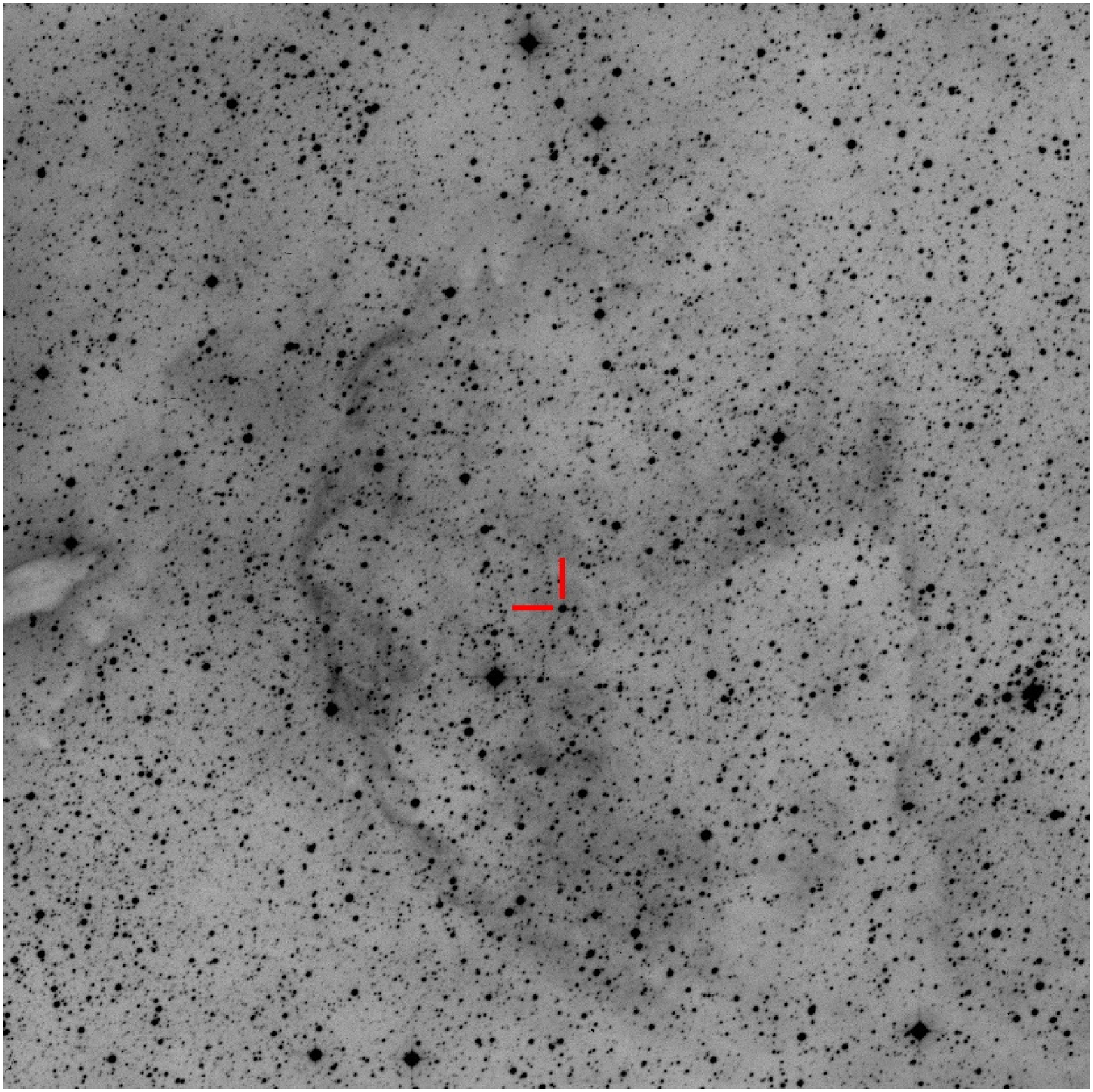}~
\includegraphics[width=0.5\linewidth]{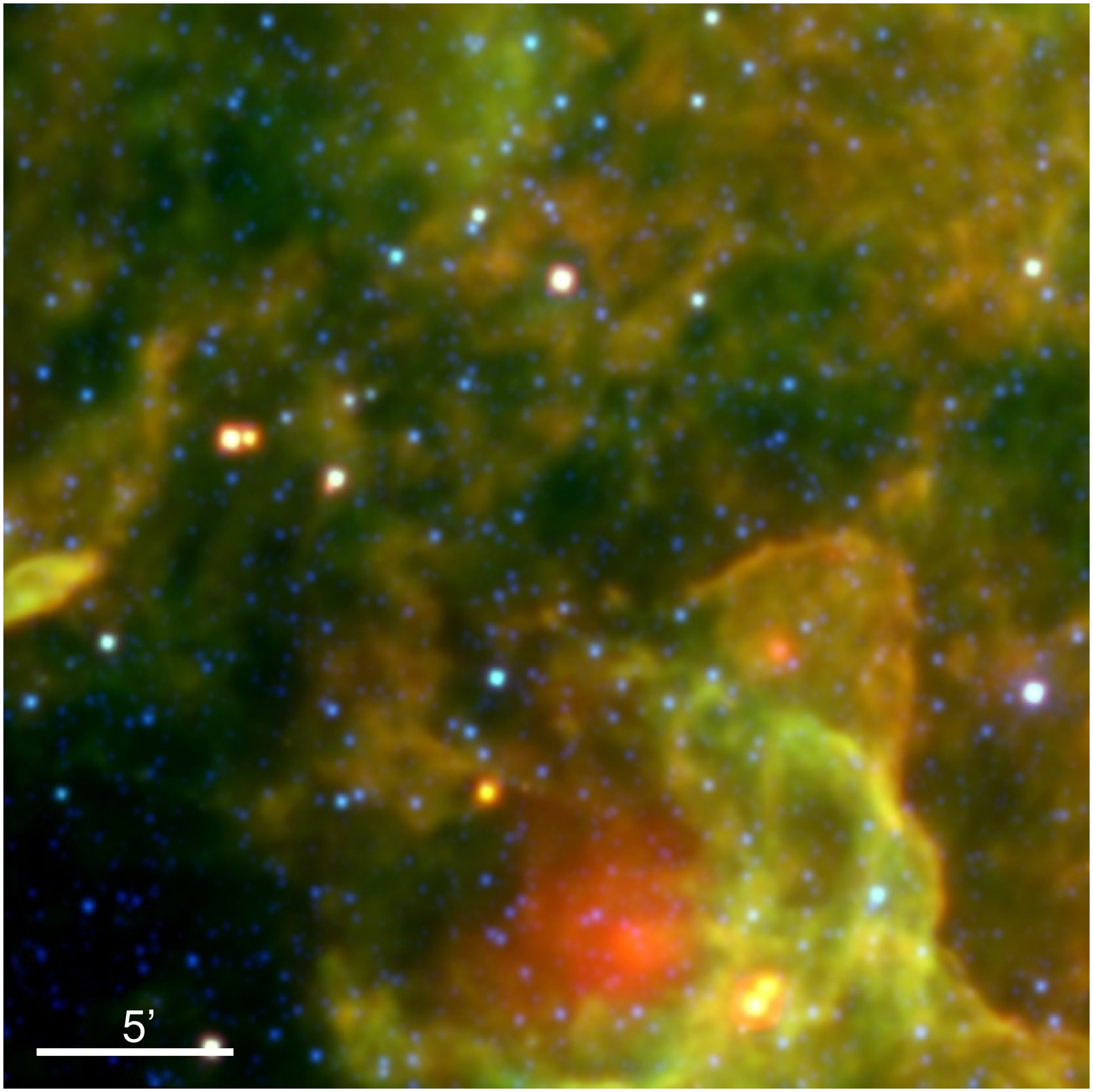}
\end{center}
\caption{Same as Fig.~\ref{fig:WR7} for WR\,68 (G320.5$-$1.4).}
\label{fig:WR68}
\end{figure}

\begin{figure}
\begin{center}
\includegraphics[width=0.5\linewidth]{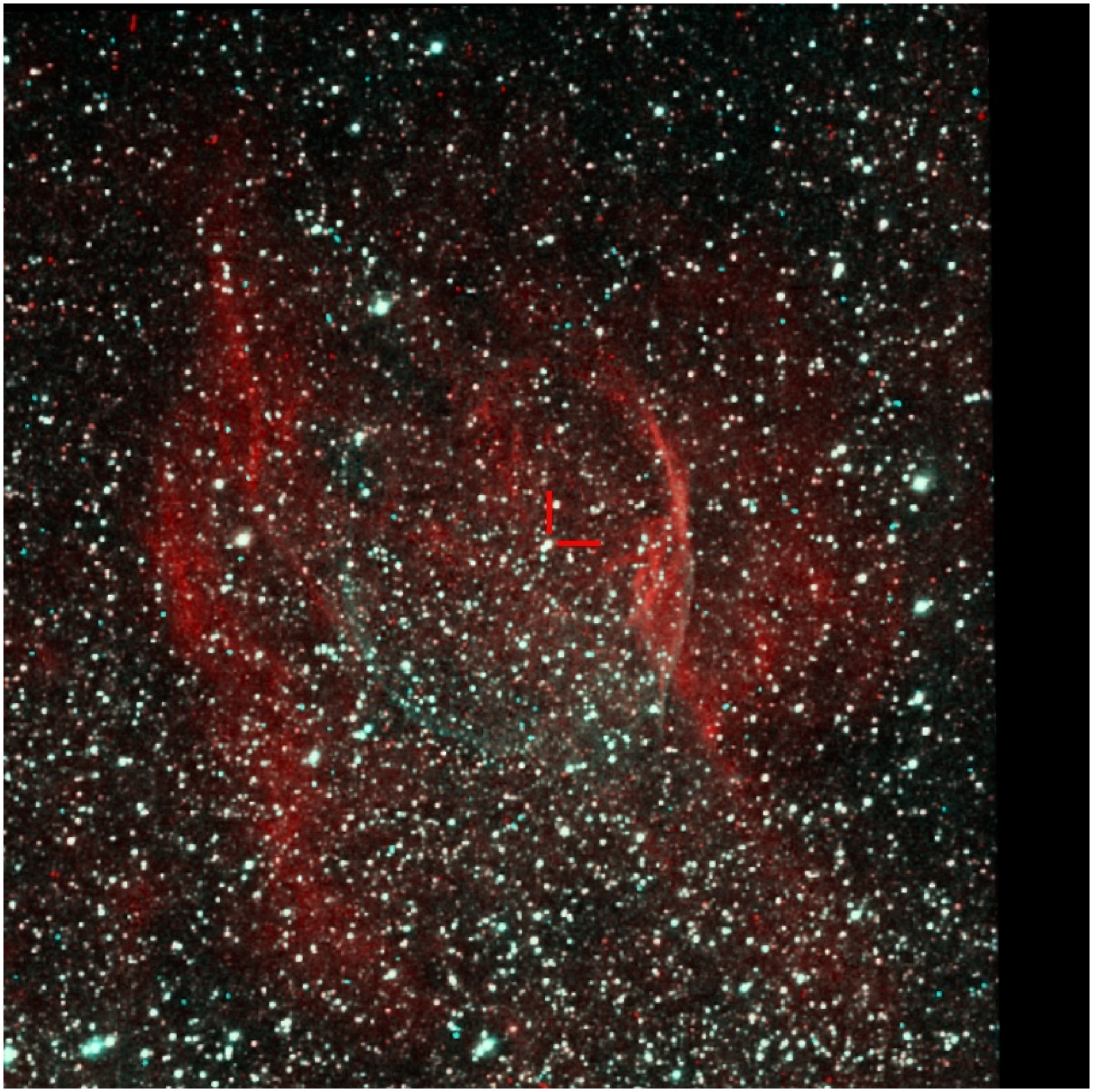}~
\includegraphics[width=0.5\linewidth]{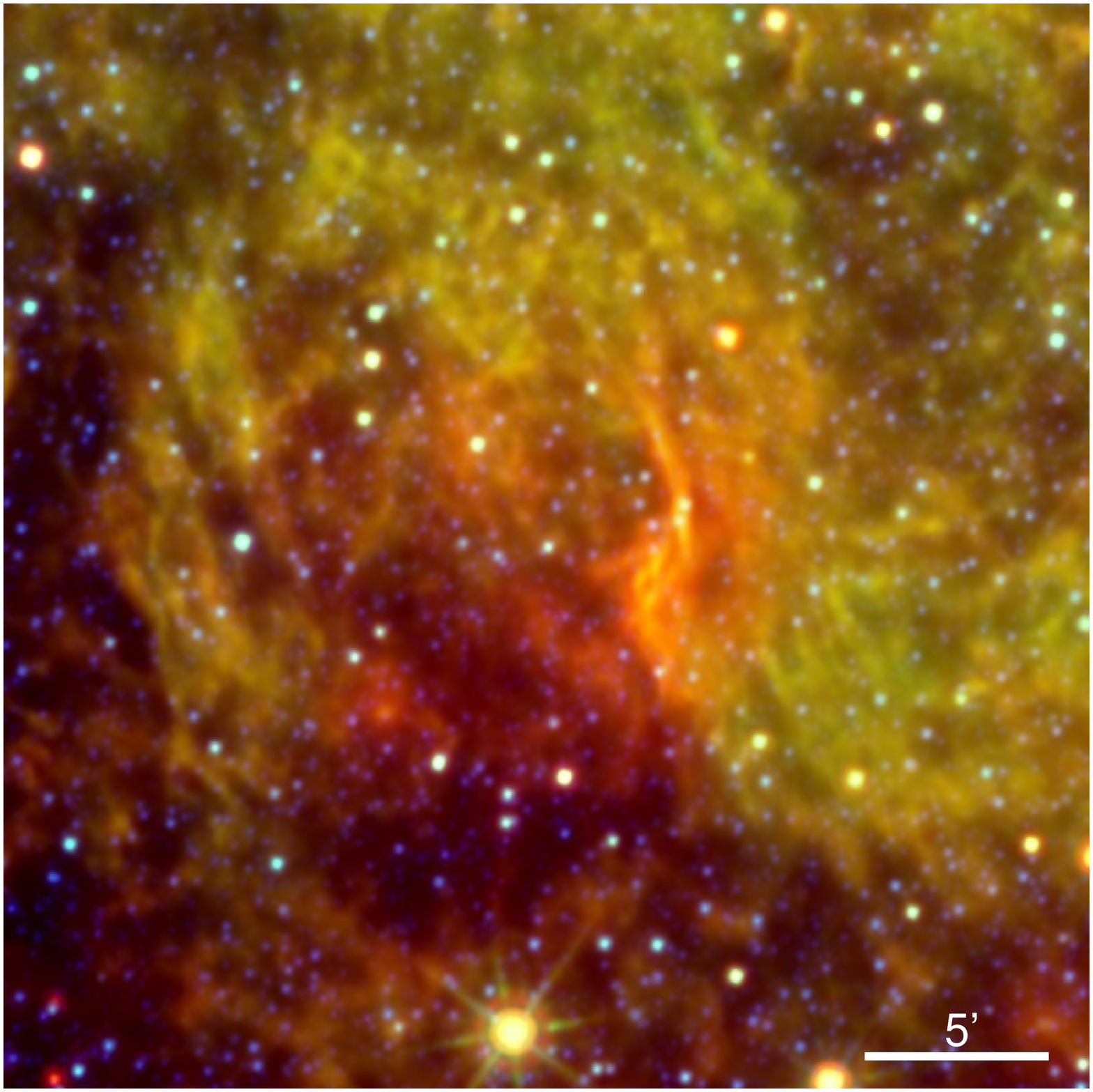}
\end{center}
\caption{Same as Fig.~\ref{fig:WR7} for WR\,75 (RCW\,104).}
\label{fig:WR75}
\end{figure}

\begin{figure}
\begin{center}
\includegraphics[width=0.5\linewidth]{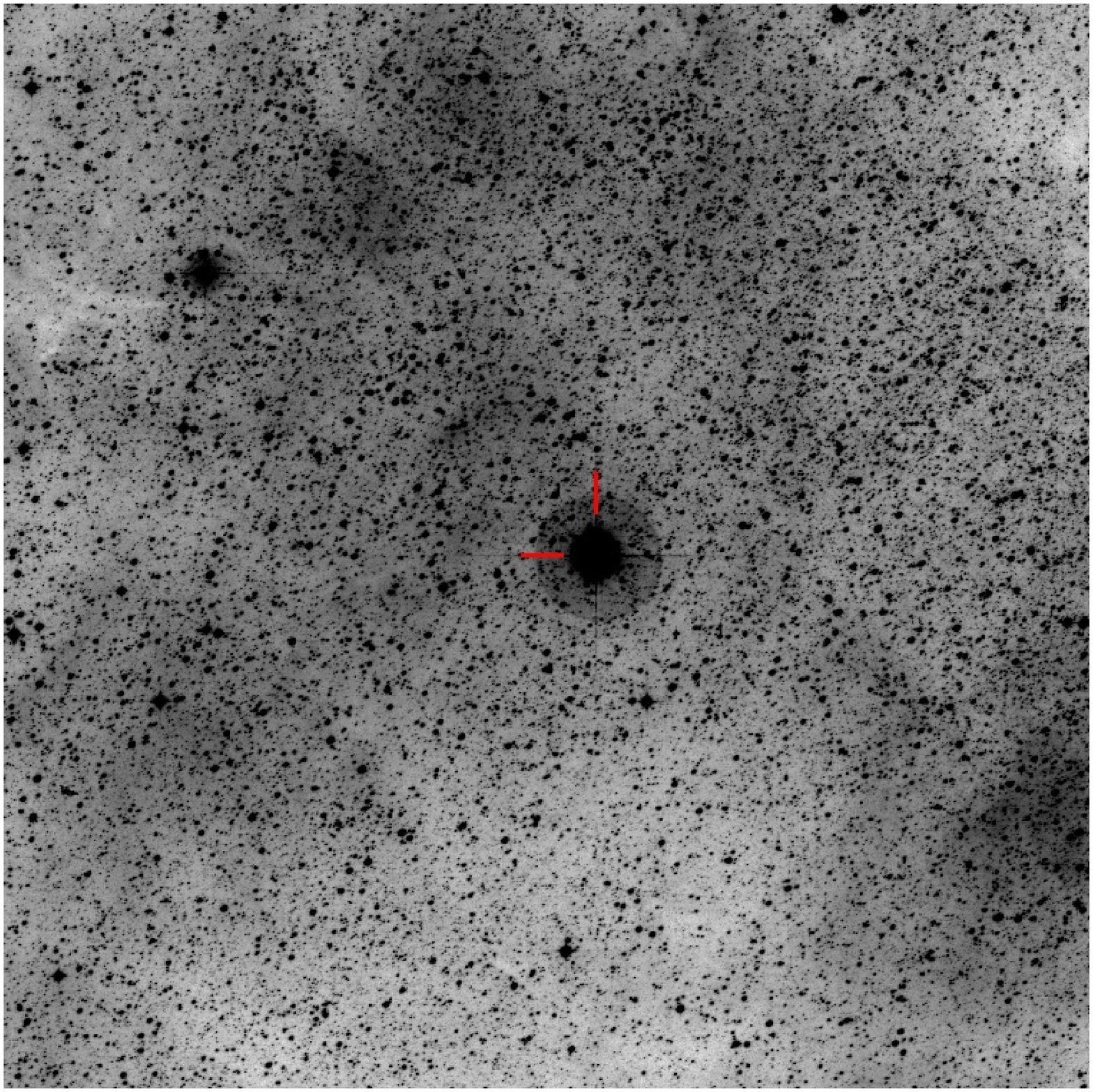}~
\includegraphics[width=0.5\linewidth]{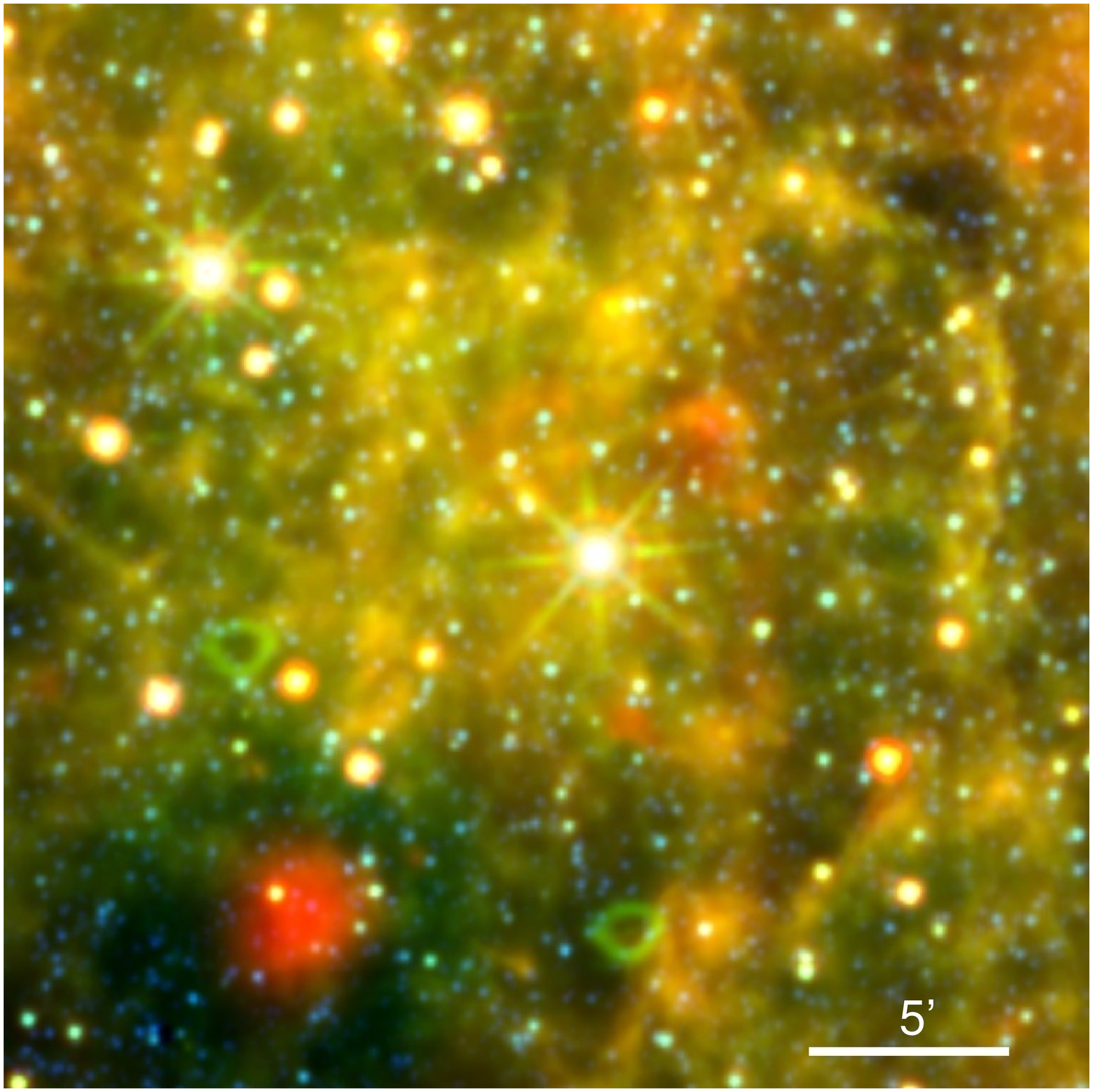}
\end{center}
\caption{Same as Fig.~\ref{fig:WR7} for WR\,85 (RCW\,118).}
\label{fig:WR85}
\end{figure}

\begin{figure}
\begin{center}
\includegraphics[width=0.5\linewidth]{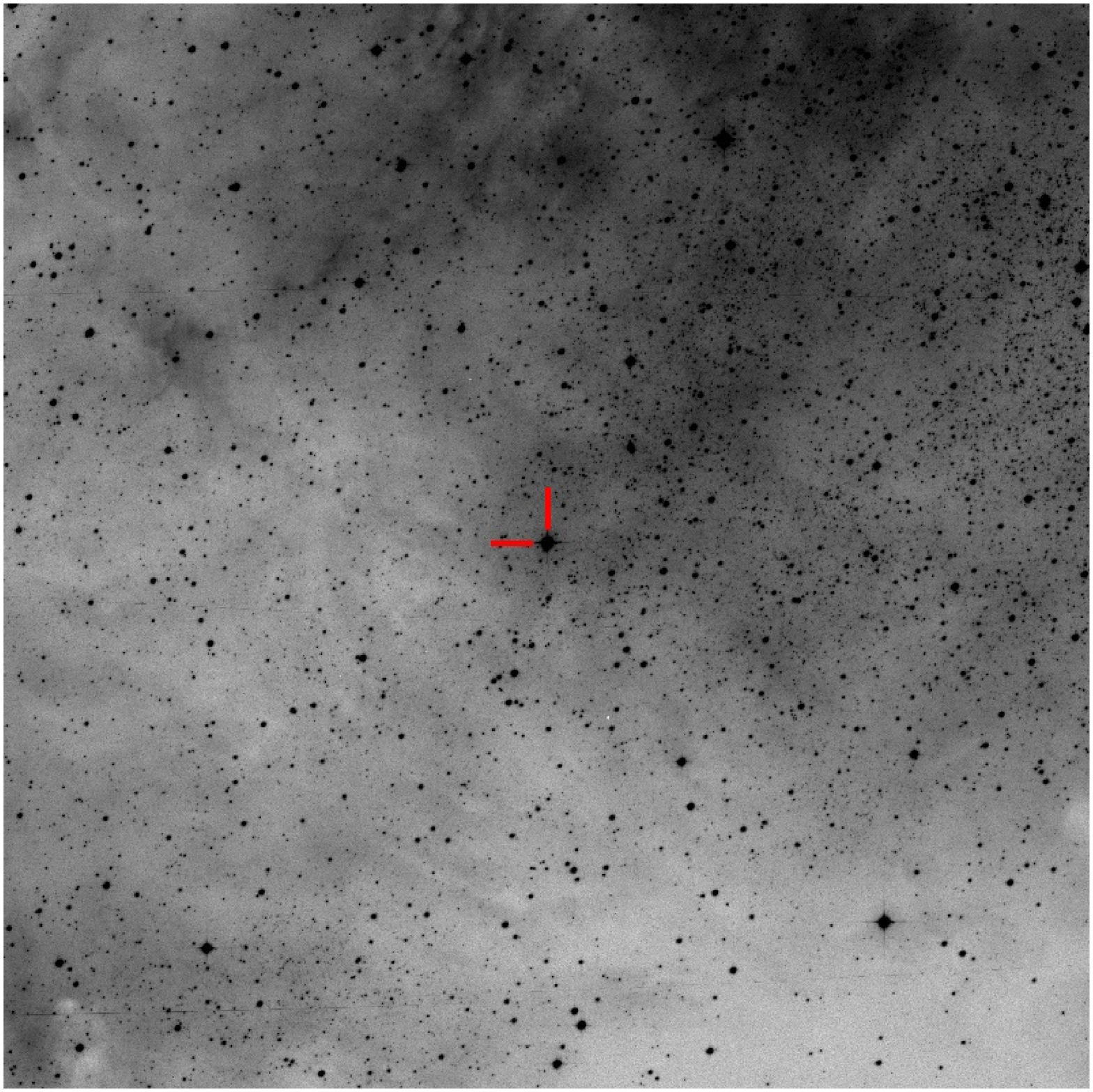}~
\includegraphics[width=0.5\linewidth]{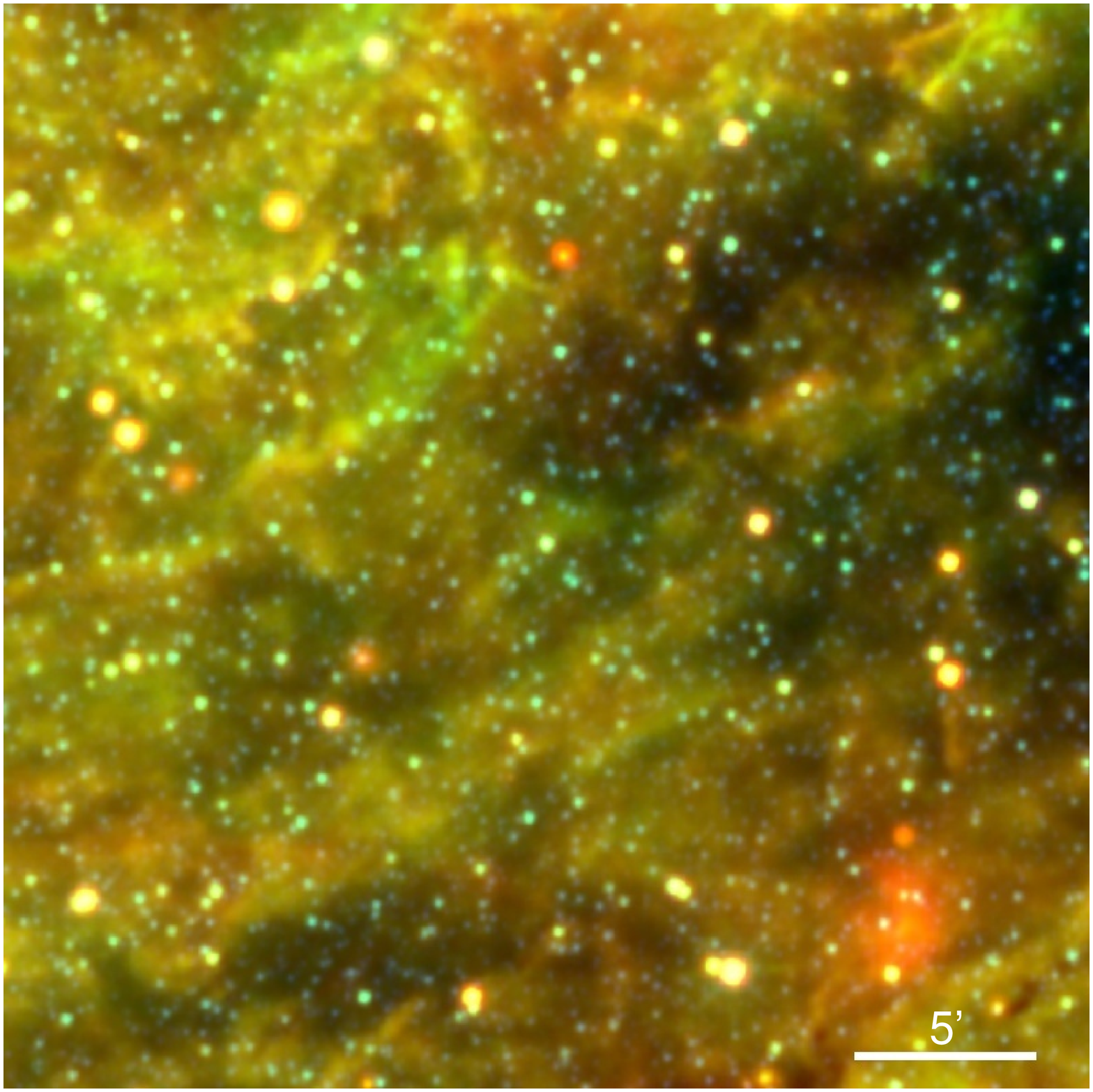}
\end{center}
\caption{Same as Fig.~\ref{fig:WR7} for WR\,86 (RCW\,130).} 
\label{fig:WR86}
\end{figure}

\begin{figure}
\begin{center}
\includegraphics[width=0.5\linewidth]{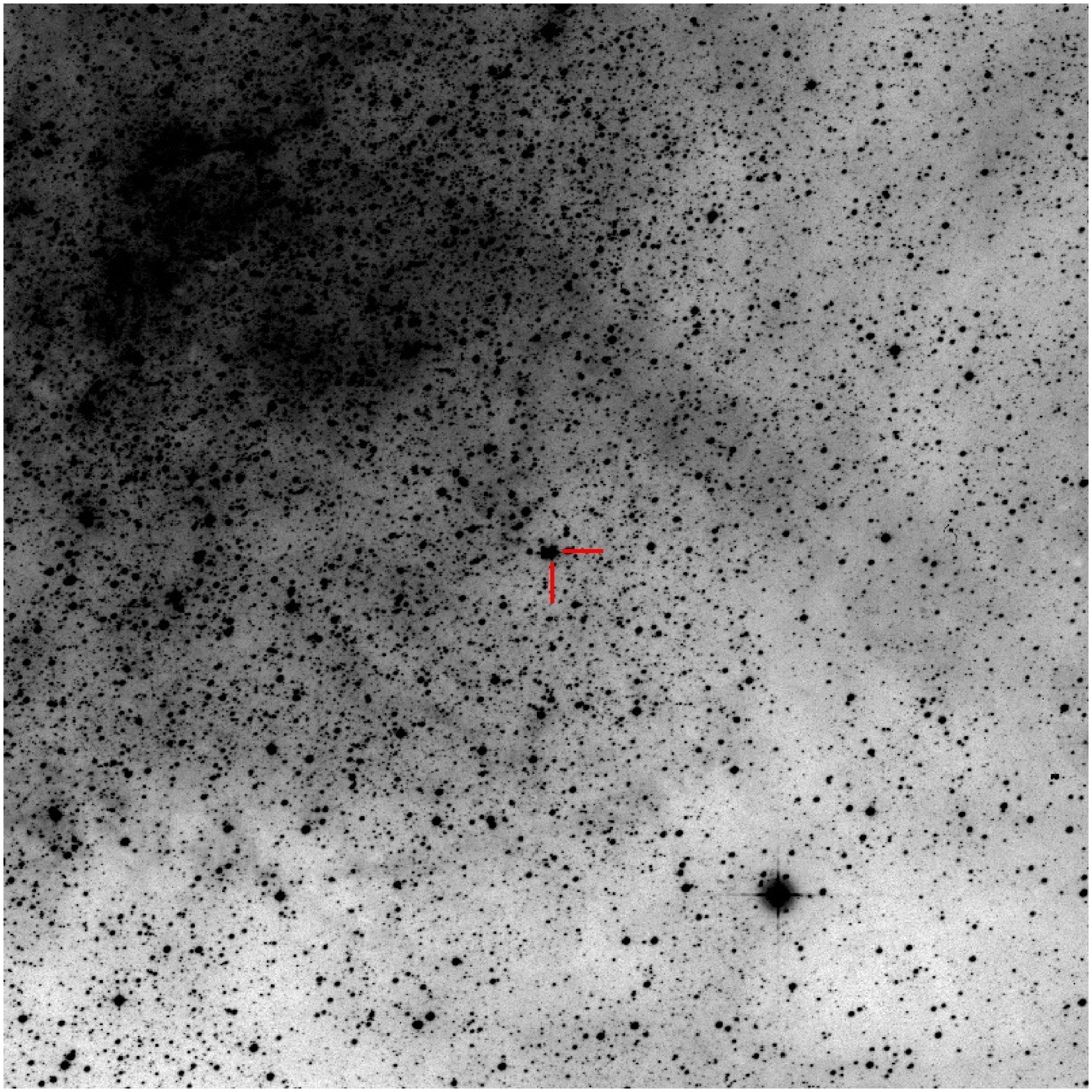}~
\includegraphics[width=0.5\linewidth]{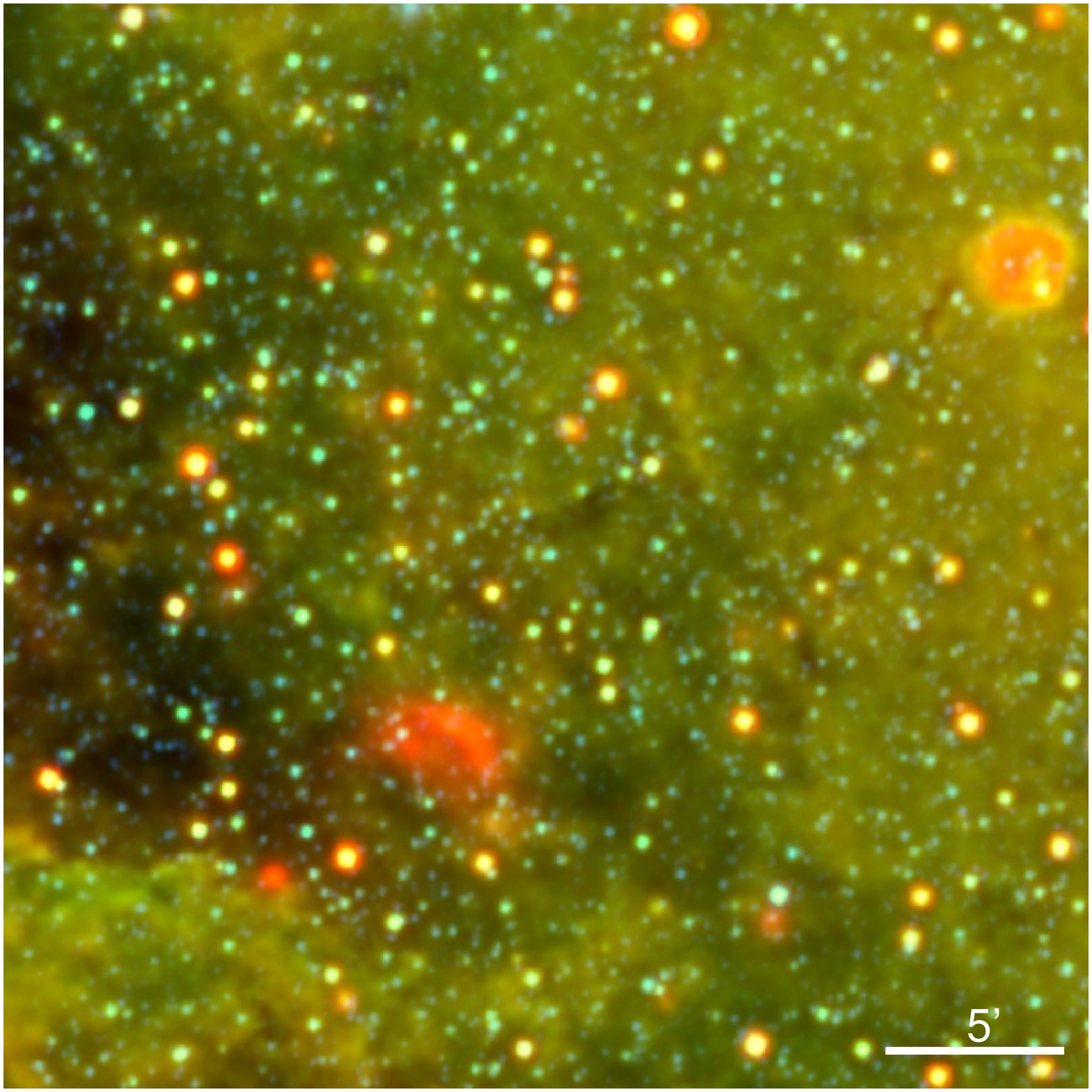}
\end{center}
\caption{Same as Fig.~\ref{fig:WR7} for WR\,94 (Anon)}
\label{fig:WR94}
\end{figure}

\begin{figure}
\begin{center}
\includegraphics[width=0.5\linewidth]{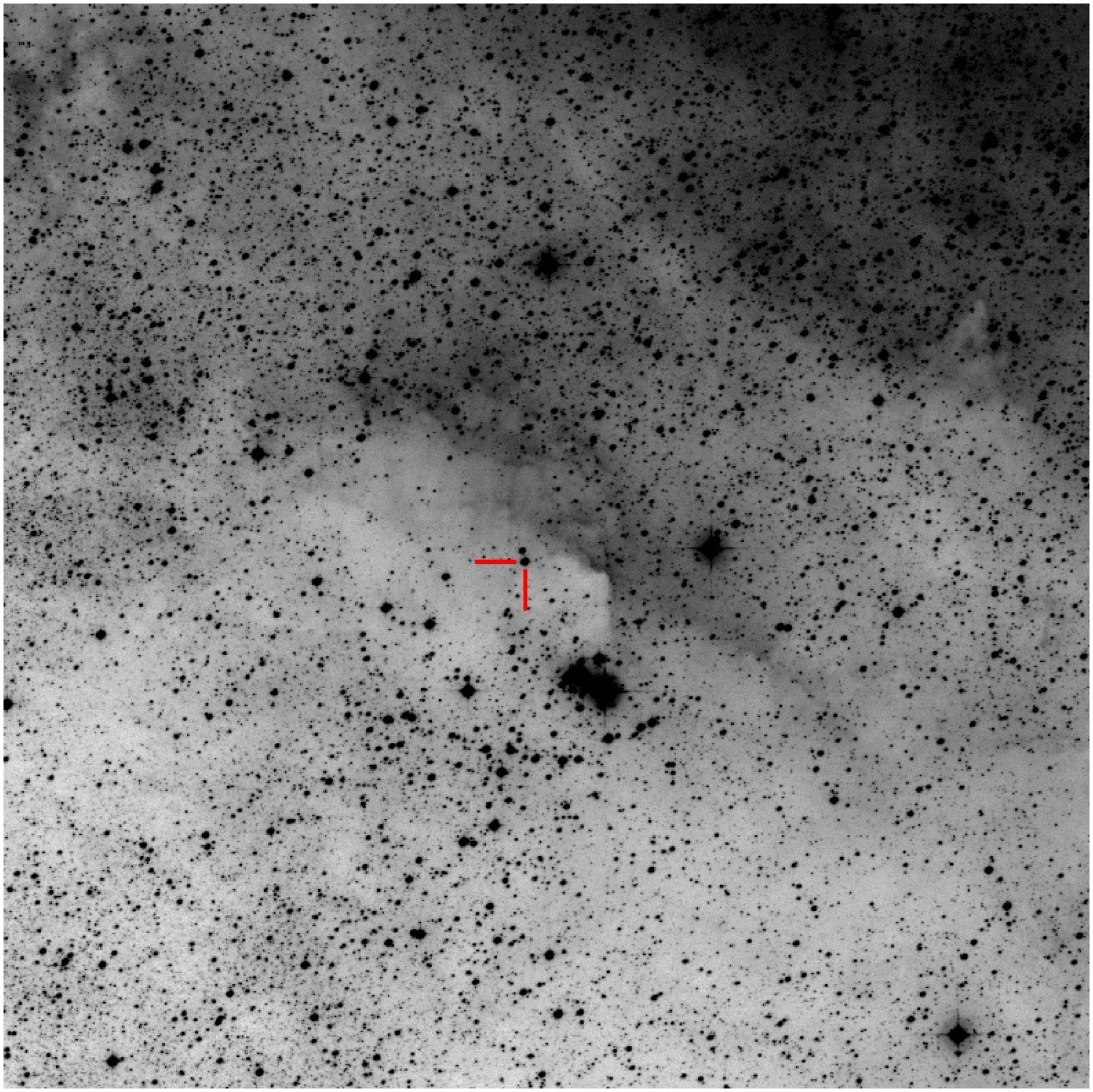}~
\includegraphics[width=0.5\linewidth]{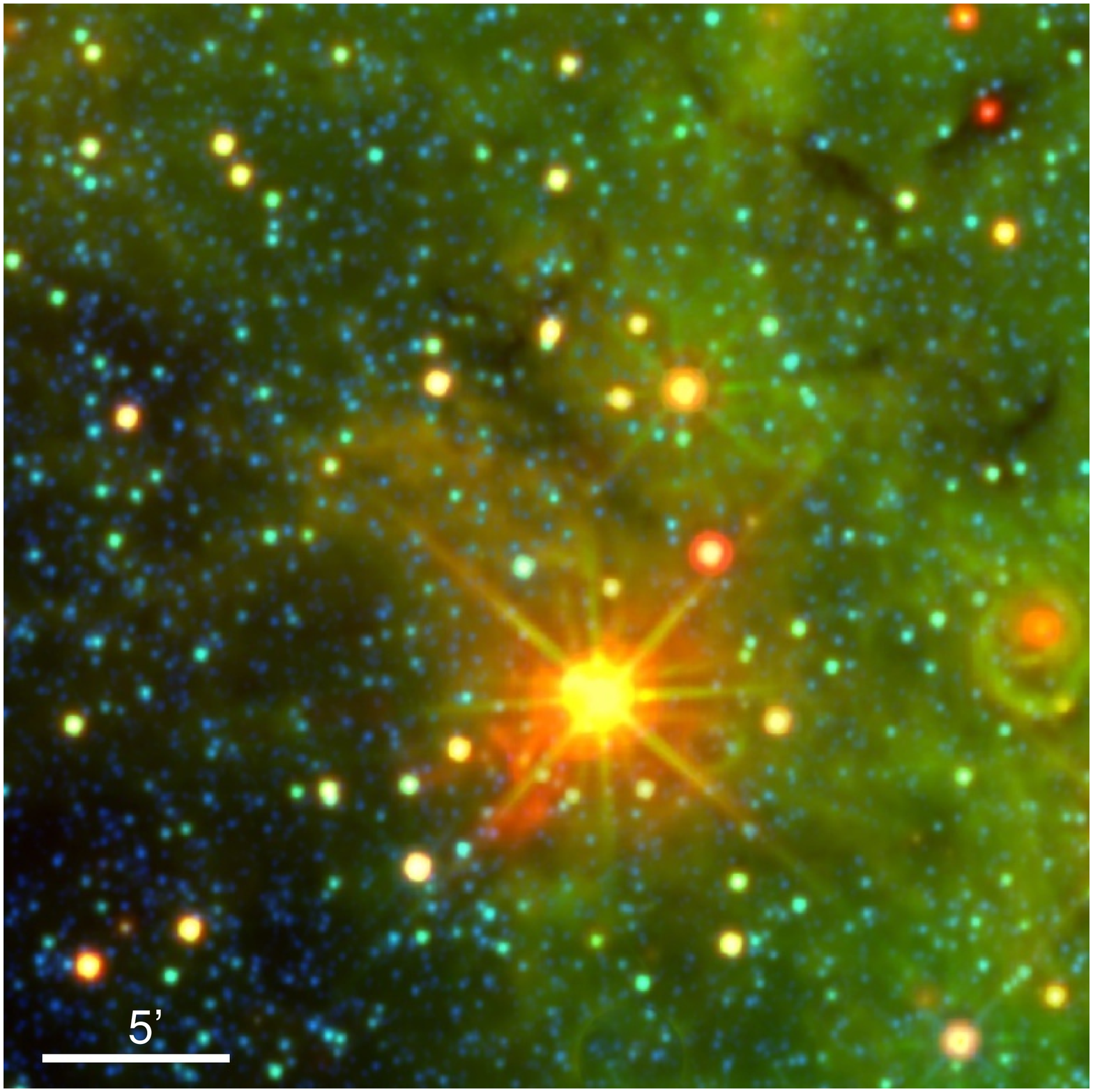}
\end{center}
\caption{Same as Fig.~\ref{fig:WR7} for WR\,95.}
\label{fig:WR95}
\end{figure}

\begin{figure}
\begin{center}
\includegraphics[width=0.5\linewidth]{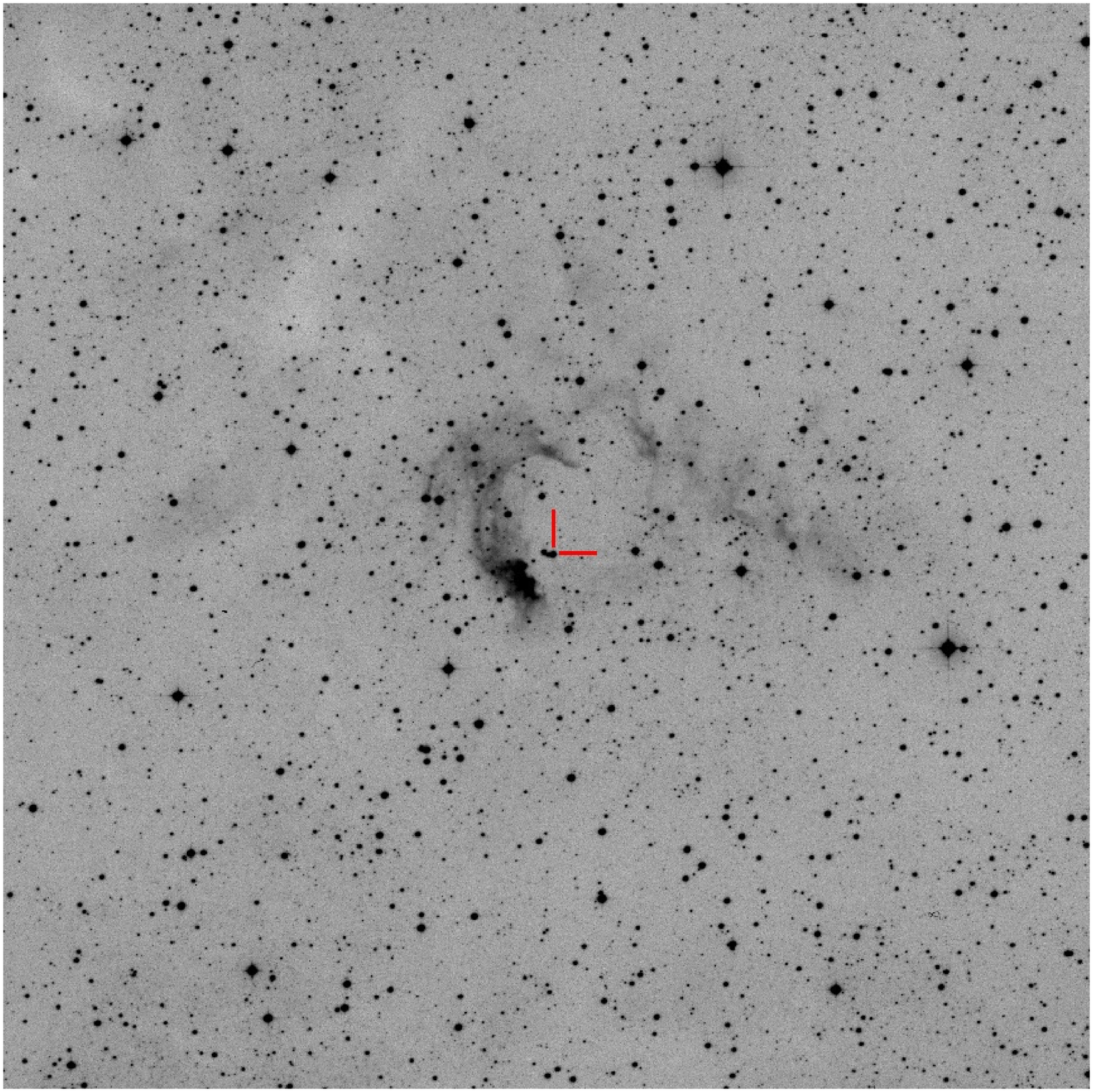}~
\includegraphics[width=0.5\linewidth]{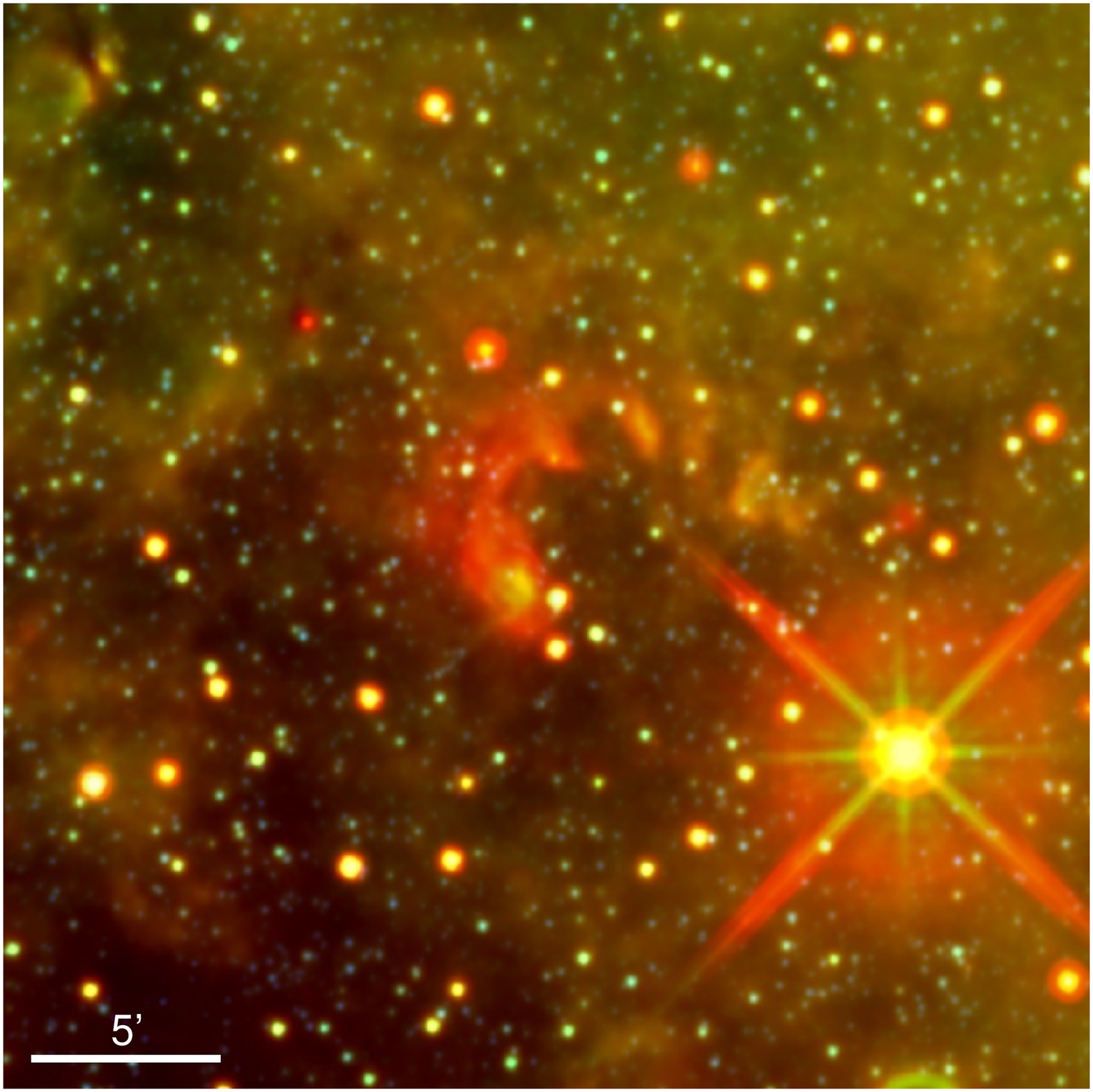}
\end{center}
\caption{Same as Fig.~\ref{fig:WR7} for WR\,101 (Anon).}
\label{fig:WR101}
\end{figure}

\begin{figure}
\begin{center}
\includegraphics[width=0.5\linewidth]{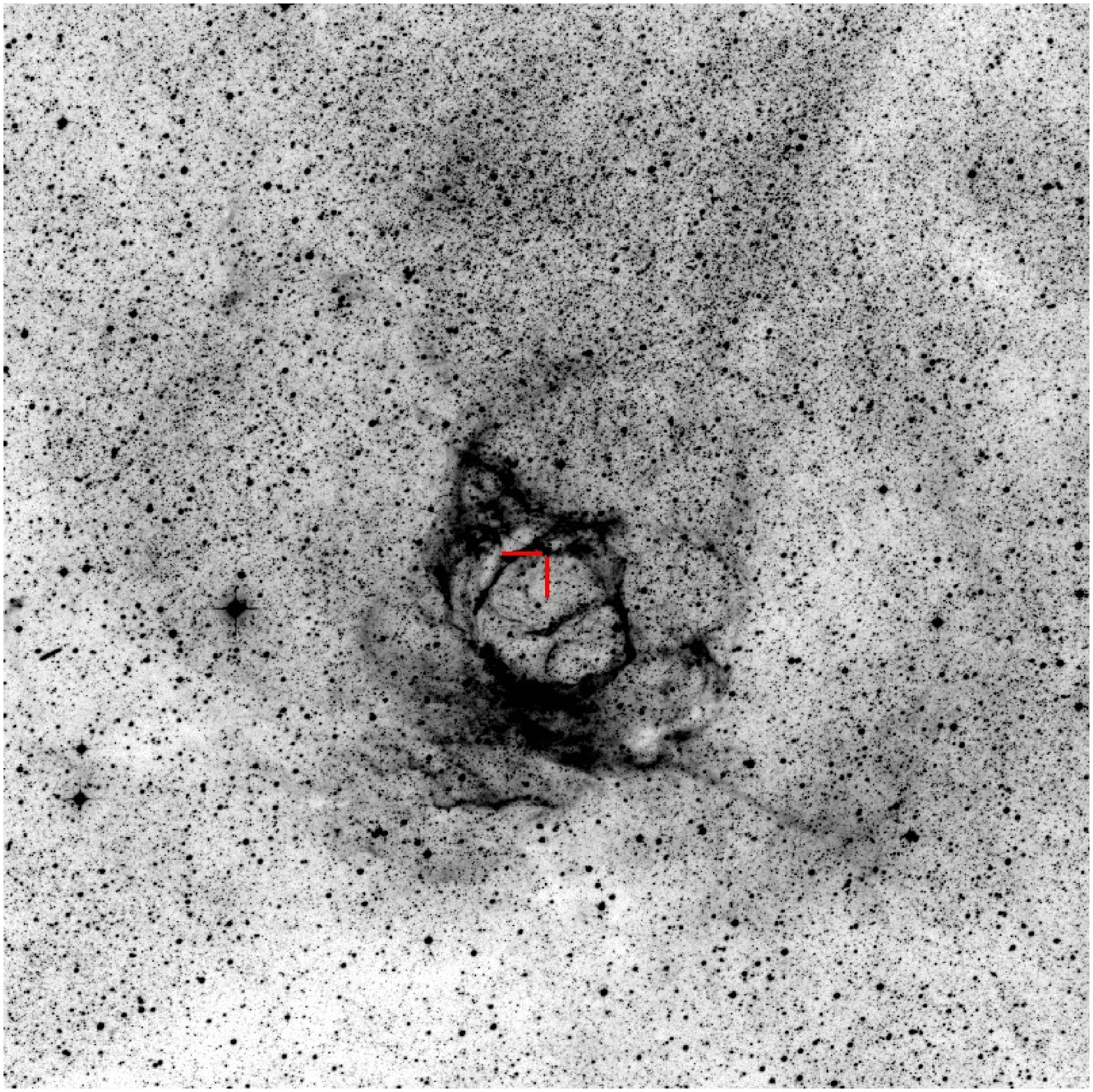}~
\includegraphics[width=0.5\linewidth]{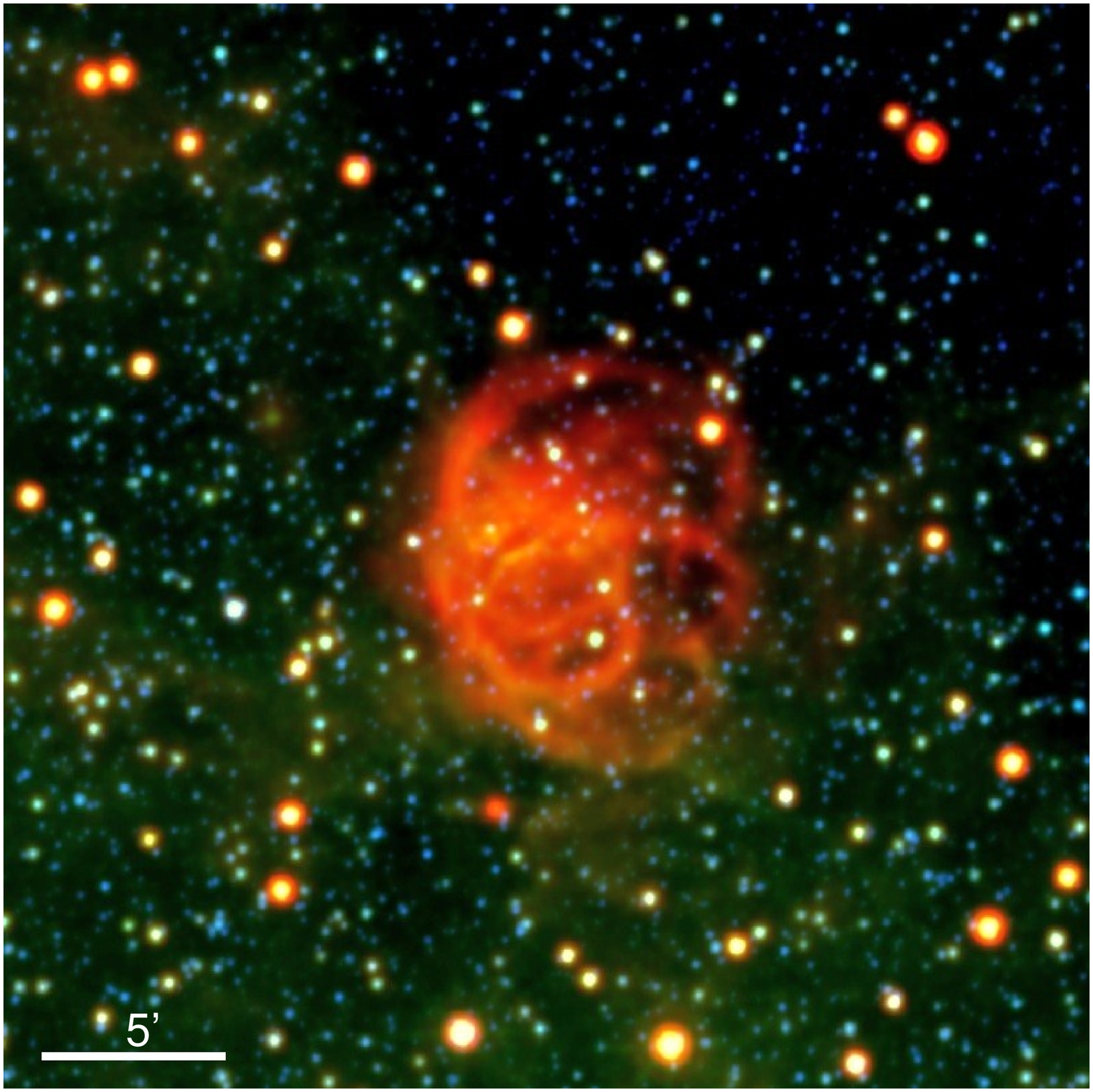}
\end{center}
\caption{Same as Fig.~\ref{fig:WR7} for WR\,102 (G2.4$+$1.4).}
\label{fig:WR102}
\end{figure}

\begin{figure}
\begin{center}
\includegraphics[width=0.5\linewidth]{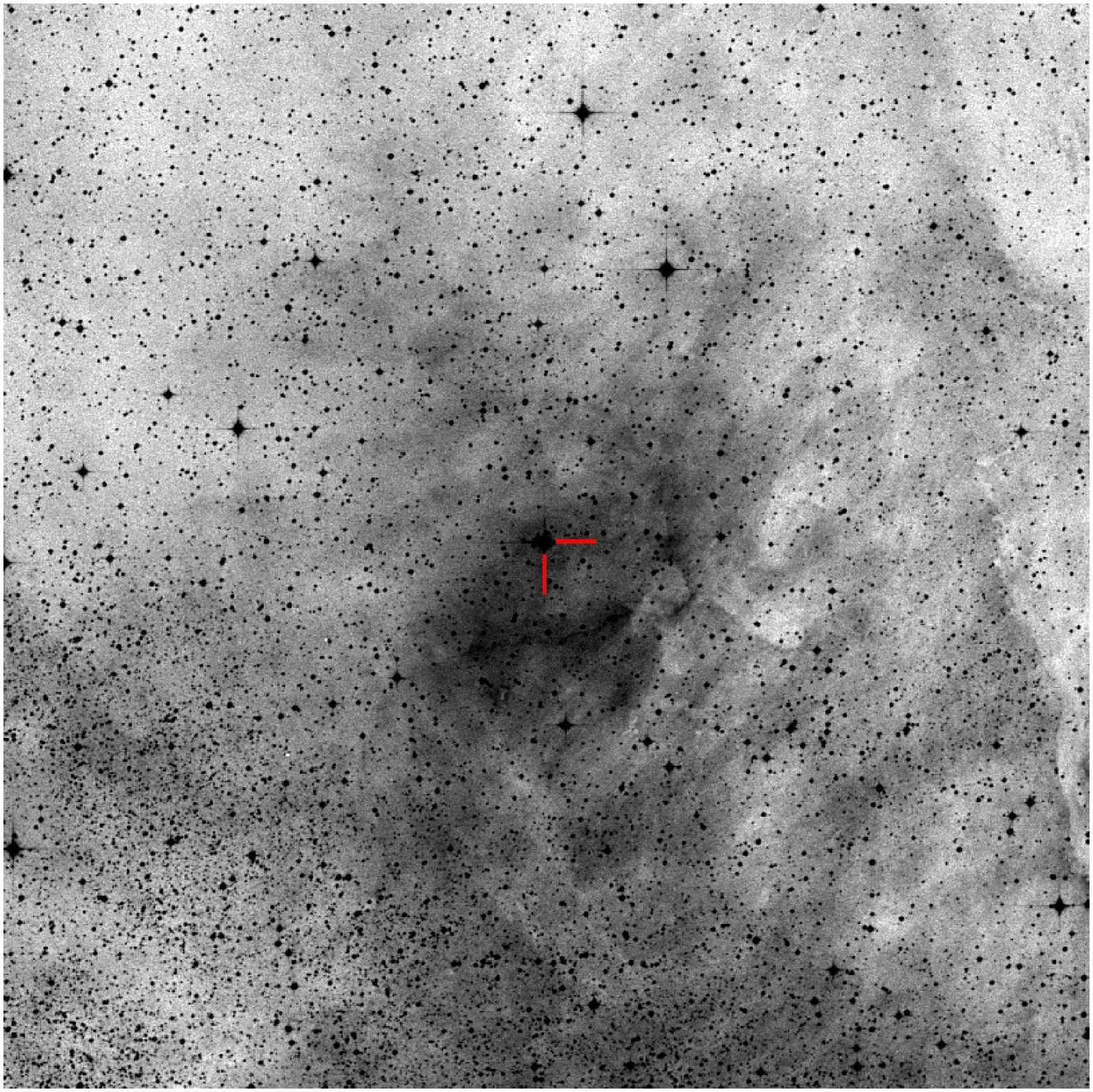}~
\includegraphics[width=0.5\linewidth]{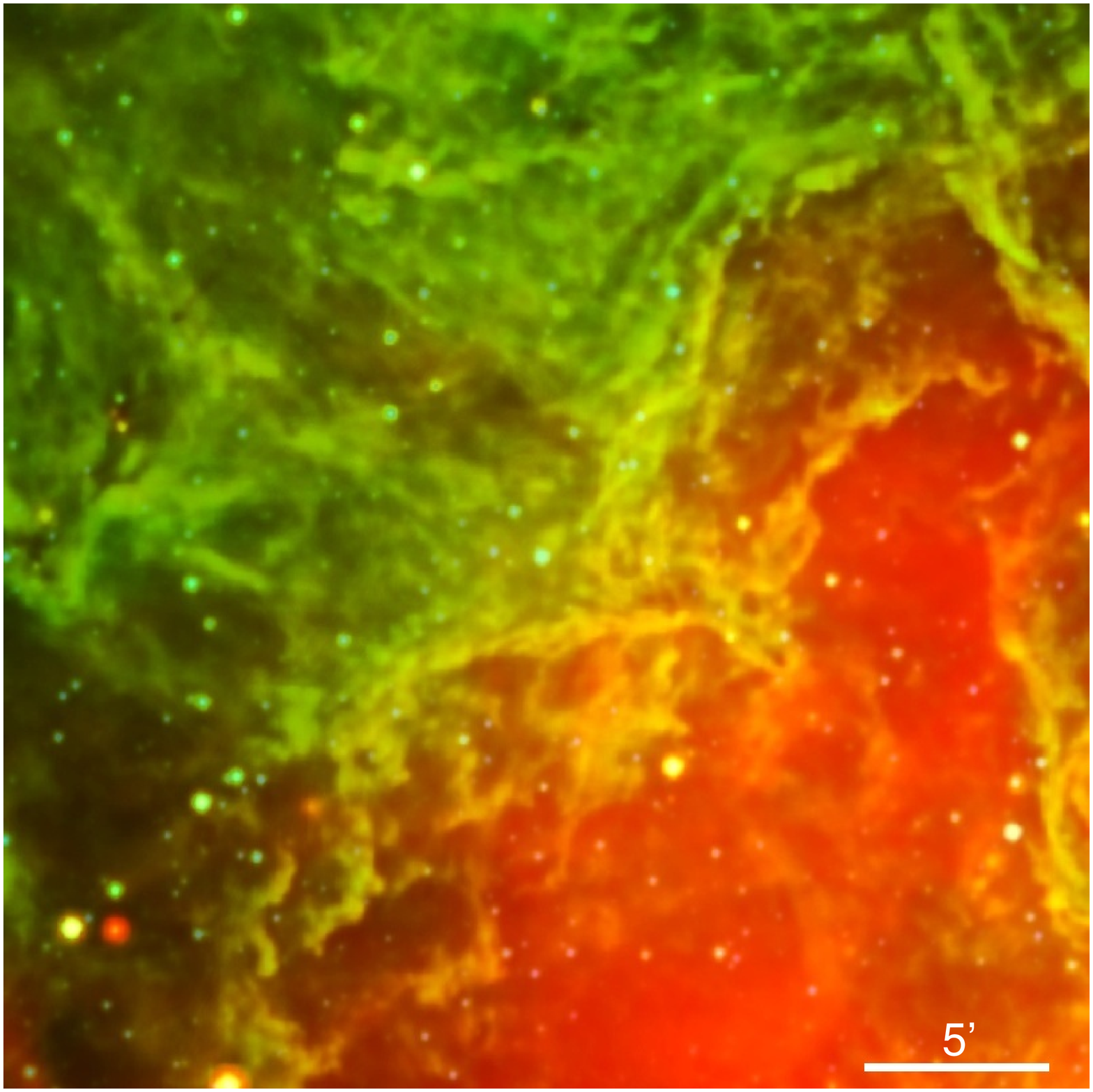}
\end{center}
\caption{Same as Fig.~\ref{fig:WR7} for WR\,113 (RCW\,167).}
\label{fig:WR113}
\end{figure}

\begin{figure}
\begin{center}
\includegraphics[width=0.5\linewidth]{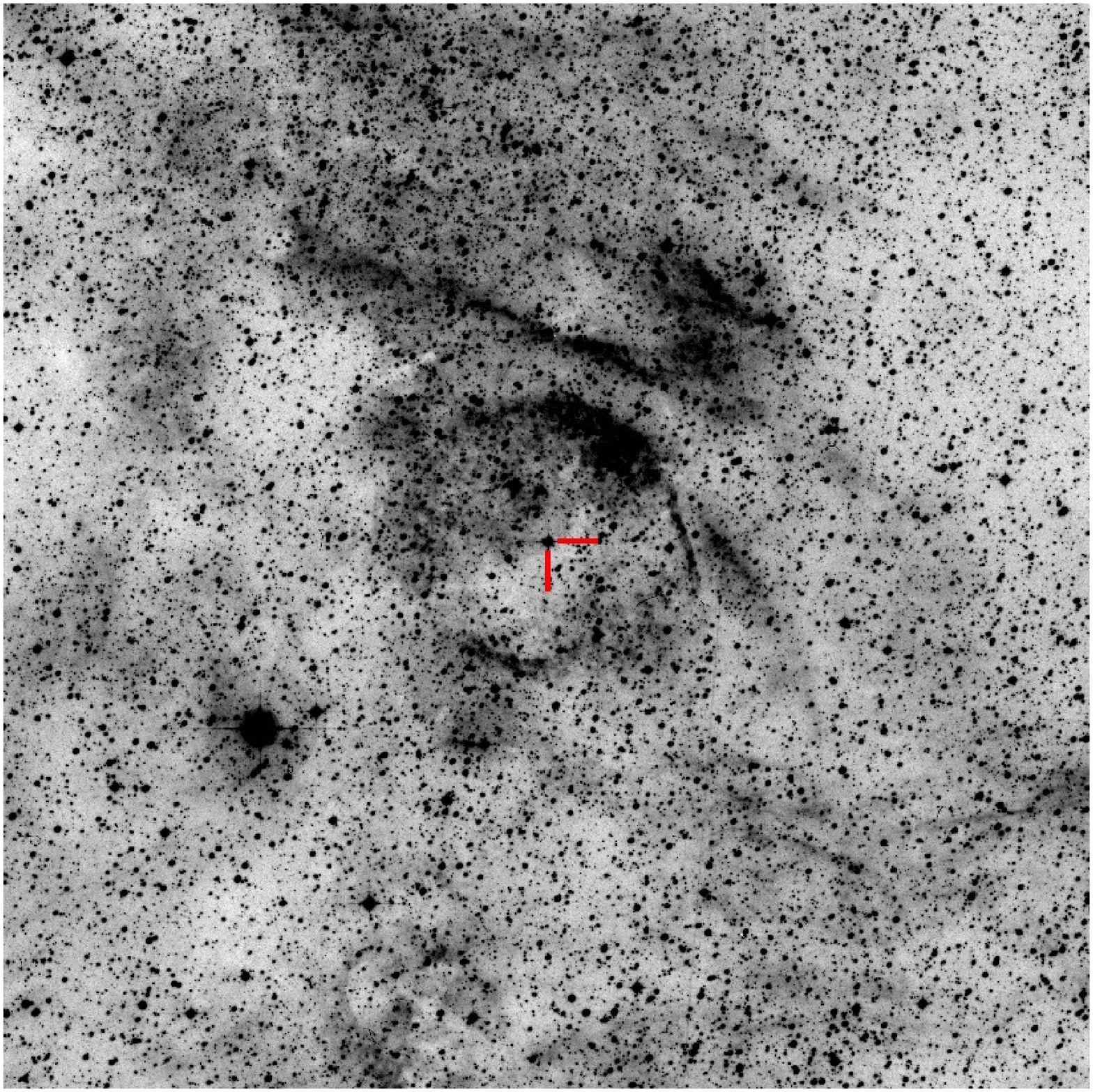}~
\includegraphics[width=0.5\linewidth]{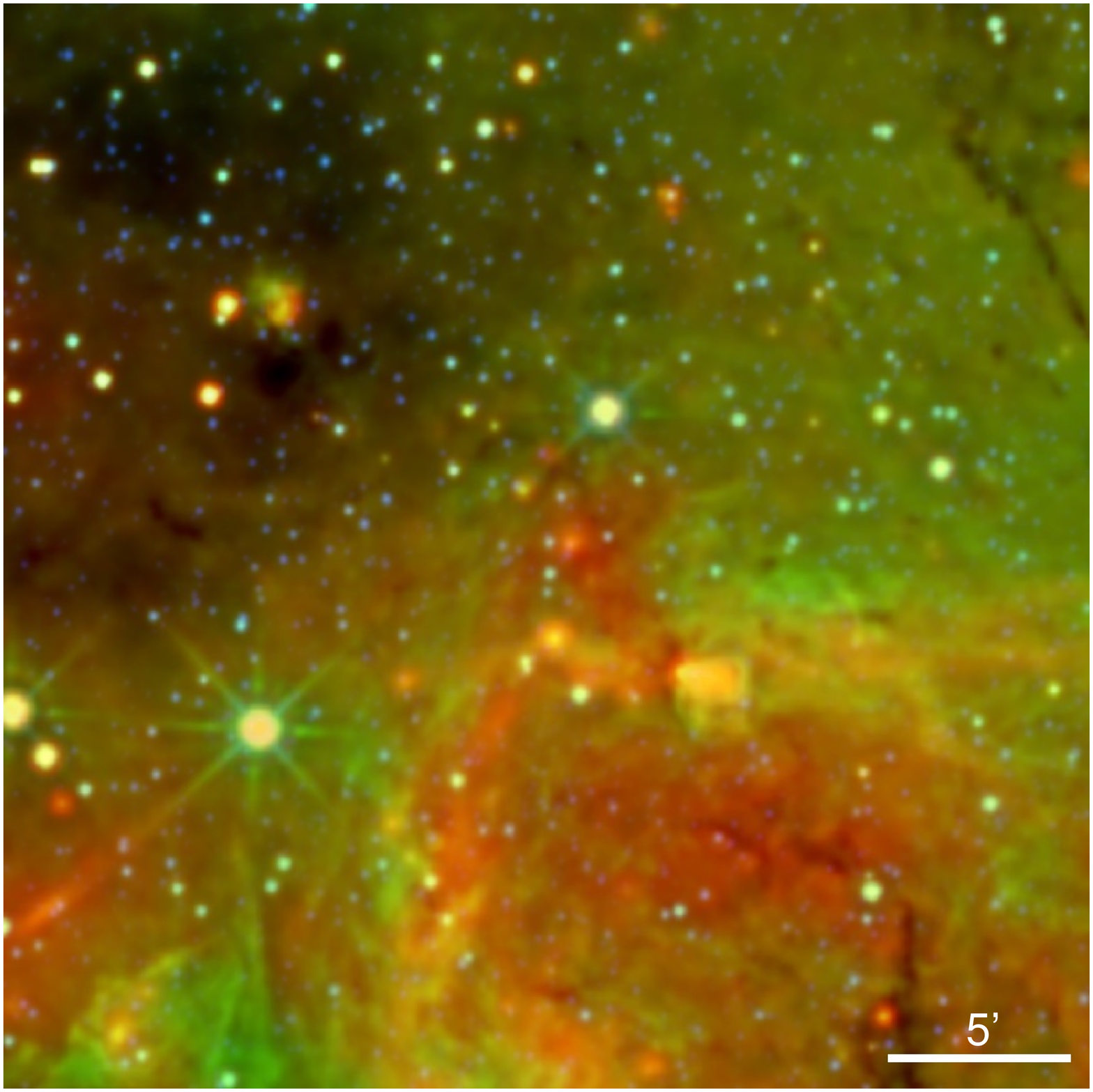}
\end{center}
\caption{Same as Fig.~\ref{fig:WR7} for WR\,116 (Anon).}
\label{fig:WR116}
\end{figure}

\begin{figure}
\begin{center}
\includegraphics[width=0.5\linewidth]{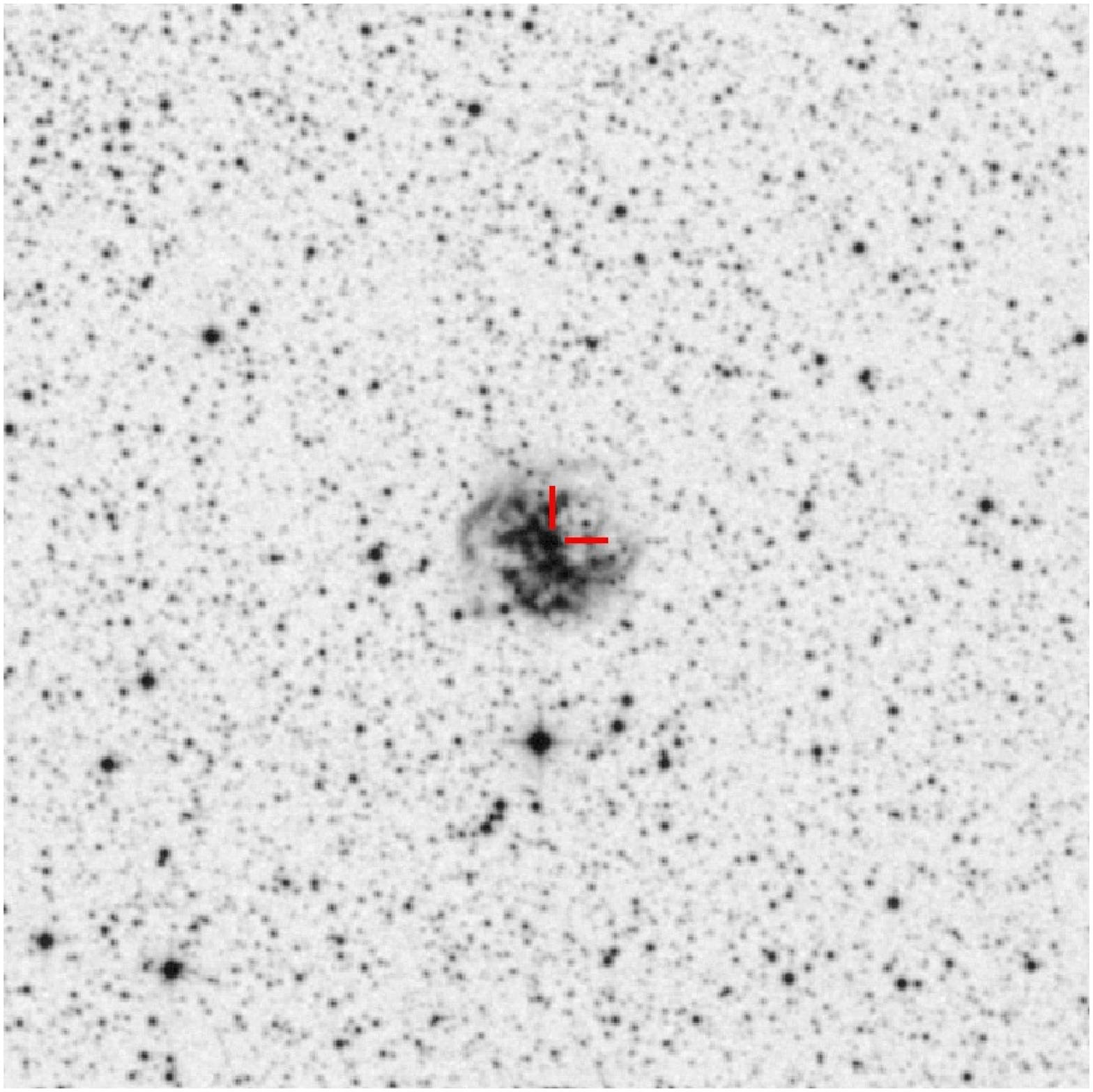}~
\includegraphics[width=0.5\linewidth]{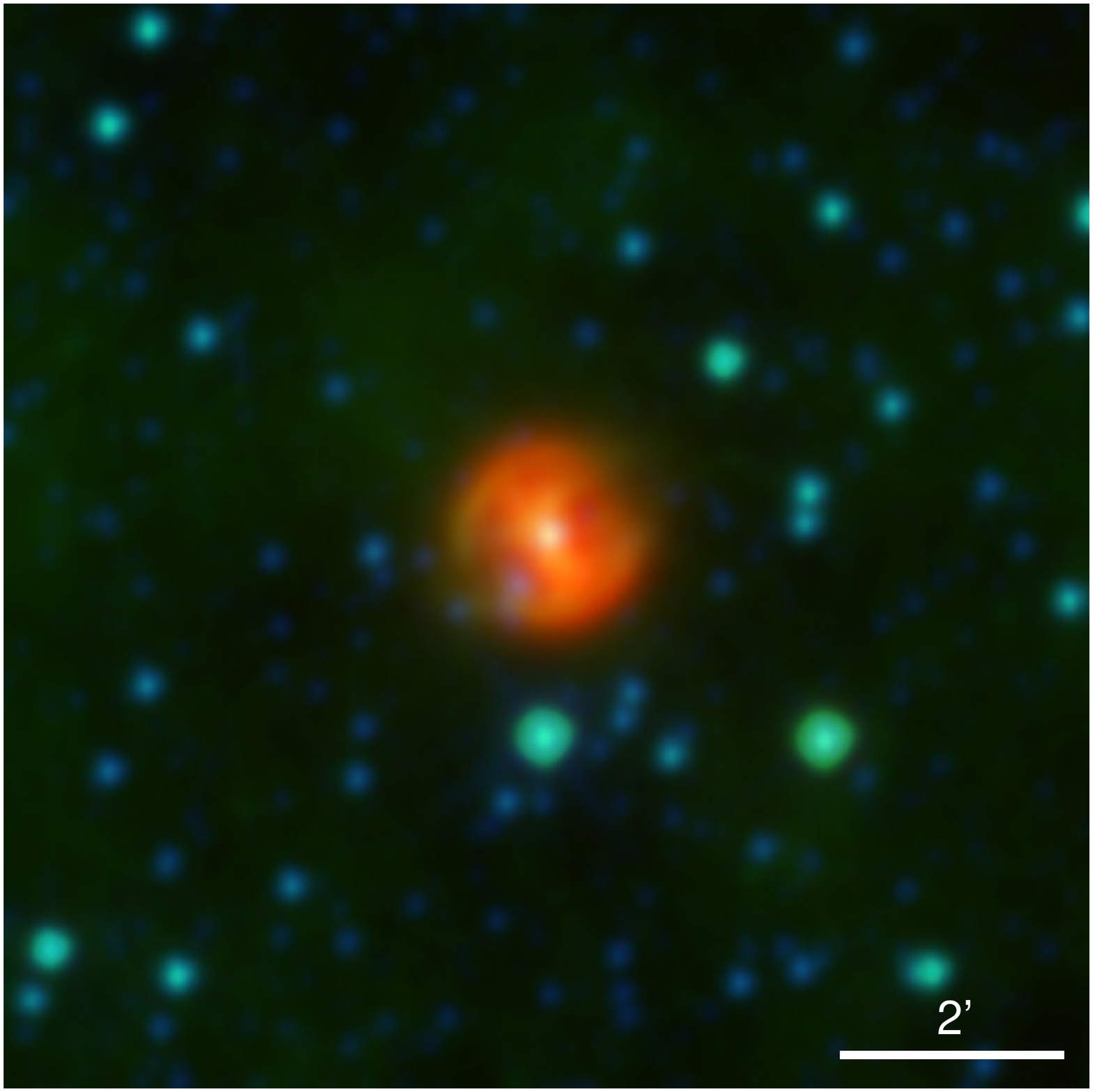}
\end{center}
\caption{Same as Fig.~\ref{fig:WR7} for WR\,124 (M1-67)}
\label{fig:WR124}
\end{figure}

\begin{figure}
\begin{center}
\includegraphics[width=0.5\linewidth]{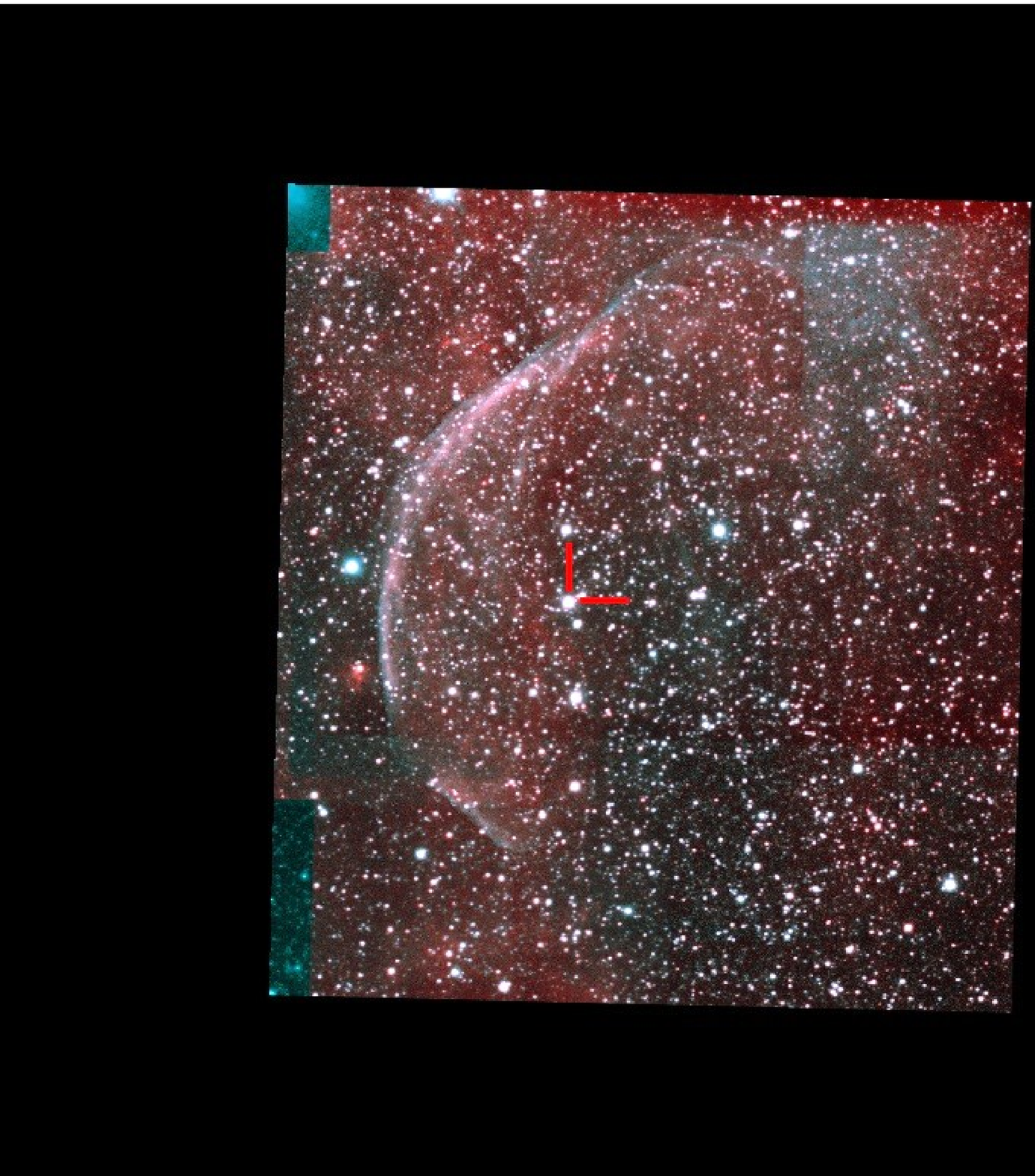}~
\includegraphics[width=0.5\linewidth]{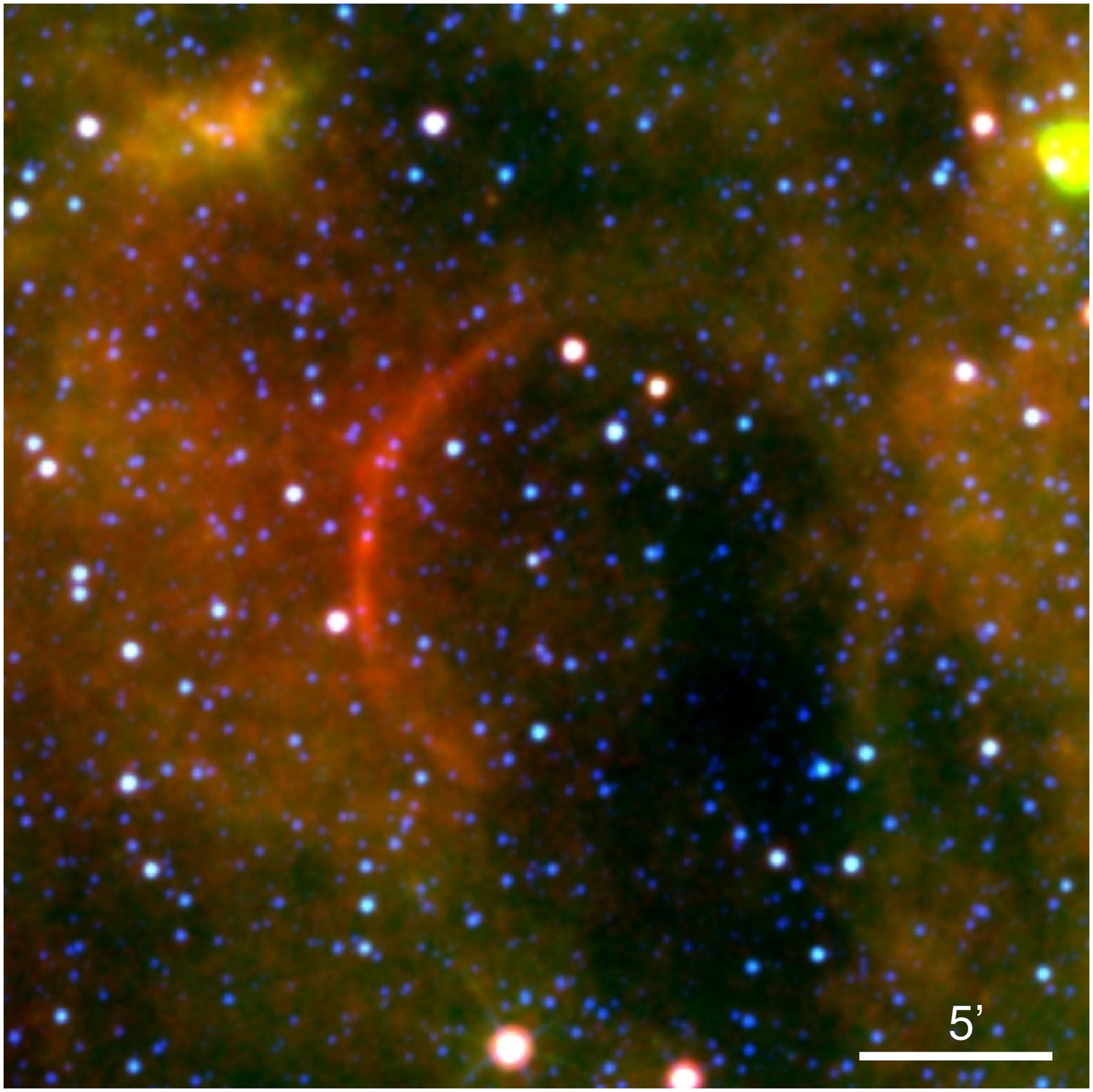}
\end{center}
\caption{Same as Fig.~\ref{fig:WR7} for WR\,128 (Anon).}
\label{fig:WR128}
\end{figure}

\begin{figure}
\begin{center}
\includegraphics[width=0.5\linewidth]{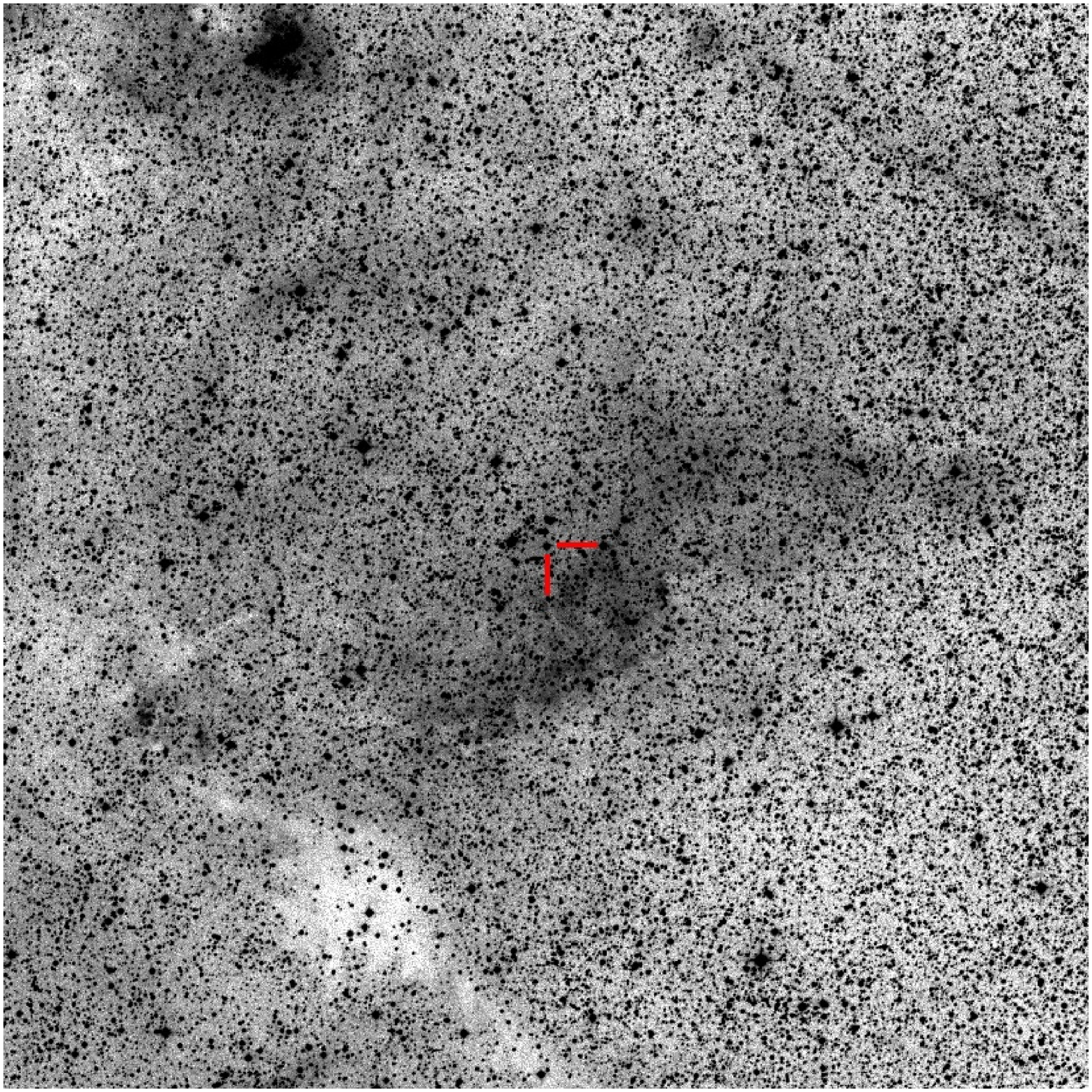}~
\includegraphics[width=0.5\linewidth]{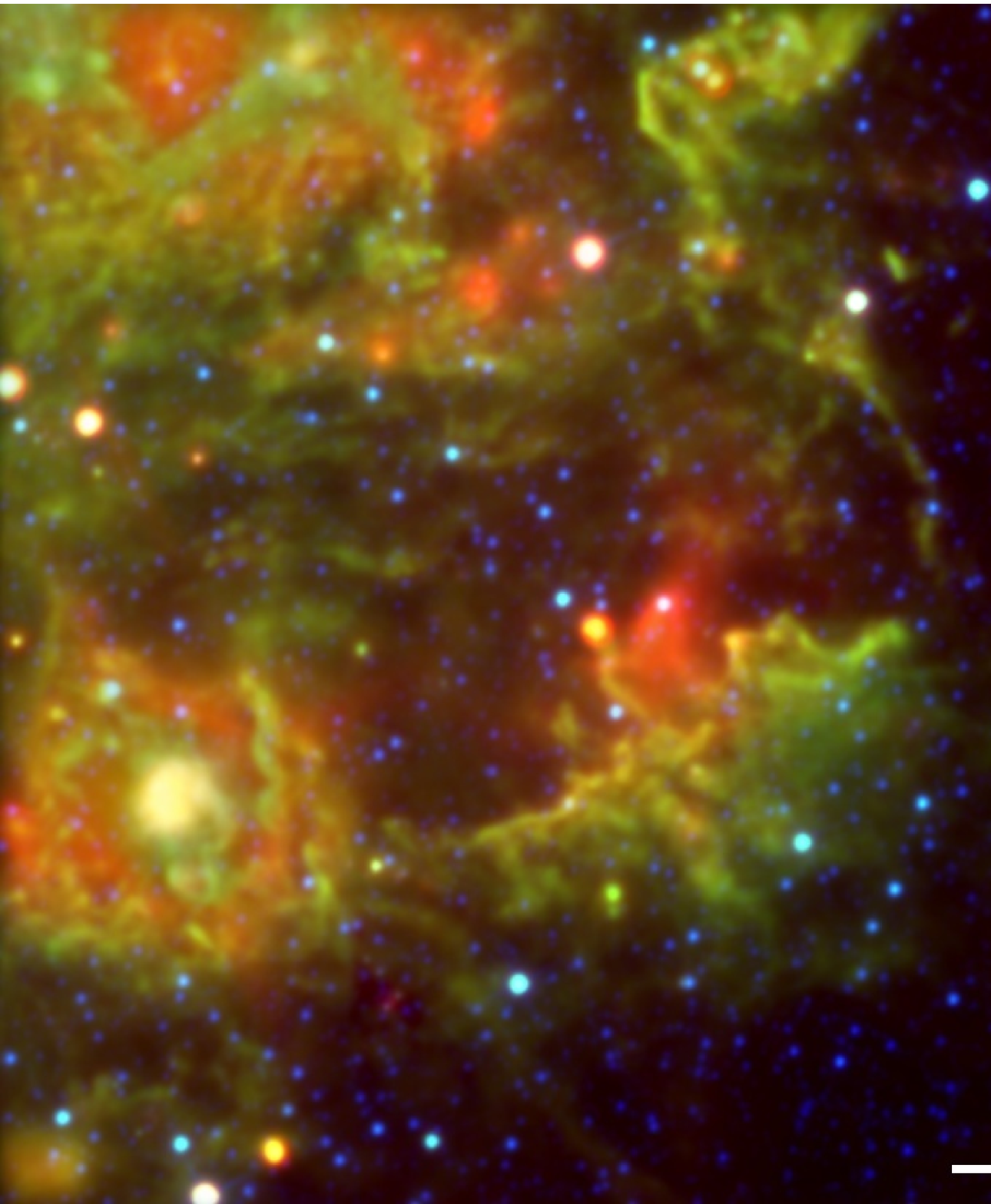}
\end{center}
\caption{Same as Fig.~\ref{fig:WR7} for WR\,131. }
\label{fig:WR131}
\end{figure}

\begin{figure}
\begin{center}
\includegraphics[width=0.5\linewidth]{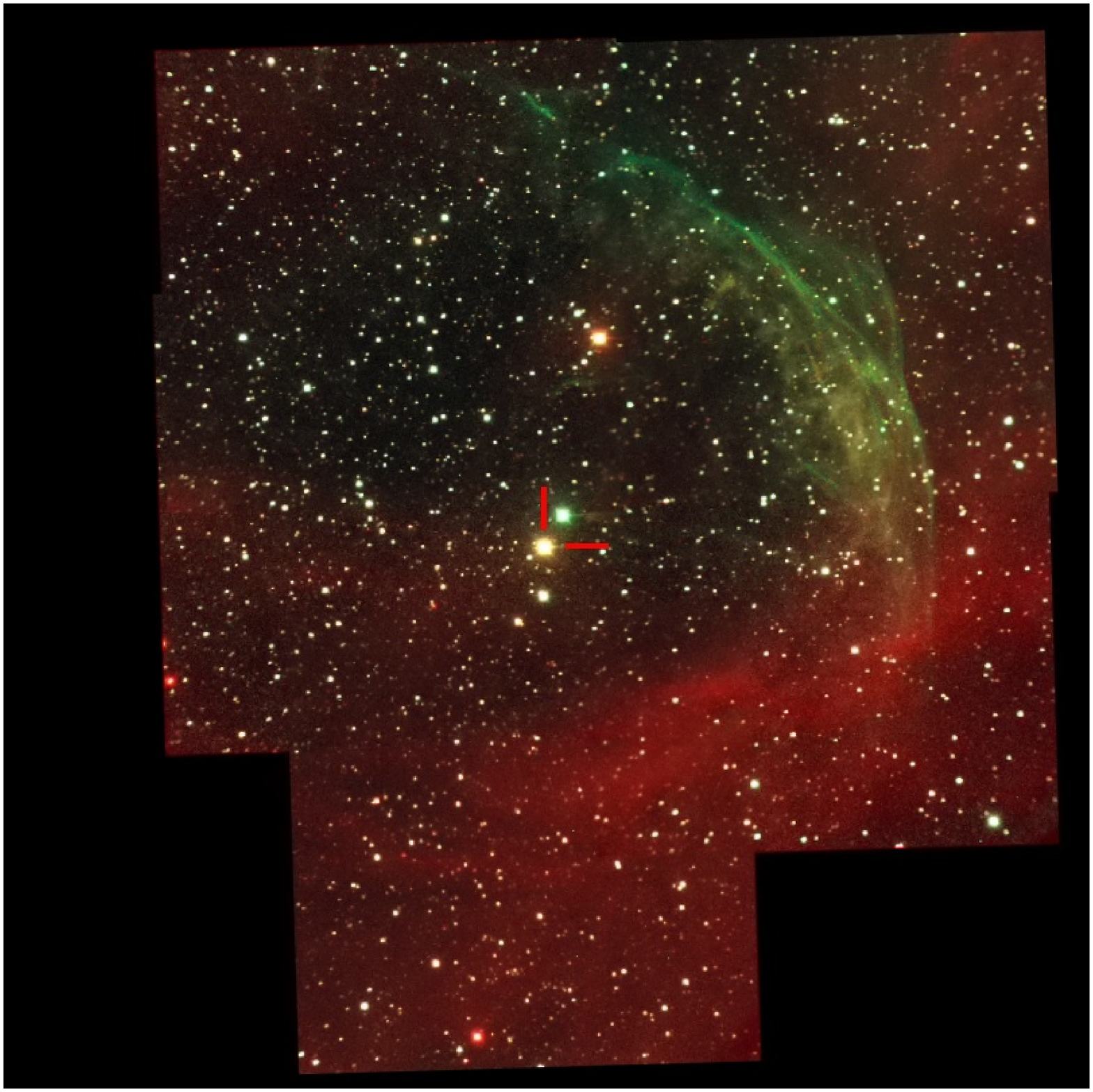}~
\includegraphics[width=0.5\linewidth]{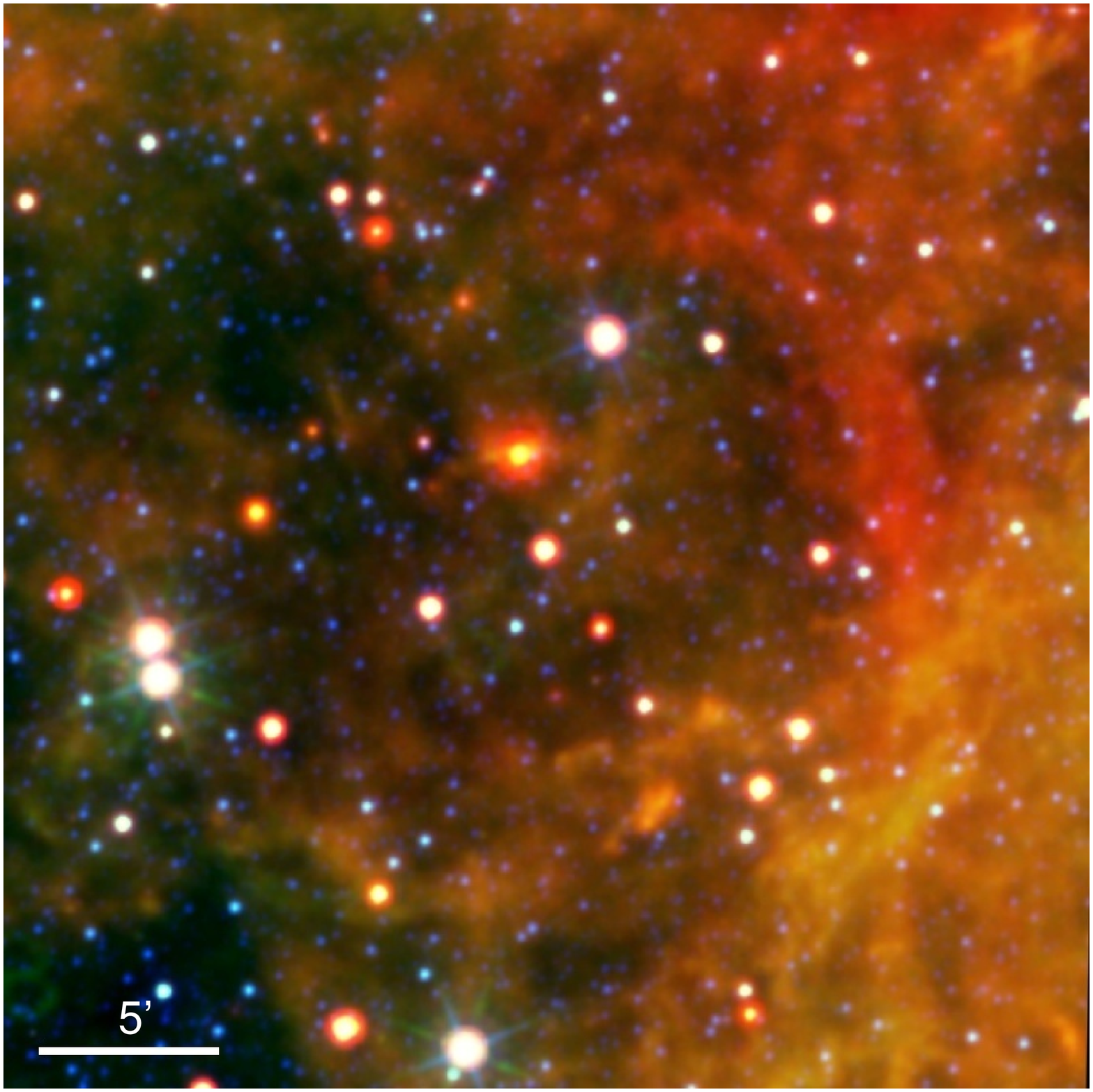}
\end{center}
\caption{Same as Fig.~\ref{fig:WR7} for WR\,134 (Anon).}
\label{fig:WR134}
\end{figure}

\begin{figure}
\begin{center}
\includegraphics[width=0.5\linewidth]{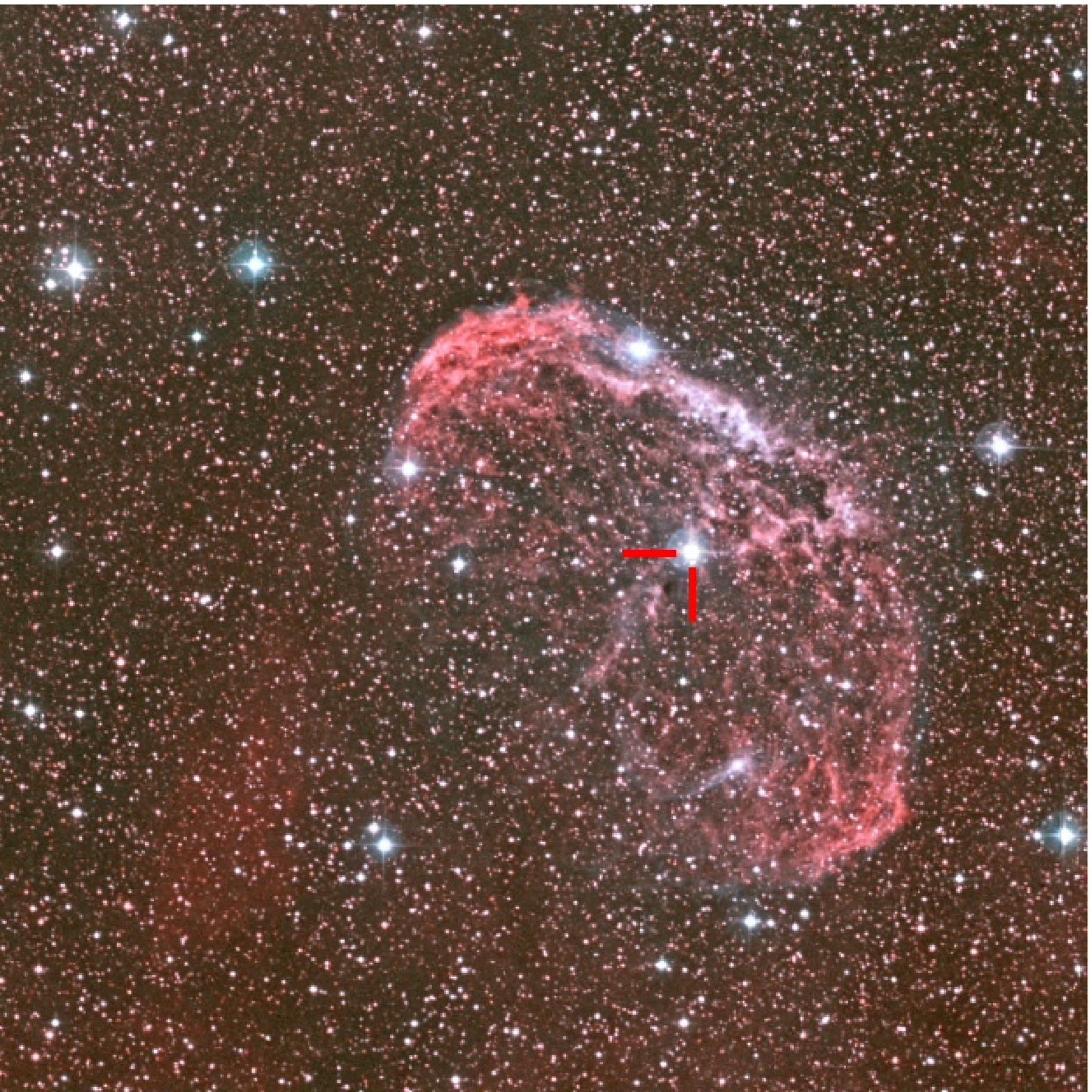}~
\includegraphics[width=0.5\linewidth]{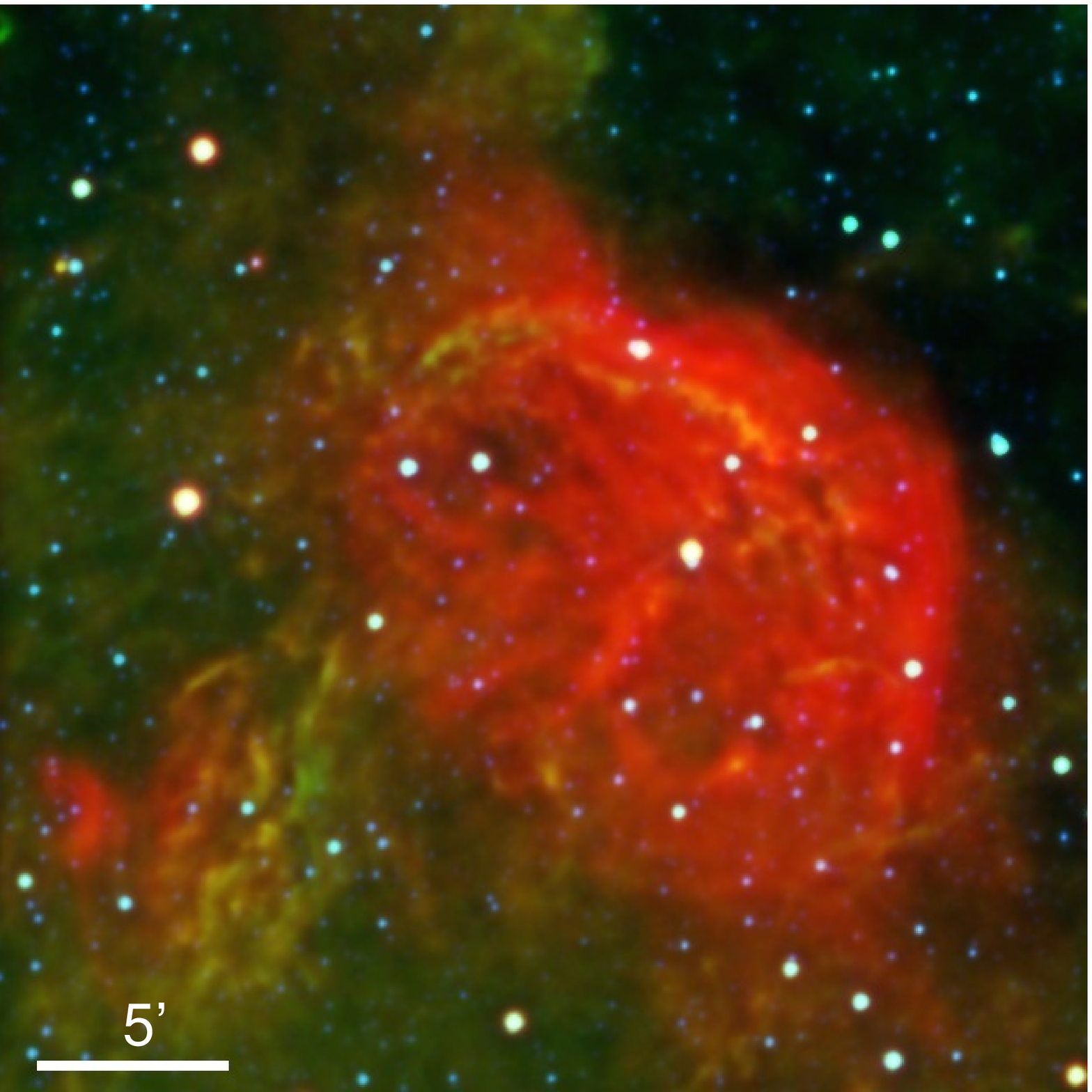}
\end{center}
\caption{Same as Fig.~\ref{fig:WR7} for WR\,136 (NGC\,6888).}
\label{fig:WR136}
\end{figure}


\begin{thebibliography}{}

\bibitem[Arnal et al.(1999)]{Arnal1999} Arnal, E.~M., Cappa,
  C.~E., Rizzo, J.~R., \& Cichowolski, S.\ 1999, AJ, 118, 1798
\bibitem[Arnal \& Cappa(1996)]{Arnal1996} Arnal, E.~M., \&
  Cappa, C.~E.\ 1996, MNRAS, 279, 788
\bibitem[Arthur(2007a)]{Arthur2007a} Arthur, S.~J.\ 2007a, Revista
  Mexicana de Astronomia y Astrofisica Conference Series, 30, 64
\bibitem[Arthur(2007b)]{Arthur2007b} Arthur, S.~J.\ 2007b, Diffuse
  Matter from Star Forming Regions to Active Galaxies, 183
\bibitem[Bochkarev(1988)]{B1988} Bochkarev, N.~G.\ 1988,
  Nature, 332, 518
\bibitem[Bruhweiler et al.(1981)]{Bruhweiler_etal1981} 
Bruhweiler, F.~C., Gull, T.~R., Henize, K.~G., \& Cannon, R.~D.\ 1981, 
ApJ, 251, 126 
\bibitem[Burgemeister et al.(2013)]{Burgemeister2013} Burgemeiste, S.,
  Gvaramadze, V.~V., Stringfellow, G.~S., et al., 2013, MNRAS, 429, 3305
\bibitem[Cappa et al.(2002)]{Cappa2002} Cappa, C.~E., Goss,
  W.~M., \& Pineault, S.\ 2002, AJ, 123, 3348
\bibitem[Cappa et al.(2009)]{Cappa2009} Cappa, C.~E., Rubio,
  M., Mart{\'{\i}}n, M.~C., \& Romero, G.~A.\ 2009, A\&A, 508, 759
\bibitem[Cappa et al.(2008)]{Cappa2008} Cappa, C.~E.,
  Vasquez, J., Arnal, E.~M., Cichowolski, S., \& Pineault, S.\ 2008,
  Mass Loss from Stars and the Evolution of Stellar Clusters, 388, 151
\bibitem[Chu(1981)]{Chu1981} Chu, Y.-H.\ 1981, ApJ, 249, 195
\bibitem[Chu(1982)]{Chu1982} Chu, Y.-H.\ 1982, ApJ, 254, 578
\bibitem[Chu(1991)]{Chu1991} Chu, Y.-H.\ 1991, Wolf-Rayet Stars and
  Interrelations with Other Massive Stars in Galaxies, 143, 349
\bibitem[Chu(2003)]{Chu2003} Chu, Y.-H.\ 2003, A Massive Star
  Odyssey: From Main Sequence to Supernova, 212, 585
\bibitem[Chu et al.(1982)]{Chu_etal82}
Chu, Y.-H., Gull, T.R., Treffers, R.R., \& Kwitter, K.B.\ 1982, 
ApJ, 254, 562
\bibitem[Chu \& Treffers(1981a)]{CT81a} Chu, Y.-H., \& Treffers,
  R.~R.\ 1981a, ApJ, 249, 586
\bibitem[Chu \& Treffers(1981b)]{CT81b} Chu, Y.-H., \& Treffers,
  R.~R.\ 1981b, ApJ, 250, 615
\bibitem[Chu et al.(1983)]{Chu1983} Chu, Y.-H., Treffers, R.~R., \&
  Kwitter, K.~B.\ 1983, ApJs, 53, 937
\bibitem[Chu et al.(2003)]{Chu_etal2003} Chu, Y.-H., Guerrero, M.~A.,
  Gruendl, R.~A., Garc\'\i a-Segura, G., \& Wendker, H.~J.\ 2003, ApJ,
  599, 1189
\bibitem[Chu et al.(2009)]{Chu2009} Chu, Y.-H., Gruendl, R.~A.,
  Guerrero, M.~A., et al.\ 2009, AJ, 138, 691
\bibitem[Dufour(1989)]{Dufour89} Dufour, R.~J.\ 1989, RMxA\&A, 18, 87
\bibitem[Dufour et al.(1988)]{DPH1988} Dufour, R.~J., Parker,
  R.~A.~R., \& Henize, K.~G.\ 1988, ApJ, 327, 859
\bibitem[Dwarkadas(2007)]{Dwarkadas2007} Dwarkadas, V.~V.\ 2007, ApJ,
  667, 226
\bibitem[Ekstr{\"o}m et al.(2012)]{Eks2012} Ekstr\"om, S., Georgy, C.,
  Eggenberger, P., et al.\ 2012, A\&A, 537, A146
\bibitem[Fern{\'a}ndez-Mart{\'{\i}}n et
  al.(2013)]{Fernandez-Martin2013} Fern{\'a}ndez-Mart{\'{\i}}n, A.,
  V{\'{\i}}lchez, J.~M., P{\'e}rez-Montero, E., et al.\ 2013, \aap,
  554, A104
\bibitem[Fern{\'a}ndez-Mart{\'{\i}}n et
  al.(2012)]{Fernandez-Martin2012} Fern{\'a}ndez-Mart{\'{\i}}n, A.,
  Mart{\'{\i}}n-Gord{\'o}n, D., V{\'{\i}}lchez, J.~M., et al.\ 2012,
  \aap, 541, AA119
\bibitem[Fesen \& Milisavljevic(2010)]{Fesen2010} Fesen, R.~A., \&
  Milisavljevic, D.\ 2010, \aj, 139, 2595
\bibitem[Flagey et al.(2011)]{Flagey2011} Flagey, N., Noriega-Crespo,
  A., Billot, N., \& Carey, S.~J.\ 2011, ApJ, 741, 4
\bibitem[Freyer et al.(2003)]{Freyer2003} Freyer, T.,
  Hensler, G., \& Yorke, H.~W.\ 2003, ApJ, 594, 888
\bibitem[Freyer et al.(2006)]{Freyer2006} Freyer, T.,
  Hensler, G., \& Yorke, H.~W.\ 2006, ApJ, 638, 262
\bibitem[Garc\'{i}a-Segura \& Mac Low(1995)]{GSML1995b}
  Garc\'{i}a-Segura, G., \& Mac Low, M.-M.\ 1995, ApJ, 455, 160
\bibitem[Garc\'{i}a-Segura et al.(1996a)]{GS1996a}
  Garc\'{i}a-Segura, G., Mac Low, M.-M., \& Langer, N.\ 1996a, A\&A, 305,
  229
\bibitem[Garc\'{i}a-Segura et al.(1996b)]{GS1996b}
  Garc\'{i}a-Segura, G., Langer, N., \& Mac Low, M.-M.\ 1996b, A\&A, 316,
  13
\bibitem[Georgy et al.(2012)]{Georgy2012} Georgy, C., Ekstr{\"o}m, S.,
  Meynet, G., et al.\ 2012, A\&, 542, AA29
\bibitem[Gruendl et al.(2000)]{Gruendl2000} Gruendl, R.~A., Chu,
  Y.-H., Dunne, B.~C., \& Points, S.~D.\ 2000, AJ, 120, 2670
\bibitem[Gvaramadze et al.(2009)]{Gv2009} Gvaramadze, V.~V., Fabrika,
  S., Hamann, W.-R., et al.\ 2009, MNRAS, 400, 524
\bibitem[Gvaramadze et al.(2010a)]{Gv2010} Gvaramadze, V.~V., Kniazev,
  A.~Y., \& Fabrika, S.\ 2010a, MNRAS, 405, 1047
\bibitem[Gvaramadze et al.(2010b)]{Gv2010b} Gvaramadze, V.~V., Kniazev,
  A.~Y., Fabrika, S., et al.\ 2010b, MNRAS, 405, 520
\bibitem[Gvaramadze et al.(2010c)]{Gv2010c} Gvaramadze, V.~V.,
  Kniazev, A.~Y., Hamann, W.-R., et al.\ 2010c, MNRAS, 403, 760
\bibitem[Gvaramadze et al.(2014)]{Gv2014} Gvaramadze, V.~V.,
  Chen\'{e}, A.-N., Kniazev, A.~Y., et al., 2014, MNRAS, 442, 929
\bibitem[Hamann et al.(2006)]{HGL06} Hamann, W.-R., Gr{\"a}fener, G.,
  \& Liermann, A.\ 2006, A\&A, 457, 1015
\bibitem[Humphreys(2010)]{H2010} Humphreys, R.~M.\ 2010, Hot and Cool:
  Bridging Gaps in Massive Star Evolution, 425, 247
\bibitem[Maeder(1991)]{Maeder1991} Maeder, A.\ 1991, A\&A, 242, 93
\bibitem[Maeder(1997)]{Maeder97} Maeder, A.\ 1997, IAU Symposium, 189,
  313
\bibitem[Marston et al.(1994a)]{MCG1994} Marston, A.~P., Chu, Y.-H.,
  \& Garc\'\i a-Segura, G.\ 1994a, ApJs, 93, 229
\bibitem[Marston et al.(1994b)]{Marston1994} Marston, A.~P., Yocum,
  D.~R., Garc\'\i a-Segura, G., \& Chu, Y.-H.\ 1994b, ApJs, 95, 151
\bibitem[Marston(2001)]{Marston2001} Marston, A.~P.\ 2001, \apj, 563,
  875
\bibitem[Mauerhan et al.(2010)]{Mauerhan2010} Mauerhan, J.~C.,
  Wachter, S., Morris, P.~W., Van Dyk, S.~D., \& Hoard, D.~W.\ 2010,
  ApJL, 724, L78
\bibitem[Meynet \& Maeder(2003)]{MM2003} Meynet, G., \&
  Maeder, A.\ 2003, A\&A, 404, 975
\bibitem[Meynet \& Maeder(2005)]{MM2005} Meynet, G., \& Maeder, A.\
  2005, A\&A, 429, 581
\bibitem[Moffat(1995)]{Moffat95} Moffat, A.~F.~J.\ 1995,
  Wolf-Rayet Stars: Binaries; Colliding Winds; Evolution, 163, 213
\bibitem[Morris et al.(2006)]{Morris2006} Morris, P.~W., Stolovy, S.,
  Wachter, S., et al.\ 2006, \apjl, 640, L179
\bibitem[Parker et al.(2005)]{Parker2005} Parker, Q.~A.,
  Phillipps, S., Pierce, M.~J., et al.\ 2005, MNRAS, 362, 689
\bibitem[Sander et al.(2012)]{SHT12} Sander, A., Hamann, W.-R., \&
  Todt, H.\ 2012, A\&A, 540, A144
\bibitem[Smith et al.(2007)]{Smith2007} Smith, J.~D.~T., Armus, L.,
  Dale, D.~A., et al.\ 2007, PASP, 119, 1133
\bibitem[Solf \& Carsenty(1982)]{SC1982} Solf, J., \& Carsenty, U.\
  1982, A\&A, 116, 54
\bibitem[Stock et al.(2011)]{Stock2011} Stock, D.~J., Barlow, M.~J.,
  \& Wesson, R.\ 2011, MNRAS, 418, 2532
\bibitem[Stock \& Barlow(2010)]{Stock2010} Stock, D.~J., \& Barlow,
  M.~J.\ 2010, MNRAS, 409, 1429
\bibitem[Stringfellow et al.(2012)]{Stringfellow2012} Stringfellow,
  G.~S., Gvaramadze, V.~V., Beletsky, Y., \& Kniazev, A.~Y.\ 2012,
  Proceedings of a Scientific Meeting in Honor of Anthony
  F.~J.~Moffat, 465, 514
\bibitem[Toal{\'a} et al.(2015)]{Toala2015} Toal{\'a}, J.~A.,
  Guerrero, M.~A., Chu, Y.-H., \& Gruendl, R.~A.\ 2015, MNRAS, 446,
  1083
\bibitem[Toal{\'a} \& Arthur(2011)]{Toala2011} Toal{\'a}, J.~A., \&
  Arthur, S.~J.\ 2011, ApJ, 737, 100
\bibitem[Toal{\'a} et al.(2012)]{Toala2012} Toal\'a, J.~A., Guerrero,
  M.~A., Chu, Y.-H., et al.\ 2012, ApJ, 755, 77
\bibitem[Treffers \& Chu(1982a)]{Treffers1982} Treffers, R.~R., \&
  Chu, Y.-H.\ 1982b, ApJ, 254, 569
\bibitem[Treffers \& Chu(1982b)]{Treffers1982a} Treffers, R.~R., \&
  Chu, Y.-H.\ 1982a, ApJ, 254, 132
\bibitem[van Buren \& McCray(1988)]{vanBuren1988} van Buren,
  D., \& McCray, R.\ 1988, ApJl, 329, L93
\bibitem[van der Hucht(2001)]{vdH2001} van der Hucht,
  K.~A.\ 2001, New A Rev., 45, 135
\bibitem[van Marle et al.(2005)]{vMarle2005} van Marle, A.~J., Langer,
  N., \& Garc{\'{\i}}a-Segura, G.\ 2005, A\&A, 444, 837
\bibitem[van Marle et al.(2007)]{vMarle2007} van Marle, A.~J., Langer,
  N., \& Garc{\'{\i}}a-Segura, G.\ 2007, A\&A, 469, 941
\bibitem[Wachter et al.(2010)]{Wachter2010} Wachter, S., Mauerhan,
  J.~C., Van Dyk, S.~D., et al.\ 2010, AJ, 139, 2330
\bibitem[Wachter et al.(2011)]{Wachter2011} Wachter, S.,
  Cohen, M., \& Leisawitz, D.\ 2011, Bulletin of the American
  Astronomical Society, 43, \#333.10
\bibitem[Wrigge et al.(1994)]{Wrigge1994} Wrigge, M.,
  Wendker, H.~J., \& Wisotzki, L.\ 1994, A\&A, 286, 219
\bibitem[Wrigge(1999)]{Wrigge1999} Wrigge, M.\ 1999, A\&A, 343, 599
\bibitem[Wrigge \& Wendker(2002)]{Wrigge2002} Wrigge, M., \&
  Wendker, H.~J.\ 2002, A\&A, 391, 287
\bibitem[Wrigge et al.(2005)]{Wrigge2005} Wrigge, M., Chu,
  Y.-H., Magnier, E.~A., \& Wendker, H.~J.\ 2005, ApJ, 633, 248
\bibitem[Wright et al.(2010)]{WISE2010} Wright, E.~L.,
  Eisenhardt, P.~R.~M., Mainzer, A.~K., et al.\ 2010, AJ, 140, 1868
\bibitem[Zhekov \& Park(2011)]{Zhekov2011} Zhekov, S.~A., \&
  Park, S.\ 2011, ApJ, 728, 135

\end{thebibliography}
\end{document}